\documentclass[prd,aps,nofootinbib,showpacs,floats,letterpaper,floatfix,eqsecnum]{revtex4}
\usepackage{dcolumn,epsfig}
\usepackage{amssymb,amsmath}
\def\be{\begin{equation}}
\def\ee{\end{equation}}
\def\beq{\begin{eqnarray}}
\def\eeq{\end{eqnarray}}

\def\IL{\relax{\rm I\kern-.18em L}}

\begin{document}

\title{Eccentric binary black-hole mergers: \\
       The transition from inspiral to plunge in general relativity}

\author{Ulrich Sperhake$^{1,}
\footnote{Electronic address: ulrich.sperhake@uni-jena.de}$,
Emanuele Berti$^{2,3}$, Vitor Cardoso$^{4,5}$, \\
Jos\'e A. Gonz{\'a}lez$^{1,6}$, Bernd Br{\"u}gmann$^{1}$, Marcus Ansorg$^{7}$\\}

\affiliation{${^1}$ Theoretisch Physikalisches Institut, Friedrich Schiller
  Universit\"at, 07743 Jena, Germany}
\affiliation{${^2}$ Jet Propulsion Laboratory, California Institute of
  Technology, Pasadena, CA 91109, USA}
\affiliation{${^3}$ McDonnell Center for the Space Sciences, Department of
  Physics, Washington University, St.~Louis, MR 63130, USA}
\affiliation{${^4}$ Department of Physics and Astronomy, The University of
  Mississippi, University, MS 38677-1848, USA}
\affiliation{${^5}$ Centro Multidisciplinar de Astrof\'{\i}sica - CENTRA,
  Departamento de F\'{\i}sica, Instituto Superior T\'ecnico, Av. Rovisco Pais
  1, 1049-001 Lisboa, Portugal}
\affiliation{${^6}$ Instituto de F\a'{\i}sica y Matem\a'aticas, Universidad
  Michoacana de San Nicol\a'as de Hidalgo, Edificio C-3, Cd. Universitaria. C.
  P. 58040 Morelia, Michoac\a'an, M\a'exico}
\affiliation{${^7}$ Max-Planck-Institut f\"ur Gravitationsphysik,
  Albert-Einstein-Institut, 14476 Golm, Germany}

\date{\today}

\begin{abstract}
  We study the transition from inspiral to plunge in general relativity by
  computing gravitational waveforms of non-spinning, equal-mass black-hole
  binaries. We consider three sequences of simulations, starting with a
  quasi-circular inspiral completing 1.5, 2.3 and 9.6 orbits, respectively,
  prior to coalescence of the holes. For each sequence, the binding energy of
  the system is kept constant and the orbital angular momentum is
  progressively reduced, producing orbits of increasing eccentricity and
  eventually a head-on collision. We analyze in detail the radiation of energy
  and angular momentum in gravitational waves, the contribution of different
  multipolar components and the final spin of the remnant, comparing numerical
  predictions with the post-Newtonian approximation and with extrapolations of
  point-particle results. We find that the motion transitions from inspiral to
  plunge when the orbital angular momentum $L=L_{\rm crit}\simeq 0.8M^2$. For
  $L<L_{\rm crit}$ the radiated energy drops very rapidly. Orbits with
  $L\simeq L_{\rm crit}$ produce our largest dimensionless Kerr parameter for
  the remnant, $j=J/M^2\simeq 0.724\pm0.13$ (to be compared with the Kerr
  parameter $j\simeq 0.69$ resulting from quasi-circular inspirals). This
  value is in good agreement with the value of $0.72$ reported in
  \cite{Hinder2007}.  These conclusions are quite insensitive to the initial
  separation of the holes, and they can be understood by extrapolating point
  particle results.  Generalizing a model recently proposed by Buonanno,
  Kidder and Lehner \cite{Buonanno:2007sv} to eccentric binaries, we
  conjecture that (1) $j\simeq 0.724$ is close to the maximal Kerr parameter
  that can be obtained by any merger of non-spinning holes, and (2) no binary
  merger (even if the binary members are extremal Kerr black holes with spins
  aligned to the orbital angular momentum, and the inspiral is highly
  eccentric) can violate the cosmic censorship conjecture.
\end{abstract}


%%%%%%%%%%%%%%%%%%%%%%%%%%%%%%%%%%%%%%%%%%%%%%%%%%%%%%%%%%%%%%%%%%%%%%%%%%%%%%%
\maketitle
%%%%%%%%%%%%%%%%%%%%%%%%%%%%%%%%%%%%%%%%%%%%%%%%%%%%%%%%%%%%%%%%%%%%%%%%%%%%%%%

%\tableofcontents

\section{Introduction}

The research area of gravitational wave (GW) physics has reached a very
exciting stage, both experimentally and theoretically.  Earth-based
laser-interferometric detectors, including LIGO \cite{Abramovici:1992ah},
GEO600 \cite{Luck:1997hv} and TAMA \cite{Ando:2001ej}, are collecting data at
design sensitivity, searching for GWs in the frequency range $\sim
10-10^3$~Hz. VIRGO \cite{Caron:1997hu} should reach design sensitivity within
one year, and the space-based interferometer LISA is expected to open an
observational window at low frequencies ($\sim 10^{-4}-10^{-1}$~Hz) within the
next decade \cite{Danzmann:1998}.

The last two years have also seen a remarkable breakthrough in the simulation
of the strongest expected GW sources, the inspiral and coalescence of
black-hole binaries \cite{Pretorius2005a,Campanelli2006, Baker2006}.  Several
groups have now generated independent numerical codes for such simulations
\cite{Pretorius2006, Baker2006a, Campanelli2006a, Sperhake2006, Scheel2006,
  Bruegmann2006a, Herrmann2007, Pollney2007, Etienne2007} and studied various
aspects of binary black hole mergers.
% including the resulting gravitational recoil
%\cite{Herrmann2007,Gonzalez2007,Koppitz2007,Gonzalez2007a,Campanelli2007v2,
%  Baker2007,Tichy2007,Herrmann2007c,Schnittman:2007ij,Lousto:2007db,Sopuerta:2006wj,Sopuerta:2006et}
%spin precession and spin flips \cite{Campanelli2007b,Campanelli2007v2},
In the context of analyzing the resulting gravitational waveforms, these
include in particular the comparisons of numerical results with post-Newtonian
(PN) predictions \cite{Baker2006c, Schnittman2007, Hannam2007, Buonanno2006,
  Berti:2007fi, Boyle2007}, multipolar analyses of the emitted radiation
\cite{Buonanno2006, Berti:2007fi, Schnittman:2007ij}, the use of numerical
waveforms in data analysis \cite{Ajith:2007qp, Pan2007, Vaishnav2007,
  Ajith:2007kx} and gravitational wave emission from systems of three black
holes \cite{Campanelli:2007ea}.

Despite this progress, a comprehensive analysis of binary black hole inspirals
remains a daunting task, mainly because of the large dimensionality of the
parameter space.  In geometrical units, the total mass of the binary is just
an overall scale factor. The source parameters to be explored by numerical
simulations (sometimes called ``intrinsic'' parameters in the GW data analysis
literature) include the mass ratio $q=M_2/M_1$, the eccentricity $e$ of the
orbit and six parameters for the magnitude of the individual black hole spins
and their direction with respect to the binary's orbital angular momentum.

In this paper we present results from numerical simulations of non-spinning,
equal-mass black-hole binaries, and we focus on the effect of the orbital
eccentricity on the merger waveforms. We consider three sequences, starting with
quasi-circular inspirals that complete $\sim 1.5$, $\sim 2.3$ and
$\sim 9.6$ orbits, respectively, prior to coalescence of the holes.
By fixing the binding energy
of the system and progressively reducing the orbital angular momentum, we
produce a sequence of orbits of increasing eccentricity and eventually a
head-on collision. For each of these simulations we analyze in detail the
radiation of energy and angular momentum in GWs, the contribution of different
multipolar components and the final spin of the remnant, comparing numerical
predictions with the PN approximation and with extrapolations of
point-particle results.

Non-eccentric inspirals are usually considered the most interesting cases for
GW detection. For an isolated binary evolving under the effect of
gravitational radiation reaction, the eccentricity decreases by roughly a
factor of 3 when the orbital semimajor axis is halved \cite{Peters:1963ux}.
For most conceivable formation mechanisms of solar-mass black hole binaries,
the orbit will usually be circular by the time the GW signal enters the
best-sensitivity bandwidth of Earth-based interferometers.
However, we wish to stress that our simulations could be of interest for GW
detection. For example, according to some astrophysical scenarios, eccentric
binaries may be potential GW sources for Earth-based detectors.  In globular
clusters, the inner binaries of hierarchical triplets undergoing Kozai
oscillations can merge under gravitational radiation reaction, and $\sim 30\%$
of these systems can have eccentricity $\sim 0.1$ when GWs enter the
detectors' most sensitive bandwidth at $\sim 10$~Hz \cite{Wen:2002km}.
%In globular clusters, binaries with essentially a thermal distribution of
%eccentricities may exist \cite{Benacquista2002}.
Massive black hole binaries to be observed by LISA could also have significant
eccentricity in the last year of inspiral.
%Analytical calculations and $N$-body simulations show that, in purely
%collisionless spherical backgrounds, the expected equilibrium distribution of
%eccentricities is skewed towards high $e\simeq 0.6-0.7$, and that dynamical
%friction does not play a major role in modifying such a distribution
%(\cite{1999ApJ...525..720C}, in particular Fig.~5).
Recent simulations using smoothed particle hydrodynamics follow the dynamics
of binary black holes in massive, rotationally supported circumnuclear discs
\cite{2006MNRAS.367..103D, 2007MNRAS.379..956D, Mayer:2007vk}. In these
simulations, a primary black hole is placed at the center of the disc and a
secondary black hole is set initially on an eccentric orbit in the disc plane.
By using the particle splitting technique, the most recent simulations follow
the binary's orbital decay down to distances $\sim 0.1$~pc. Dynamical friction
is found to circularize the orbit if the binary {\it corotates} with the disc
\cite{2007MNRAS.379..956D}.  However, if the orbit is {\it counterrotating}
with the disc the initial eccentricity does not seem to decrease, and black
holes may still enter the GW emission phase with high eccentricity
\cite{2006MNRAS.367..103D}.

Complementary studies show that eccentricity evolution may still occur in
later stages of the binary's life, because of close encounters with single
stars and/or gas-dynamical processes.  Three-body encounters with background
stars have been studied mainly in spherical backgrounds. These studies find
that stellar dynamical hardening can lead to an increase of the eccentricity,
acting against the circularization driven by the large-scale action of the
gaseous and/or stellar disc, possibly leaving the binary with non-zero
eccentricity when gravitational radiation reaction becomes dominant
\cite{1996NewA....1...35Q,Aarseth:2002ie,Berczik:2005ff,Berczik:2006tz,Matsubayashi:2005eg}.
It has also been suggested that the gravitational interaction of a binary with
a circumbinary gas disc could increase the binary's eccentricity. The
transition between disc-driven and GW-driven inspiral can occur at small
enough radii that a small but significant eccentricity survives, typical
values being $e\sim 0.02$ (with a lower limit $e\simeq 0.01$) one year prior
to merger (cf.  Fig.~5 of \cite{Armitage:2005xq}). If the binary has an
``extreme'' mass ratio $q\lesssim 0.02$ the residual eccentricity predicted by
this scenario can be considerably larger, $e\gtrsim 0.1$. Numerical
simulations should be able to test these predictions in the near future.  As
shown by Sopuerta, Yunes and Laguna, eccentricity could significantly increase
the recoil velocity resulting from the merger of non-spinning black-hole
binaries \cite{Sopuerta:2006et}.

Independently of the presence of eccentricity in astrophysical binary mergers,
the problem we consider here has considerable theoretical interest.  Our
simulations explore the transition between gravitational radiation from a
quasi-circular inspiral (the expected final outcome in most astrophysical
scenarios) and the radiation emitted by a head-on collision, where the binary
has maximal symmetry. Our work should provide some guidance for analytical
studies of the ``transition from inspiral to plunge''. The first analytical
study of this problem in the context of PN theory was carried out by Kidder,
Will and Wiseman \cite{Kidder:1993}. The transition between the adiabatic
phase and the plunge was studied in \cite{Buonanno2000} using nonperturbative
resummed estimates of the damping and conservative parts of the two-body
dynamics, i.e. the so-called ``Effective One Body'' (EOB) model.  Ori and
Thorne \cite{Ori:2000zn} provided a semi-analytical treatment of the
transition in the extreme mass ratio limit.  Waveforms comprising inspiral,
merger and ringdown for comparable-mass bodies have also been produced using
the EOB model (see eg.~\cite{Buonanno:2005xu} for extensions of the original
model to spinning binaries and for references to previous work).  Preliminary
comparisons of EOB and numerical relativity waveforms showed that improved
models of ringdown excitation \cite{Buonanno2006,Pan2007,Berti:2006wq} or
additional phenomenological terms in the EOB effective potential
\cite{Buonanno:2007pf} are needed to achieve acceptable phase differences
between the numerical and analytical waveforms.

Our study is complementary to Ref.~\cite{Pretorius:2007jn}, that considered
sequences of eccentric, equal-mass, non-spinning binary black hole evolutions
around the ``threshold of immediate merger'': a region of parameter space
separating binaries that quickly merge to form a final Kerr black hole from
those that do not merge in a short time. Similar scenarios have also been
studied in Ref.~\cite{Washik2008}, with particular regard to the maximal spin
of the final hole generated in this way. The universality of the gravitational
wave signal during the merger was analysed in Ref.~\cite{Hinder2007}, where it
was pointed out that binaries largely circularize after about 9 orbits when
starting with eccentricities below about $0.4$. The first comparison between
numerical evolutions of eccentric binaries with post-Newtonian predictions was
presented in \cite{Hinder2008}.

Our focus in this work is on the
high-eccentricity region of the parameter space, which always leads to merger.
In particular, the near-head-on limit of our study is of interest as a first
step to compute the energy loss and production cross-section of mini-black
holes in TeV-scale gravity scenarios (possibly at the upcoming LHC
\cite{Giddings:2007nr}), and trans-Planckian scattering in general
\cite{Dray:1984ha,Giddings:2007bw}.  Present semi-analytical techniques
(including a trapped surface search in the union of Aichelburg-Sexl shock
waves, close-limit approximation calculations and perturbation theory) only
give rough estimates of the emitted energy and production cross-section
\cite{Cardoso:2005jq} and do not provide much insight into the details of the
process (but see \cite{Sperhake:2008ga} for a first numerical investigation).

Our main finding is that, for all sequences we studied, the motion radically
changes character when the black holes'
% initial momentum $P\sim P_{\rm crit}\simeq 0.08-0.09M$ and the
orbital angular momentum $L\sim L_{\rm
  crit}\simeq 0.8 M^2$, turning from an eccentric inspiral into a plunge.  In
particular, for $L\lesssim L_{\rm crit}$ we observe that:

\begin{itemize}
\item The number of orbits $N_{\rm waves}$ (as estimated using the
  gravitational wave cycles) or $N_{\rm punc}$ (as computed from the
  punctures' trajectories) becomes less than one, so the motion effectively
  turns into a plunge (see Table \ref{tab: models} and Fig.~\ref{fig: traj}
  below);

\item The energy emission starts decreasing exponentially (Fig.~\ref{fig:
    El});

\item PN-based eccentricity estimates yield meaningless results (Table
  \ref{tab: models});

\item The polarization becomes linear rather than circular
      (Fig.~\ref{fig: pol});

\item The final angular momentum starts {\it decreasing}, rather than
  increasing, as $P$ and $L$ decrease (Fig.~\ref{fig: jfin}).
\end{itemize}

Binary mergers with $L\simeq L_{\rm crit}$ are those producing the largest
Kerr parameter for the final black hole observed in our simulations, $j_{\rm
  fin}\simeq 0.724$. One is led to suspect that for maximally spinning holes
having spins aligned with the orbital angular momentum, a large orbital
eccentricity may lead to violations of the cosmic censorship conjecture. Using
arguments based on the extrapolation of point-particle results (see also
\cite{Buonanno:2007sv}), we conjecture that (1) the maximal
Kerr parameter that can be obtained by any merger of non-spinning holes
is not much larger than $j\simeq 0.724$, and
(2) cosmic censorship will {\it not} be violated as a result of any merger,
even in the presence of orbital eccentricity. Further numerical simulations
are needed to confirm or disprove these conjectures.

The paper is organized as follows. We begin in Sec.~\ref{eccentricity}
discussing to what extent the Newtonian concept of eccentricity can be
generalized to characterize orbiting binaries in general relativity. For this
purpose, we introduce and compare various PN estimates of the orbital
eccentricity, and we show that these eccentricity estimates break down when
the motion turns from inspiral to plunge.  Sec.~\ref{numerics} is a brief
introduction to the numerical code used for the simulations. After a
discussion of the choice of initial data and of the code's accuracy, we show
how reducing the orbital angular momentum affects the gravitational waveforms,
the puncture trajectories and the polarization of the waves.  In
Sec.~\ref{energyj} we study the multipolar energy distribution of the
radiation and the angular momentum of the final Kerr black hole. In
Sec.~\ref{BKL} we show that the salient features of our simulations can be
understood using extrapolations of point-particle results.  Sec.~\ref{QNMs} is
devoted to fits of the ringdown waveform and to estimates of the energy
radiated in ringdown waves. We conclude by considering possible future
extensions of our investigation.

%\clearpage

%=============================================================================
\section{Post-Newtonian estimates of the eccentricity}
\label{eccentricity}

%Our two sequences of simulations start with a quasi-circular configuration
%completing $\sim 2.3$ and $1.5$ orbits respectively, prior to coalescence of
%the holes.
In Newtonian dynamics, the shape of a binary's orbital configuration is
determined by two parameters, the semi-major axis and
the eccentricity. These parameters are intimately tied to the
binding energy and orbital angular momentum of the binary and our
construction of sequences of binaries with increasing eccentricity
is based on this Newtonian intuition. Specifically, we fix
%Newtonian intuition tells us that, by fixing
the binding energy of
the system, progressively reduce the orbital angular momentum and thus
produce a sequence of orbits of increasing eccentricity.
Before doing so, however, we need to address a conceptual difficulty,
namely, how to quantify eccentricity in general relativity.

It turns out, unfortunately, that there exists no unique,
unambiguous definition of eccentricity in fully non-linear general relativity.
For this reason, in the following we will use PN arguments to quantify the
initial eccentricity (or rather, {\em eccentricities}) of the simulations. We
will consider in detail two different generalizations of the Newtonian
eccentricity: the 3PN extension \cite{Memmesheimer:2004cv} of a
quasi-Keplerian parametrization originally proposed by Damour and Deruelle
\cite{Damour:1985}, and a definition in terms of observable quantities
recently introduced by Mora and Will \cite{Mora:2003wt}.

\subsection{Quasi-Keplerian parametrization}

A quasi-Keplerian parametrization of eccentric orbits of objects with
mass $M_1$ and $M_2$ has been derived at 1PN
order in harmonic coordinates by Damour and Deruelle \cite{Damour:1985},
extended to 2PN order in ADM coordinates by Damour, Sch{\"a}fer and Wex
\cite{Damour:1988mr, Schaefer:1993} and completed to 3PN order by Memmesheimer
{\it et al.} \cite{Memmesheimer:2004cv}.
This 3PN parametrization gives the relative separation vector
${\bf r}=(r\sin~\phi, r\cos~\phi,0)$
of the compact objects and the mean anomaly
$l$ as
\begin{subequations}
\label{quasikepler}
\begin{eqnarray}
  r &=& a_r(1-e_r \cos u), \\
  \frac{2\pi (\phi-\phi_0)}{\Phi} &=& v + \left( f_{4\phi} + f_{6\phi} \right)
       \sin~2v + \left( g_{4\phi} + g_{6\phi}\right) \sin~3v \nonumber \\
        && + i_{6\phi} \sin 4v + h_{6\phi} \sin 5v, \\
  l \equiv \frac{2\pi(t-t_0)}{T} &=& u-e_t \sin~u
       + \left( g_{4t} + g_{6t} \right)(v-u) \nonumber \\
        &&+ \left( f_{4t} + f_{6t} \right) \sin~v
        + i_{6t} \sin 2v + h_{6t}\sin 3v,
\end{eqnarray}
\end{subequations}
where $u$ is the eccentric anomaly,
$v=2\arctan\{[(1+e_{\phi})/(1-e_{\phi})]^{1/2} \tan (u/2) \}$ and $T$ is the
orbital period. The key element in the parametrization, that makes it useful
for comparisons with numerical results, is that the auxiliary functions $a_r$,
$e_r$, $\Phi$, $f_{4\phi}$, $f_{6\phi}$, $g_{4\phi}$, $g_{6\phi}$,
$i_{6\phi}$, $h_{6\phi}$, $n=2\pi/T$, $e_t$, $g_{4t}$, $g_{6t}$, $f_{4t}$,
$f_{6t}$, $i_{6t}$, $h_{6t}$ and $e_{\phi}$ can be expressed exclusively in
terms of the binding energy $E_{\rm b}$, the total angular momentum $L$ and
the symmetric mass ratio $\eta$ of the binary system. The complete expressions
in terms of the dimensionless quantities $E \equiv E_{\rm b}/\mu$ and $h
\equiv L/(\mu M)$ are listed in Eqs.~(20) and Eqs.~(25) of
\cite{Memmesheimer:2004cv} for ADM-type and harmonic coordinates,
respectively. Here $M=M_1 + M_2$ and $\mu = M_1 M_2 / (M_1 + M_2)$ are
the total and reduced mass of the system, respectively.

A comparison with the Newtonian accurate Keplerian parametrization
\begin{subequations}
\label{kepler}
\begin{eqnarray}
  r &=& a(1-e\cos u), \\
  \phi-\phi_0 &=& 2\arctan \left[ \left(\frac{1+e}{1-e} \right)^{1/2}
      \tan\frac{u}{2} \right], \\
  l &=& u-e\sin u\,,
\end{eqnarray}
\end{subequations}
illustrates that the concept of eccentricity is much more complex in general
relativity and a single number, such as the Newtonian eccentricity $e$, no
longer suffices to parametrize the shape of the orbit. Nevertheless, the
similarity of the Newtonian and 3PN expressions suggest that the numbers
$e_t$, $e_r$ and $e_{\phi}$ represent some measure of the deviation of the
binary's orbit from quasi-circularity. This becomes particularly clear if we
plot these quantities as functions of the orbital angular momentum $L/M^2$ for
fixed binding energy $E_{\rm b}/M$ and mass ratio $\eta$.

\begin{figure}[t]
\centering
\includegraphics[height=7.2cm,angle=-90]{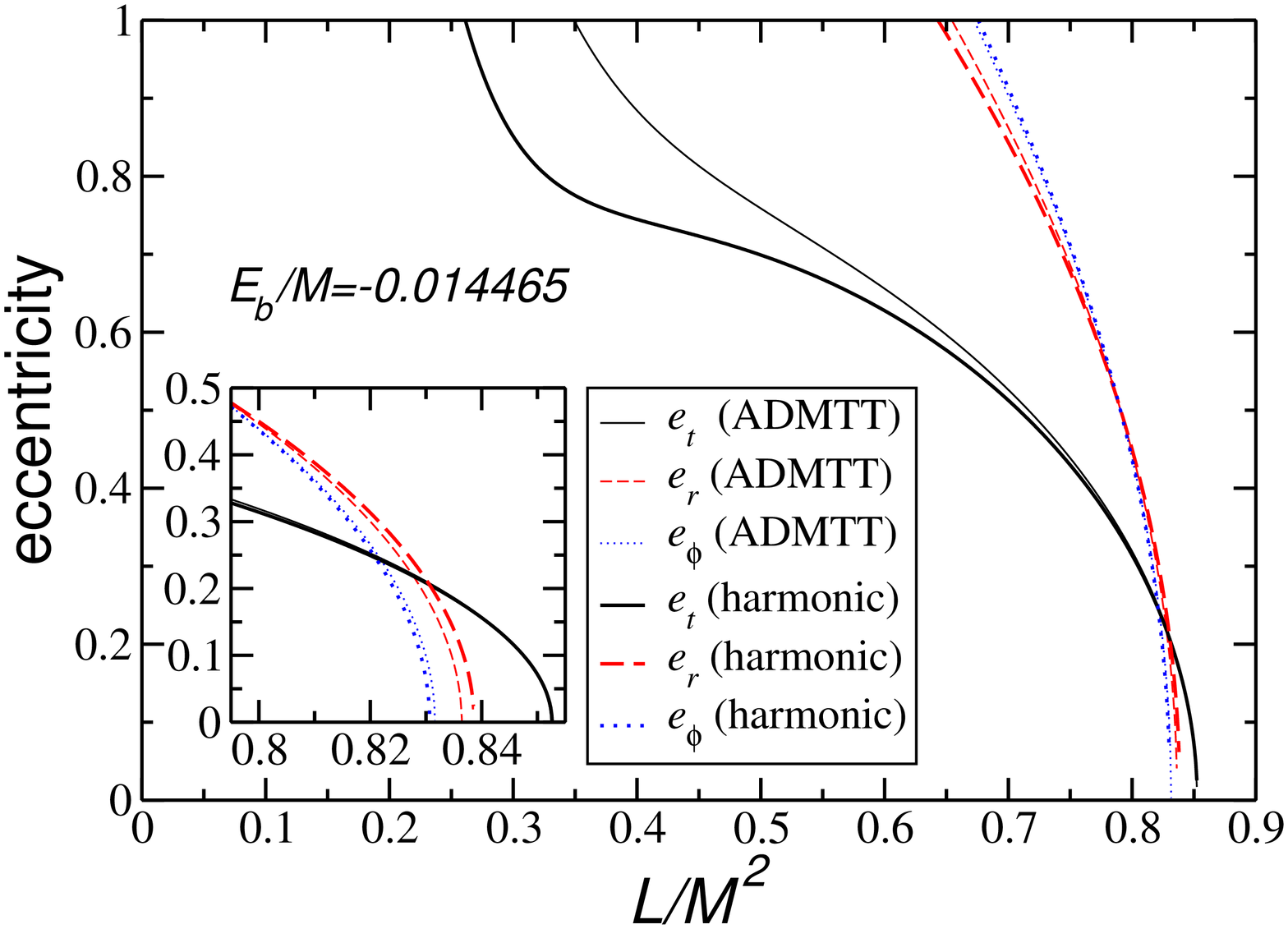}
\includegraphics[height=7.2cm,angle=-90]{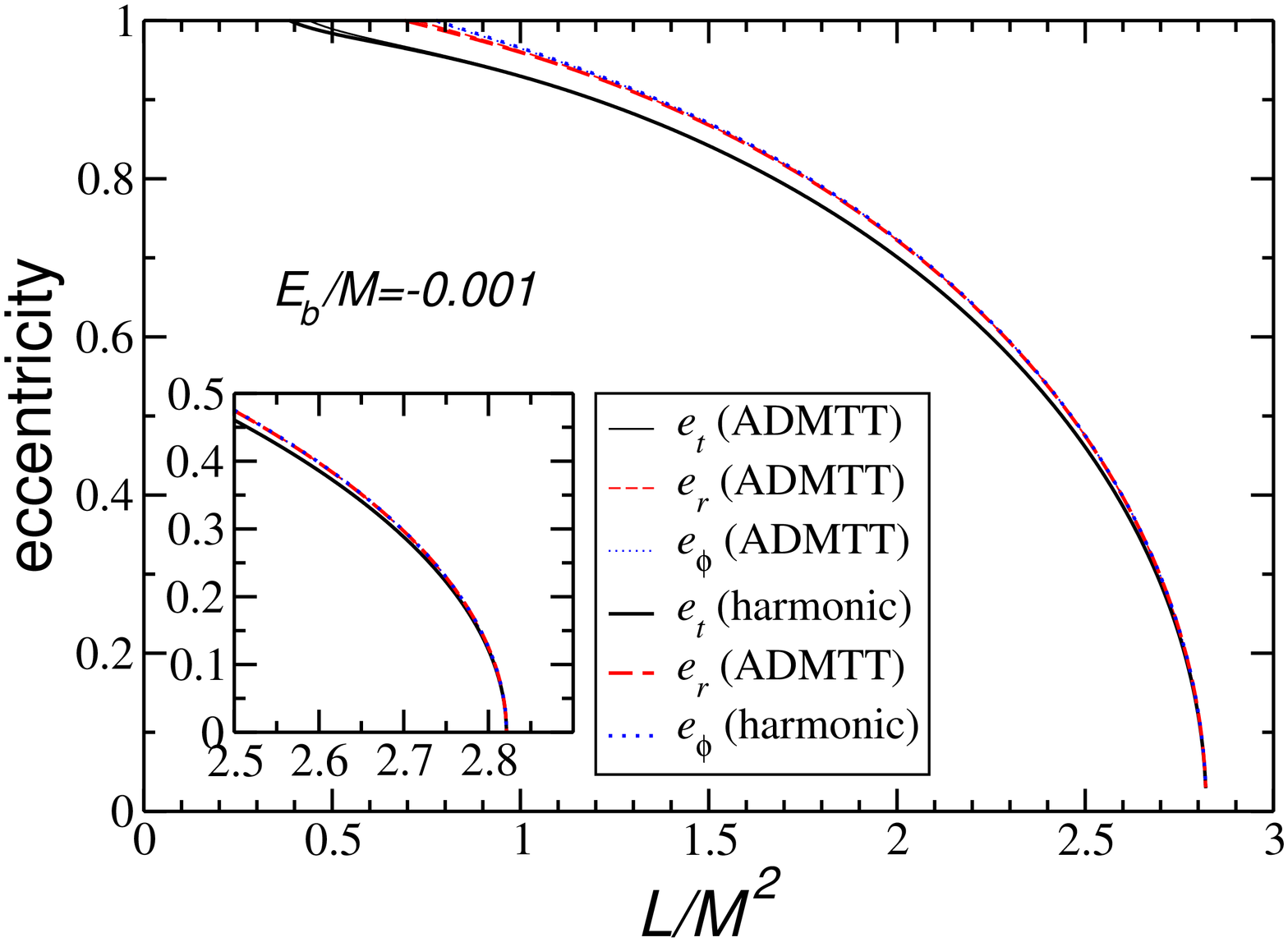}
\caption{The PN eccentricity parameters $e_t$, $e_r$ and $e_{\phi}$ for an
  equal mass binary with binding energy $E_{\rm b}$ are shown as
  functions of the orbital angular momentum $L/M^2$ for ADM-type coordinates
  and harmonic coordinates.  The left panel shows the result for a binding
  energy corresponding to our sequence 1, the right panel that obtained for a
  much smaller binding energy.  }
\label{fig: e_trphi}
\end{figure}

The result obtained for our sequence 1 models is shown in the left panel of
Fig.~\ref{fig: e_trphi}.  Several features of this plot are noteworthy.
First, all eccentricity parameters diverge in the limit of a head-on collision.
%(as opposed to the Newtonian $e$).
This is an artifact of the appearance of
$1/(-2Eh^2)$ terms in the PN expressions for $e_t$, $e_r$ and $e_{\phi}$ in
Eqs.~(20) and (25) of Ref.~\cite{Memmesheimer:2004cv}. We note that the limit
$L\rightarrow 0$ also plays a special role in the Newtonian case. The usual
distinction between the range $0\le e <1$, corresponding to bound elliptic
orbits, and $e\ge1$, corresponding to unbound parabolic or hyperbolic
trajectories, no longer applies in the case of vanishing angular momenta.
Since in Newtonian theory
\begin{equation}
e^2 = 1 + f(M_1,M_2) E_b L^2,
\end{equation}
where $f(M_1,M_2)$ is a function of the masses, in the head-on limit
we would formally have $e=1$, irrespective of the sign of the
binding energy.  Indeed such trajectories only have one degree of
freedom, the energy, and the concept of eccentricity no longer
applies. In this sense, it is not surprising that the PN formalism
fails to provide meaningful values for $e_t$, $e_r$ and $e_{\phi}$
in the head-on limit.

The second observation to be made is the steep gradient of all eccentricity
parameters close to the circular limit of vanishing $e_t$, $e_r$ and
$e_{\phi}$. The strong sensitivity of these parameters to the orbital angular
momentum $L$ results in finite values of the three eccentricities (of about
$0.1$) even when using quasi-circular parameters, as obtained from Eq.~(65) of
Ref.~\cite{Bruegmann2006a}.  Similarly, we observe that $e_t$, $e_r$ and
$e_{\phi}$ do not vanish for the same values of $L$ (see the inset in the left
panel of Fig.~\ref{fig: e_trphi}). Instead, the values of $e_t$ and $e_r$
corresponding to the orbital angular momentum where $e_{\phi}$ vanishes are of
the order of 0.1. A similar uncertainty results from comparing the PN values
obtained in harmonic and Arnowitt-Deser-Misner-Transverse-Traceless
(ADMTT) coordinates (cf.~the results in the two gauges
in the left panel).  We thus take $0.1$ as an approximate lower limit for
these eccentricity parameters obtainable for such relatively large binding
energies using the 3PN Keplerian parametrization. This is also approximately
the value of $e_t$, $e_r$ and $e_{\phi}$ obtained for the quasi-circular
configurations of Table \ref{tab: models}.

A third noteworthy feature of the ``quasi-Keplerian'' PN parametrization
(\ref{quasikepler}) is its breakdown for close, near-merger binary orbits. For
example, if we tried to compute $e_r$ and $e_\phi$ for the ``almost circular''
parameters we use in this paper (as listed in Table \ref{tab: models}) they
would turn out to be imaginary when, roughly speaking, $P/M\gtrsim 0.10$ (or
$L/M^2\gtrsim 0.83$).  This is easy to understand by looking at the inset of
the left panel of Fig.~\ref{fig: e_trphi}. There we see that these
eccentricities have a zero crossing for values of $L/M^2$ which are {\it
  smaller} than those specified in our quasi-circular simulations: in both
ADMTT and harmonic coordinates, for the specified value of the binding energy
$e_r$ goes to zero when $L/M^2 \simeq 0.84$, and $e_\phi$ goes to zero when
$L/M^2 \simeq 0.83$.  For $L/M^2$ larger than this ``critical'' value $e_r^2$
and $e_\phi^2$ become negative, so that $e_r$ and $e_\phi$ are imaginary.  In
the case of sequences 2 and 3, we observe the same behaviour.
% we similarly obtain negative $e_r^2$ and $e_{\phi}^2$
%at $L/M^2 \simeq 0.87$ and $0.86$ for the ADMTT and harmonic gauges,
%respectively.
This is just a sign that we should not trust the PN
approximation for these highly relativistic configurations, so our
eccentricity estimates should be taken with a grain of salt.

An eccentricity plot using the binding energy of sequence 2 or 3
would look almost
indistinguishable from the plot for sequence 1, as shown in the left panel of
Fig.~\ref{fig: e_trphi}, so we decided not to display them here.  Instead, in
the right panel of Fig.~\ref{fig: e_trphi} we show the eccentricities computed
for a much smaller binding energy, $E_{\rm b}/M=-0.001$. This binding energy
corresponds to a binary with much larger separation and smaller orbital
velocity, that should be described with much higher accuracy by the
quasi-Keplerian PN parametrization.  In fact, in the Newtonian limit the three
eccentricities should agree with each other, reducing to the Newtonian
definition at large separations. For example, to leading PN order $e_t$ and
$e_\phi$ are related by
\begin{equation}
e_\phi=e_t[1-(4-\eta)(2\pi M/T)^{2/3}],
\end{equation}
where $T$ is the orbital period (see eg.~\cite{Tessmer:2006fc}). The relation
between the different eccentricities at higher PN orders can be found in
Eq.~(21) of Ref.~\cite{Memmesheimer:2004cv}.

From the right panel of Fig.~\ref{fig: e_trphi} we see that the three
eccentricity parameters do agree much better, as expected, when the binary
members are far apart, and that differences resulting from the use of harmonic
or ADMTT coordinates become negligible. We still see the breakdown of the
formalism in the head-on limit.  However, now all six curves are much closer
to the expected Newtonian behavior, with vanishing eccentricity in the
circular limit (where $L$ approaches the maximum allowed value) and $e\approx
1$ for smaller angular momenta.  Unfortunately, it is currently prohibitively
costly from a computational point of view to start numerical simulations from
such low binding energies.  For this reason, in this paper we focus on the
merger and ringdown signals resulting from eccentric binaries, rather than on
detailed comparisons with PN predictions for the emission of GWs during the
inspiral.

\subsection{Mora-Will parametrization}

An alternative estimate of the binary's initial eccentricity can be obtained
using the PN diagnostic formalism developed by Mora and Will
(\cite{Mora:2003wt}, henceforth MW). Instead of imposing a quasi-Keplerian
parametrization with different eccentricities for $t$, $r$ and $\phi$, MW
define a {\it single} eccentricity parameter $e_{\rm MW}$ and a PN expansion
parameter $\zeta$ as follows:
\begin{eqnarray}
e_{\rm MW} &\equiv& \frac{ \sqrt{\Omega_p} - \sqrt{\Omega_a} }
    { \sqrt{\Omega_p} + \sqrt{\Omega_a} } \,,
\nonumber \\
\zeta &\equiv& \left ( \frac{\sqrt{M\Omega_p} +
\sqrt{M\Omega_a}}{2} \right )^{4/3} \,.
    \label{ezeta}
\end{eqnarray}
Here $\Omega_p$ and $\Omega_a$ are the orbital angular frequencies when the
binary passes through a local maximum (pericenter) and through the next local
minimum (apocenter), respectively.

The MW eccentricity parameter $e_{\rm MW}$ has several advantages: it is
defined in terms of observable quantities, it reduces to its Newtonian
counterpart for small orbital frequencies, and it is gauge invariant through
first PN order. The binding energy and angular momentum of the system can be
expressed as functions of $e_{\rm MW}$ and $\zeta$. The relevant equations for
black hole binaries are given by Eqs.~(2) and (3) in \cite{Berti:2006bj}.
In Ref.~\cite{Grigsby:2007fu} these same PN equations
have been used to study truly eccentric black hole binary initial data, and to
point out some interesting features of the resulting effective potentials.

The PN expansion parameter $\zeta$ is related to the frequencies at periastron
and apoastron by
\begin{eqnarray}\label{periapo}
\zeta=
\frac{(M\Omega_a)^{2/3}}{(1-e_{\rm MW})^{4/3}}=
\frac{(M\Omega_p)^{2/3}}{(1+e_{\rm MW})^{4/3}}.
\end{eqnarray}
To estimate the eccentricity, we can assume that the binary's orbit is (say)
at apoastron, so that $\Omega=\Omega_a$. Then the binding energy and angular
momentum can be expressed as functions of $e_{\rm MW}$ and $M\Omega$. We can
equate these functions to the binding energies and angular momenta listed in
Table \ref{tab: models}:
\begin{subequations}
\label{systemMW}
\begin{eqnarray}
\label{systemMWE}
E_b^{\rm MW}(e_{\rm MW},M\Omega)&=&E_b,\\
\label{systemMWL}
L^{\rm MW}(e_{\rm MW},M\Omega)&=&L.
\end{eqnarray}
\end{subequations}
If our assumption is correct, and the orbit is indeed at apoastron, this
system will have a solution $(e_{\rm MW},M\Omega)$ with eccentricity $e_{\rm
  MW}>0$. It should be obvious from Eq.~(\ref{periapo}) that a solution with
$e_{\rm MW}<0$ would simply correspond to the binary being at periastron.

In Ref.~\cite{Berti:2006bj} it was shown that, when comparing numerically
computed quasiequilibrium binary black hole initial data against the MW PN
diagnostic predictions {\it for a given orbital frequency $M\Omega$}, energies
are usually in better agreement than angular momenta. Unfortunately, our
initial data sets specify the orbital angular momentum and the binding energy,
but not the orbital frequency. Therefore, to estimate the binary's
eccentricity we adopt two different procedures. The first procedure is based
on numerically solving the system (\ref{systemMW}) for the two unknowns
$(e_{\rm MW},M\Omega)$, which are both considered as free parameters.  This
procedure is quite general.  However it can involve a significant amount of
systematic error, since the angular momentum can deviate by a large amount
from numerical initial data even for small eccentricities and/or large
separations, and we are trying to solve for both energy and angular momentum
simultaneously.

The second procedure is based on an assumption of quasi-circularity, and it
consists of two steps. We first assume that $e_{\rm MW}=0$ and solve
Eq.~(\ref{systemMWE}) for $M\Omega$. For sequence 1, the orbital frequency
obtained by imposing $E_b^{\rm MW}(e_{\rm MW}=0,M\Omega)=E_b$ is $M\Omega =
0.0513$; for sequence 2, we find $M\Omega=0.0431$; for sequence 3, we find
$M\Omega=0.0212$. Then we substitute this value of $M\Omega$ in
Eq.~(\ref{systemMWL}) and solve for the eccentricity. We call this solution
$e_{\rm MW,circ}$, because it is obtained assuming that the energy function is
very close to the circular prediction, and that deviations of the eccentricity
from zero only affect the angular momentum.

Table \ref{tab: models} lists the eccentricities $e_t$ obtained using the
definition by Memmesheimer {\it et al.} and the two eccentricities estimated
from the MW diagnostic. Dashed entries mean that no solutions to the system
(\ref{systemMW}) exist for physically relevant values of the parameters.  The
MW eccentricity estimate $e_{\rm MW}$ is consistent with the quasi-Keplerian
parameter $e_t$: in fact, the two definitions are roughly in agreement, within
factors of order unity. It is also encouraging that the minima of $e_t$ and
$e_{\rm MW}$ correspond to the same simulations along each sequence: these
variables seem to provide a reasonably accurate measure of deviations from
circularity, even for binaries which are very close to merger.  The second
procedure (by construction) should become inconsistent for large
eccentricities, but $e_{\rm MW,circ}$ is much closer to the eccentricity
estimates ($\sim 0.01$) obtained in \cite{Berti:2006bj}, and measured in
longer quasi-circular inspiral simulations by different groups
\cite{Husa2007,Pfeiffer:2007yz,Baker:2006kr,Buonanno2006}.

In the following we will use (somewhat arbitrarily) the parameter $e_t$
obtained in harmonic coordinates as a measure of deviations from circularity
of the initial binary configuration. We should still bear in mind that this
parameter deviates significantly from the eccentricity $e$ of a Keplerian
ellipse, in particular for small angular momenta.

%=============================================================================
\section{\label{numerics}Numerical simulations}

The simulations presented in this work have been performed with the {\sc Lean}
code \cite{Sperhake2006}, which is based on the {\sc Cactus} computational
toolkit \cite{Cactusweb}.  The code employs the so-called moving-puncture
method, i.e.~it evolves initial data of puncture type \cite{Brandt:1997tf}
using the Baumgarte-Shapiro-Shibata-Nakamura (BSSN) formulation of the
Einstein equations and gauge conditions which allow the black holes to move
across the computational domain.  In contrast to the original version of the
BSSN system, where the conformal factor $\psi$ is evolved in the form of the
logarithmic variable $\phi=\ln \psi$, we here evolve the variable
$\chi=e^{-4\phi}$,
%%
%\begin{equation}
%  \chi = e^{-4\phi},
%\end{equation}
%%
as originally introduced in Ref.~\cite{Campanelli2006}.  These equations are
numerically approximated using fourth-order spatial discretization
for sequences 1 and 2, and sixth-order spatial discretization
\cite{Husa:2007hp} for sequence 3.
Integration in time is performed with the method of lines using the fourth
order accurate Runge-Kutta (RK)
scheme. Using the notation of Table 1 in
Ref.~\cite{Sperhake2006}, this corresponds to schemes labelled RK${\chi_4}$
and RK${\chi_6}$, respectively. We note, however, that neither version
of the code is genuinely fourth or sixth order accurate, as second
order ingredients are used at the refinement and outer boundaries
(see \cite{Sperhake2006} for details).

The {\sc Lean} code facilitates use of mesh-refinement via the {\sc Carpet}
package \cite{Schnetter:2003rb, Carpetweb}.  Specifically, the computational
grid is represented by a nested set of Cartesian boxes with resolution
successively increasing by factors of two.  The innermost refinement levels
are typically split into two components centered around either hole and
following the black hole motion.  Once the black hole separation decreases
below a threshold value, the two components merge into one centered around the
origin.  A more detailed description of the {\sc Lean} code is given in
Ref.~\cite{Sperhake2006}.

GWs are extracted using the Newman-Penrose formalism.  The Newman-Penrose
scalar $\Psi_4$ is calculated using the electromagnetic decomposition of the
Weyl tensor, and decomposed into contributions of different multipoles using
spherical harmonics $_{-2}Y_{\ell m}$ of spin-weight $-2$ according to
\begin{equation}
  \Psi_4 = \sum_{\ell, m} \, _{-2}Y_{\ell m} \psi_{\ell m}.
\end{equation}
For all simulations presented in this work we extract GWs at different
extraction radii $r_{\rm ex}$. We use the results obtained at different radii
to estimate the uncertainty arising from the use of a finite extraction
radius.

The calculation of apparent horizons is performed using Thornburg's {\sc
  AHFinderDirect} \cite{Thornburg:1995cp, Thornburg2004} and plays an
important role in the construction of initial data sequences.

We use initial data of puncture type as provided by Ansorg's {\sc TwoPuncture}
thorn \cite{Ansorg:2004ds} in all simulations. For zero spins, an initial data
set is uniquely determined by the bare mass parameters $m_1$ and $m_2$ of the
holes, their initial coordinate separation $D$ and the Bowen-York
\cite{Bowen:1980yu} parameters ${\bf P}_1$ and ${\bf P}_2$ for the black
holes' linear momenta.  Without loss of generality we always set ${\bf
  P}\equiv {\bf P_1} =-{\bf P}_2$. The initial orbital angular momentum is
then given by ${\bf L}={\bf D}\times {\bf P}$.  Because the black holes are
initially located on the $x$-axis and orbit in the $xy$-plane, the initial
angular momentum is given by its $z$-component, which we can write as
$L=L_z=DP$ (where $P=P_y$).

We conclude this brief description of the numerical code with a summary of the
variables used in the remainder of this work. We denote by $M_1$ and $M_2$ the
initial black hole masses.
% which, in the absence of spins, coincide with the irreducible masses.
The total mass is the sum of the individual
masses, $M\equiv M_1+M_2$, and the reduced mass is $\mu\equiv M_1M_2/M$. We
measure the mass ratio using either $q=M_1/M_2$ or the ``symmetric mass
ratio'' parameter $\eta=\mu/M$ often used in PN studies.  Finally, we denote
the Arnowitt-Deser-Misner (ADM) mass by $M_{\rm ADM}$, which is close to (but
in general different from) the total black hole mass $M$.  The normalization
of dimensional quantities is performed as follows.  Initial parameters are
normalized using the total black hole mass $M$ for compatibility with the PN
relations.  In contrast, we measure radiated energy and momenta (as well as
time $t$ and radii $r$) in units of the ADM mass, as is common in numerical
studies.  All normalizations will be clear from the use of $M$ or $M_{\rm
  ADM}$ in column or axis labels; the relation between the two normalizations
can be worked out from Eq.~(\ref{Ebinding}) below and from the binding
energies of the two sequences we consider, which are given in the caption of
Table \ref{tab: models}.

\begin{table}[ht]
  \centering \caption{\label{tab: models}
    Sequences of models studied in this work. The irreducible mass of the
    individual black holes is $M_1=M_2=0.5$ in all simulations, and
    consequently $\mu=1/4$. The binding energy for sequence 1 (top 16
    models in the table) is $E_{\rm b}^{(1)}/M=-0.014465$, that for
    sequence 2 (following 8 models) is $E_{\rm b}^{(2)}/M=-0.013229$ and
    that for sequence 3 (following 9 models)
    is $E_{\rm b}^{(3)}=-0.008861$. The bottom
    line lists parameters for a model which serves for
    estimating the uncertainties of the long simulations of sequence 3
    (see Sec.~\ref{sec: accuracy})
    and has merely been included for completeness.
    $\Delta
    t_{\rm CAH}$ is the time of formation of a common apparent horizon as
    measured from the radiation peak, shifted by the extraction radius:
    $\Delta t_{\rm CAH}=t_{\rm CAH}+r_{\rm ex}-t_{\rm peak}$. For
    sequence 3, we have not calculated apparent horizons and we estimate
    this quantity to be $\Delta t_{\rm CAH}=-16~M_{\rm ADM}$. $N_{\rm waves}$
    and $N_{\rm punc}$ are the number of orbits obtained from the phase of the
    $(\ell=2~,m=2)$ multipole and the puncture's orbital motion,
    respectively. The last four columns list various PN estimates of the
    eccentricity (see Sec.~\ref{eccentricity} for a discussion).
  }
\begin{tabular}{cccc|ccc|ccc|cccc}
\hline  \hline
$\frac{P}{M}$ & $\frac{D}{M}$ & $\frac{L}{M^2}$ & $m_{1,2}$
& $\frac{10^2E_{\rm rad}}{M_{\rm ADM}}$ & $\frac{J_{\rm rad}}{M_{\rm ADM}^2}$
& $\frac{J_{\rm fin}}{M_{\rm fin}^2}$
& $\frac{\Delta t_{\rm CAH}}{M_{\rm ADM}}$ & $N_{\rm waves}$ & $N_{\rm punc}$
&$e_t$(ADMTT) &$e_t$(harm) &$e_{\rm MW} (M\Omega)$ &$e_{\rm MW,circ}$\\
\hline
0.14   & 6.054 & 0.8476  &  0.4772  & 3.437  & 0.2257  & 0.6915  &  -15.95  &  1.32 &  1.51 & 0.0983  & 0.0987    &0.1496 (0.0388)  &0.0198 \\
0.1383 & 6.131 & 0.8479  &  0.4775  & 3.451  & 0.2267  & 0.6910  &  -15.52  &  1.32 &  1.52 & 0.0956  & 0.0961    &0.1458 (0.0391)  &0.0188 \\
0.13   & 6.528 & 0.8486  &  0.4790  & 3.478  & 0.2272  & 0.6916  &  -15     &  1.33 &  1.53 & 0.0878  & 0.0886    &0.1353 (0.0398)  &0.0162 \\
0.12   & 7.064 & 0.8477  &  0.4808  & 3.494  & 0.2249  & 0.6932  &  -14.87  &  1.27 &  1.47 & 0.0977  & 0.0982    &0.1488 (0.0389)  &0.0196 \\
0.11   & 7.672 & 0.8439  &  0.4825  & 3.530  & 0.2196  & 0.6953  &  -15.74  &  1.16 &  1.37 & 0.1306  & 0.1300    &0.1948 (0.0359)  &0.0336 \\
0.10   & 8.361 & 0.8361  &  0.4842  & 3.486  & 0.2072  & 0.6994  &  -15.29  &  0.99 &  1.21 & 0.1800  & 0.1783    &0.2674 (0.0319)  &0.0634 \\
0.09   & 9.140 & 0.8226  &  0.4857  & 3.261  & 0.1857  & 0.7043  &  -14.65  &  0.82 &  1.04 & 0.2424  & 0.2393    &-                &0.1194 \\
0.08   &10.013 & 0.8010  &  0.4872  & 2.839  & 0.1570  & 0.7051  &  -15.30  &  0.68 &  0.88 & 0.3161  & 0.3112    &-                &0.2370 \\
0.07   &10.978 & 0.7685  &  0.4885  & 2.259  & 0.1240  & 0.6964  &  -16.59  &  0.56 &  0.72 & 0.3995  & 0.3921    &-                &-      \\
0.06   &12.024 & 0.7215  &  0.4898  & 1.631  & 0.0909  & 0.6718  &  -16.53  &  0.43 &  0.59 & 0.4913  & 0.4799    &-                &-      \\
0.05   &13.121 & 0.6561  &  0.4908  & 1.061  & 0.0613  & 0.6257  &  -17.01  &  0.37 &  0.47 & 0.5887  & 0.5699    &-                &-      \\
0.04   &14.217 & 0.5687  &  0.4917  & 0.621  & 0.0379  & 0.5530  &  -17.67  &  0.31 &  0.36 & 0.6894  & 0.6540    &-                &-      \\
0.03   &15.234 & 0.4570  &  0.4924  & 0.328  & 0.0213  & 0.4510  &  -18.24  &  0.28 &  0.27 & 0.8050  & 0.7202    &-                &-      \\
0.02   &16.073 & 0.3214  &  0.4930  & 0.161  & 0.0106  & 0.3206  &  -19.11  &  0.24 &  0.17 & 1.1005  & 0.8090    &-                &-      \\
0.01   &16.630 & 0.1663  &  0.4933  & 0.080  & 0.0042  & 0.1669  &  -20.02  &  0.18 &  0.09 & 3.8101  & 2.5892    &-                &-      \\
0      &16.826 & 0       &  0.4934  & 0.057  & 0       & 0       &  -19.80  &  -    &  -    & $\infty$& $\infty$  &-                &-      \\
\hline
0.1247 & 7.000 & 0.8729  &  0.4798  & 3.585  & 0.2540  & 0.6912  &  -12.74  &  2.06 &  2.30 & 0.1095  & 0.1096    &0.1529 (0.0323)  &0.0202 \\
0.12   & 7.278 & 0.8734  &  0.4807  & 3.581  & 0.2558  & 0.6897  &  -15.19  &  2.05 &  2.24 & 0.1049  & 0.1052    &0.1468 (0.0326)  &0.0186 \\
0.10   & 8.678 & 0.8678  &  0.4841  & 3.659  & 0.2532  & 0.6875  &  -15.15  &  1.79 &  1.99 & 0.1480  & 0.1472    &0.2049 (0.0295)  &0.0363 \\
0.08   &10.493 & 0.8394  &  0.4871  & 3.602  & 0.2066  & 0.7053  &  -16     &  1.00 &  1.20 & 0.2758  & 0.2725    &-                &0.1320 \\
0.06   &12.754 & 0.7653  &  0.4897  & 2.181  & 0.1186  & 0.6974  &  -16.14  &  0.56 &  0.74 & 0.4567  & 0.4485    &-                &-      \\
0.04   &15.288 & 0.6115  &  0.4917  & 0.795  & 0.0471  & 0.5902  &  -17     &  0.27 &  0.46 & 0.6681  & 0.6428    &-                &-      \\
0.02   &17.488 & 0.3498  &  0.4930  & 0.185  & 0.0122  & 0.3378  &  -18.87  &  0.27 &  0.44 & 1.0122  & 0.8078    &-                &-      \\
0      &18.398 & 0       &  0.4934  & 0.058  & 0       & 0       &  -20.16  &  -    &  -    & $\infty$& $\infty$  &-                &-      \\
\hline
0.0850 &12.000 & 1.0201  &  0.4883  & 3.914  & 0.3956  & 0.6963  &  $-16$   &  9.39 &  9.64 & 0.0869  & 0.0871    &0.1032 (0.0174)  &0.0086 \\
0.08   &12.758 & 1.0206  &  0.4890  & 3.907  & 0.3963  & 0.6960  &  $-16$   &  9.42 &  9.65 & 0.0819  & 0.0821    &0.0972 (0.0176)  &0.0077 \\
0.07   &14.441 & 1.0290  &  0.4905  & 3.912  & 0.3854  & 0.6971  &  $-16$   &  8.56 &  8.79 & 0.1542  & 0.1536    &0.1829 (0.0150)  &0.0272 \\
0.06   &16.398 & 1.0015  &  0.4917  & 3.996  & 0.3658  & 0.6898  &  $-16$   &  6.38 &  6.60 & 0.2648  & 0.2632    &0.3151 (0.0119)  &0.0815 \\
0.05   &18.619 & 0.9476  &  0.4929  & 4.186  & 0.3128  & 0.6914  &  $-16$   &  3.24 &  3.44 & 0.3968  & 0.3937    &0.4765 (0.0090)  &0.1899 \\
0.04   &21.028 & 0.8562  &  0.4938  & 3.373  & 0.1778  & 0.7236  &  $-16$   &  0.83 &  1.06 & 0.5425  & 0.5368    &-                &0.3960 \\
0.03   &23.461 & 0.7165  &  0.4946  & 1.295  & 0.0721  & 0.6602  &  $-16$   &  0.45 &  0.65 & 0.6921  & 0.6799    &-                &-      \\
0.02   &25.628 & 0.5218  &  0.4951  & 0.419  & 0.0263  & 0.4992  &  $-16$   &  0.39 &  0.42 & 0.8434  & 0.8048    &-                &-      \\
0.00   &27.710 & 0       &  0.4956  & 0.0585 & 0       & 0       &  $-16$   &  -    &  -    & $\infty$& $\infty$  &-                &-      \\
\hline
0.0808 &10.905 & 0.8807  &  0.9834, & 3.084  & 0.3037  & 0.6304  &  $-16$   &  8.10 &  8.41 & 0.1022  & 0.1022    &-                &-      \\
       &       &         &  0.4822  &        &         &         &          &       &       &         &           &                 &       \\
\hline \hline
\end{tabular}
\end{table}

\subsection{Sequence of initial data}

The construction of a sequence of non-spinning, equal-mass
binary initial configurations with
constant binding energy is a two-step process. First, we determine
a quasi-circular configuration using the 3PN accurate relation
(65) of Ref.~\cite{Bruegmann2006a}, which provides the initial linear
momentum $P/M$ for given black hole masses $M_1$ and $M_2$ and
separation $D/M$. Without loss of generality we fix the scaling
freedom of the numerical coordinates by setting $M_1=M_2=0.5$.
For fixed values of $P$ and $D$, there thus remains the task of
finding bare mass parameters $m_1=m_2$ which result in black-hole
masses of $M_1=M_2=0.5$. This is done iteratively
by using the Newton-Raphson method with initial guesses $m_i=M_i$.
In the absence of spins, the black-hole masses are given by
the irreducible masses, as calculated by the apparent horizon finder
{\sc AHFinderDirect}. For the quasi-circular model we
calculate the binding energy according to
\begin{equation}
  E_{\rm b} = M_{\rm ADM} - M.
  \label{Ebinding}
\end{equation}

The second step consists in the construction of additional configurations
with the same binding energy but different linear momentum parameter $P$
and, thus, different orbital angular momentum $L=DP$. For this purpose
we fix $P$ and demand, as before, $M_1=M_2=0.5$. We then iteratively
solve for the numerical parameters $m_1=m_2$ and $D$ which yield the
required black hole masses and binding energy.

%For our purposes, a convenient way to characterize an initial binary
%configuration is to consider the total ADM mass $M_{\rm ADM}$ of the system,
%the mass ratio $q$, the binding energy $E_{\rm b}$ and the orbital angular
%momentum $L$. These quantities are functions of the free initial parameters,
%the bare masses $m_1$, $m_2$, the linear momentum parameter $P$ and the
%coordinate separation $D$ of the punctures. Except for the angular momentum
%$L=DP$, the relation between the various quantities is complicated and not
%analytically known.  Determining the free initial parameters for given values
%of $M_{\rm ADM}$, $q$, $L$ and $E_{\rm b}$ therefore requires iterative
%solving.
%
%In practice this is performed as follows. Without loss of generality, we
%eliminate one function by choosing to fix the linear momentum parameter $P$
%instead of $L$. We then specify suitable initial guesses for the parameters
%$m_1$, $m_2$ and $D$ and evaluate $M_{\rm ADM}$, $M_1$ and $M_2$ from the
%spectral initial data solver and the horizon finder. These in turn yield the
%binding energy via
%Using the Newton-Raphson method we iteratively determine the values of $m_1$,
%$m_2$ and $D$ corresponding to the requested binding energy and ADM mass.  A
%sequence of models is then constructed by keeping the binding energy $E_{\rm
%  b}$ constant while varying the orbital angular momentum $L$.
Three sequences obtained this way
with $E_{\rm b}/M=-0.014465$, $-0.013229$ and $-0.008861$ are listed in Table
\ref{tab: models}. In the following we will refer to them as ``sequence 1'',
``sequence 2'' and ``sequence 3'',
respectively. The corresponding quasi-circular configurations
are those with linear momentum $P/M=0.1383$, $0.1247$ and $0.0850$.
% For either sequence we calculated the linear
%momentum corresponding to a quasi-circular configuration at 3PN order in the
%ADMTT gauge according to Eq.~(65) of Ref.~\cite{Bruegmann2006a}, resulting in
%$P/M=0.1383$ for sequence 1 and $P/M=0.1247$ for sequence 2.

It is worth mentioning, in this context, that our choice of fixing the
binding energy is not unique. Alternative choices in constructing
sequences include keeping constant the coordinate separations of the
punctures, or the proper horizon-to-horizon distance. We have opted
against these two options because they would result in very
short merger times for small angular momenta. At constant binding
energy, instead, smaller values of the orbital angular momentum imply
larger separations of the holes and, thus, a delay in the formation of
the common apparent horizon.

\subsection{Accuracy of the simulations}
\label{sec: accuracy}

In order to estimate the uncertainties associated with the numerically
calculated quantities of sequences 1 and 2,
we have performed a convergence analysis for the
quasi-circular configuration with $P/M=0.1247$ of sequence 2, and for the
eccentric configuration with $P/M=0.08$ of sequence 1.  We have evolved these
systems using the following grid setup:
%
%\begin{eqnarray}
$  \left\{ \left( 192,96,56,24,12\right) \times \left( 3,1.5,0.75 \right),
          h_i \right\}. \nonumber
$
%\end{eqnarray}
%
By this notation we mean that there is a total of 8 refinement levels. The 5
outer levels are centered on the origin and extend out to $x=y=z=\pm 192$,
$96$, $56$, $24$ and $12$ respectively. The 3 inner levels have 2 components
each, centered on either hole with radius $3$, $1.5$ and $0.75$. Finally, the
grid spacing is $h_i$ on the finest level (where $h_1=1/48$, $h_2=1/44$ and
$h_3=1/40$) and increases by a factor of 2 consecutively on each level.

\begin{figure}[hbt]
\centering
\includegraphics[height=7.2cm,angle=-90]{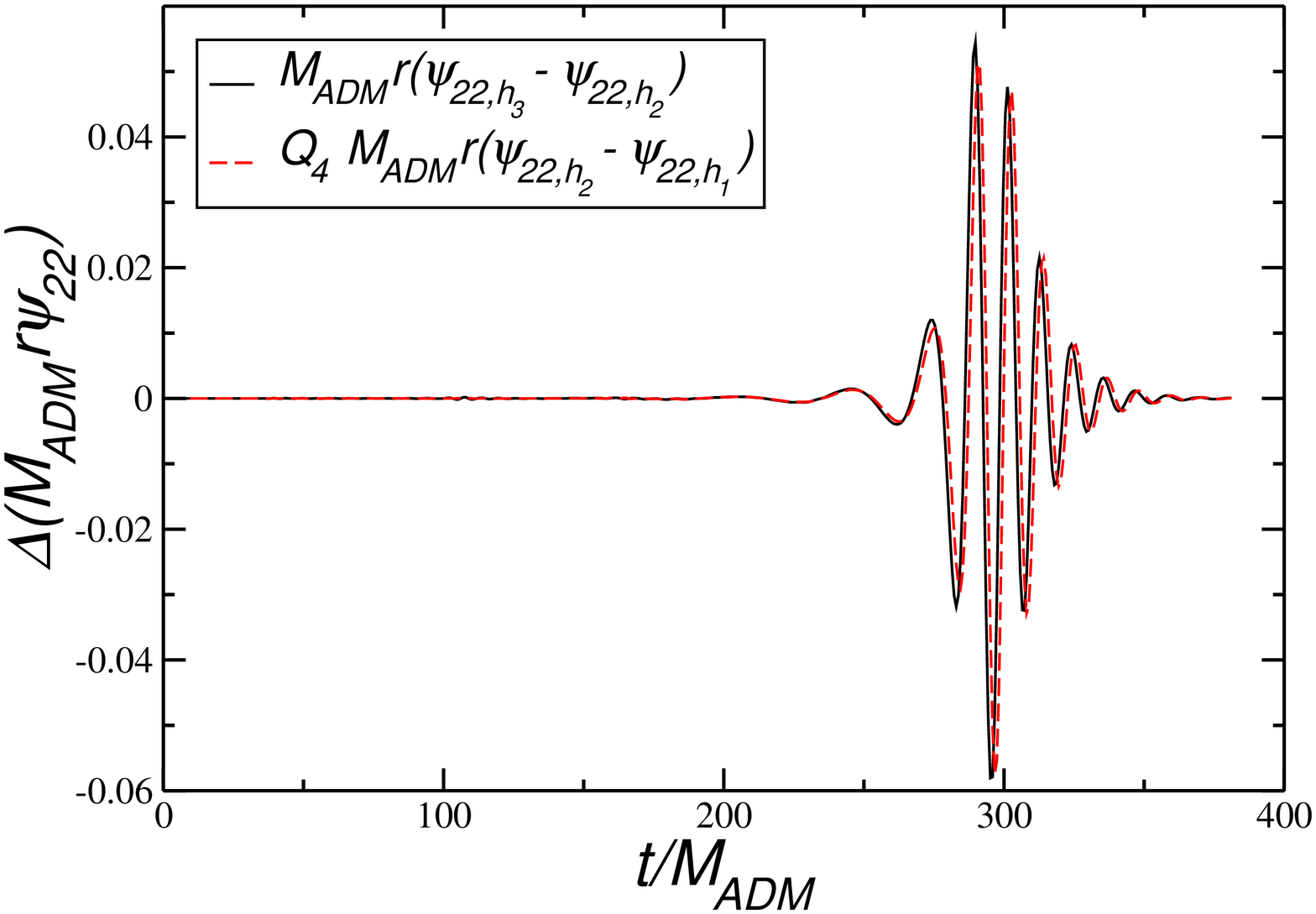}
\includegraphics[height=7.2cm,angle=-90]{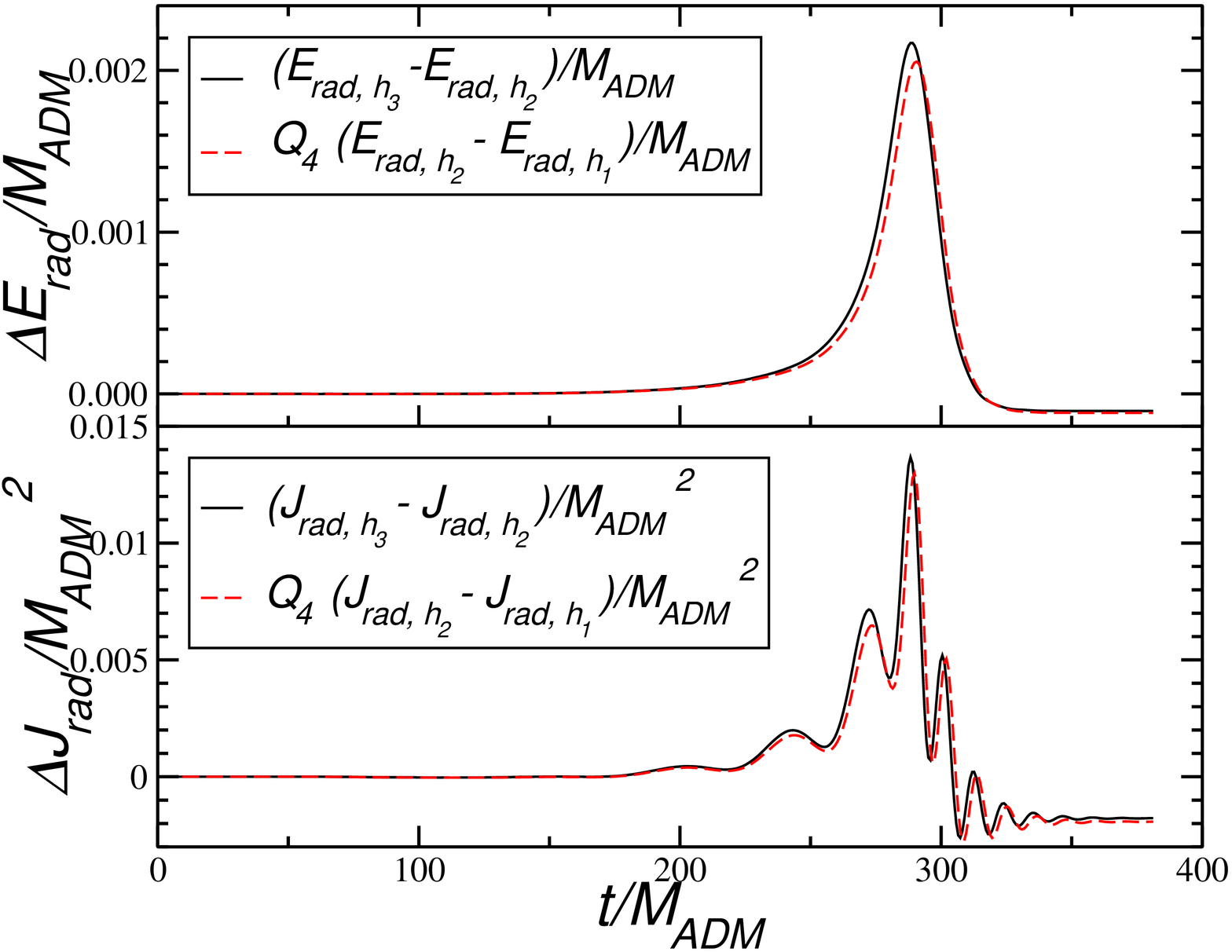}
\includegraphics[height=7.2cm,angle=-90]{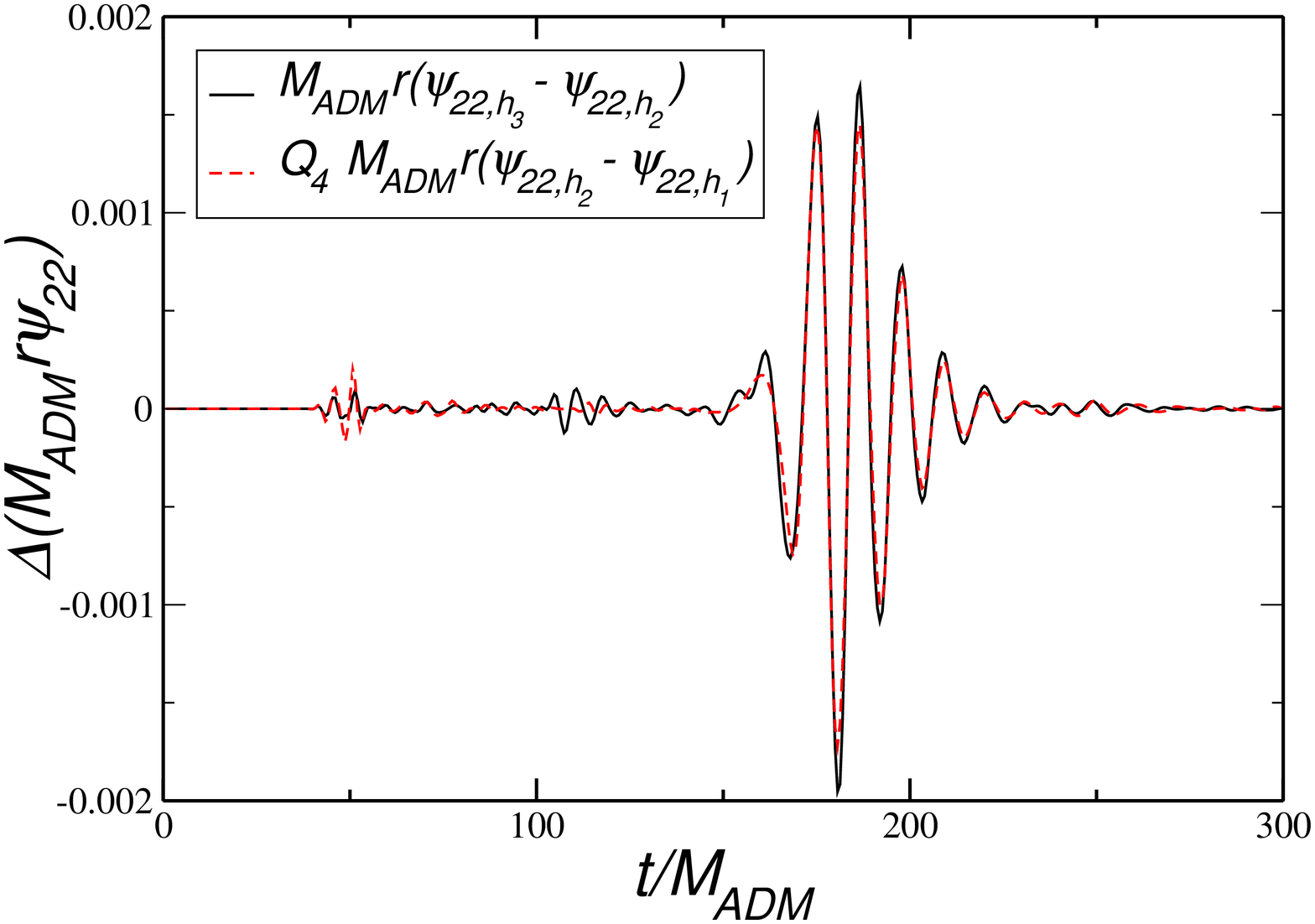}
\includegraphics[height=7.2cm,angle=-90]{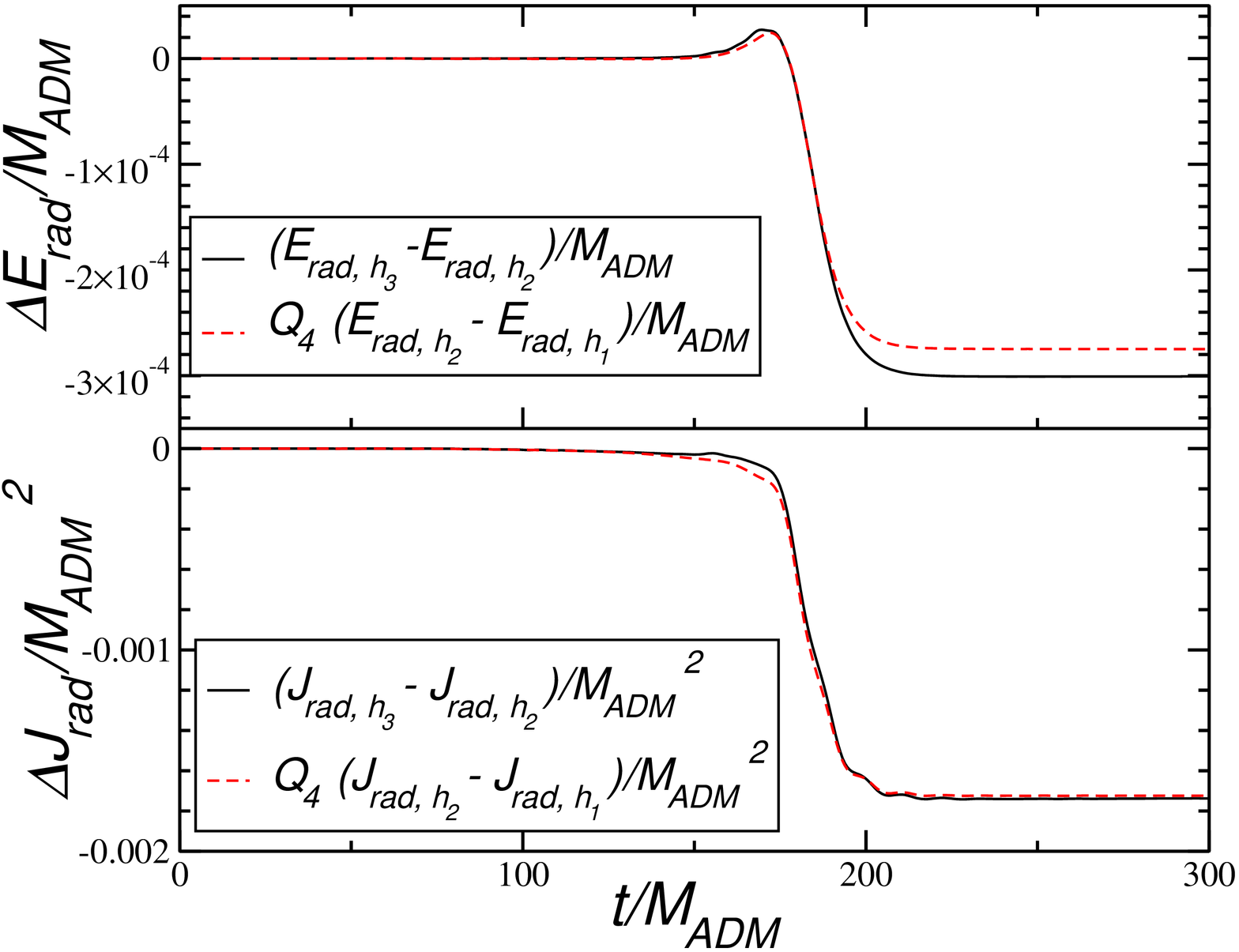}
\includegraphics[height=7.2cm,angle=-90]{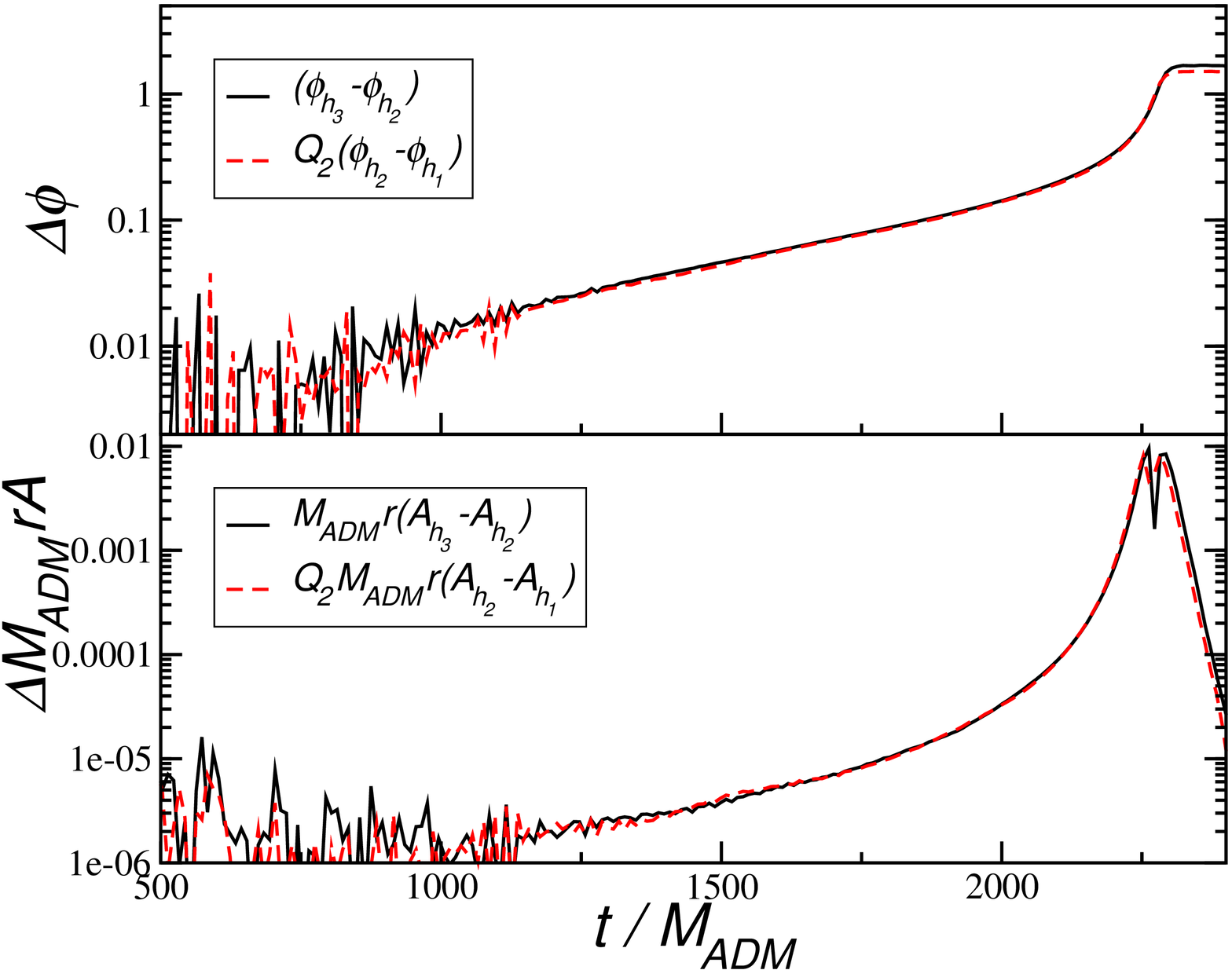}
\includegraphics[height=7.2cm,angle=-90]{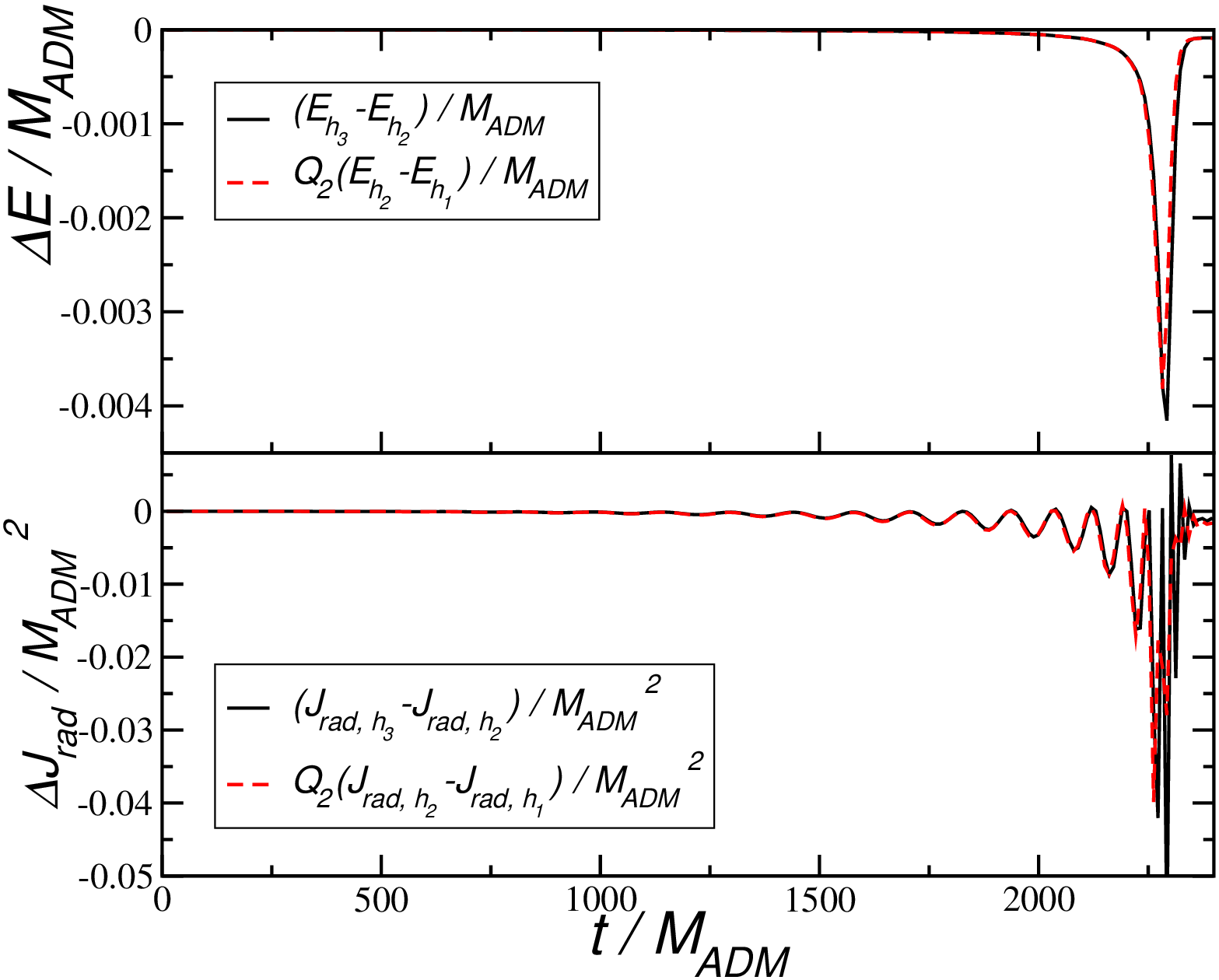}
\caption{Upper panels: convergence analysis of the $P/M=0.1247$,
  quasi-circular model of sequence 2, using resolutions $h=1/48$, $1/44$ and
  $1/40$.  The left panel shows the differences in the $(\ell=2,~m=2)$
  multipole of $M_{\rm ADM}r\psi_{22}$, the right panel the total radiated
  energy and $z$-component of the angular momentum. In both cases, the
  differences between the higher resolution simulations have been rescaled
  by $Q_4$ for
  the expected fourth order convergence. Middle panels: same, but for the
  $P/M=0.08$, eccentric model of sequence 1. Lower panels: same, but for
  the unequal mass model listed at the bottom of Table \ref{tab: models}.
  For clarity, we present the convergence of $\Psi_4$ using phase and
  amplitude instead of the real part. For this configuration, we
  observe second order convergence. The corresponding convergence
  factors for the resolutions used here, are $Q_4=1.58$ and $Q_2=0.72$.
  %{\bf [TODO: Uli, what is the value of $Q_2$ and $Q_4$? Define here.]}
}
\label{fig: convergence}
\end{figure}

Fourth-order convergence is shown in Fig.~\ref{fig: convergence} for the
$(\ell=2,~m=2)$ multipole of the Newman-Penrose scalar $\Psi_4$, the total
radiated energy $E$ and the radiated angular momentum in the $z$-direction
$J_{\rm rad}$.

Using the fourth order convergence, we apply a Richardson extrapolation to the
total radiated energy and obtain $E/M_{\rm ADM}=0.03686$, $0.03668$ and
$0.03656$ respectively at extraction radii $r_{\rm ex}/M_{\rm ADM}=50.7$,
$60.8$ and $70.9$.
% {\bf [Uli:
%everything still needs rescaling with Madm = 0.986770548344399]}.
These values correspond very well to a $1/r_{\rm ex}$ fall-off of the
uncertainty arising from the use of finite extraction radii. The total
radiated energy extrapolated to $r_{\rm ex}\rightarrow \infty$ is $E/M_{\rm
  ADM}=0.03583$.  For the medium resolution case ($h=h_2=1/44$) and using an
extraction radius $r_{\rm ex}/M_{\rm ADM}=70.9$, this analysis predicts an
uncertainty $\sim 2\%$ due to the discretization and $\sim 2.5\%$ due to the
use of finite extraction radius. The uncertainties in the radiated angular
momentum $J_{\rm rad}$ are $\sim 2\%$ for both error sources. The convergence
study of the eccentric simulation with $P/M=0.08$ of sequence 1 yields similar
error estimates. We estimate the resulting total error from quadratic error
propagation to be about $3\%$. In fact, this estimate is likely to be very
conservative because the two error sources have opposite signs: finite
resolution tends to underestimate radiated energy and momenta, while finite
extraction radius usually leads to an overestimate.

Performing a convergence analysis of simulations lasting as long as those
of sequence 3 requires vast computational resources. In order to reduce
the cost, we view these simulations as part of a wider parameter study,
to be presented elsewhere, which also involves unequal-mass binaries.
In order to assess the accuracy of those long simulations, we have focussed
on a quasi-circular configuration with $q=2$ which is listed as the last entry
in Table \ref{tab: models}. From experience, we consider unequal-mass
binaries significantly more challenging numerically than systems with
equal mass, and therefore feel justified in using the uncertainties
resulting from this model as conservative error estimates of our
sequence 3 runs.
We have evolved this configuration using a grid
$\{(384,192,128,48,24)\times (6,3,1.5,0.75),h_i\}$ with
$h_i=1/44$, $1/48$ and $1/56$.
The bottom panels of Fig.~\ref{fig: convergence} show the convergence
of the phase and amplitude of $\Psi_4$ (bottom right panel) and
the radiated energy and angular momentum (bottom left panel).
The difference between the higher resolution runs has been
rescaled for second order convergence. We observe
second order convergence for these runs except for the late stages
near merger, when the convergence increases to third order.
Similar glitches in the convergence near merger have been observed
in \cite{Hannam2007}. Using the
same technique as above, we obtain uncertainties of about $3~\%$ for
the radiated energy and $6~\%$ for the radiated angular momentum
for the simulations using $h_2=1/48$ and $r_{\rm ex}=80.7~M_{\rm ADM}$.

Because most simulations have been performed at the medium resolution only, we
cannot in general apply Richardson extrapolation.  Unless specified otherwise
we therefore present results as obtained numerically using $h=1/44$ and
$r_{\rm ex}/M_{\rm ADM}=70.9$ for sequences 1 and 2, as well as
$h=1/48$ and $r_{\rm ex}=120~M$ for sequence 3,
bearing in mind the uncertainties we have just
mentioned.
%This applies, in particular, to the list of models in Table \ref{tab: models}.

We conclude this discussion
by mentioning that the {\sc Lean} code has been demonstrated to
accurately evolve black holes with large spins in
Ref.~\cite{Marronetti:2007wz}. Specifically, methods to calculate the
spin from quasinormal ringing, balance arguments and apparent
horizon data were shown to result in excellent agreement for
Kerr parameters above $0.9$.

\subsection{Numerical waveforms}

In Ref.~\cite{Berti:2007fi} we studied the multipolar energy distribution of
unequal-mass black hole binaries in quasi-circular orbits. By projecting
$2.5$PN calculations of the inspiral gravitational waveforms onto
spin-weighted spherical harmonics $_{-2}Y_{\ell m}$, we concluded that odd-$m$
multipoles of the radiation are suppressed for equal-mass binaries. Low-$\ell$
multipoles carry most of the radiation, and within a given $l$-multiplet,
modes with $\ell=m$ are typically dominant for quasi-circular motion. This
analytical prediction was shown to agree very well with numerical simulations,
and it has recently been confirmed by more accurate PN calculations
\cite{Kidder:2007rt}. Since in this paper we study equal-mass runs, we expect
even-$m$, low-$\ell$ multipoles to be dominant, at least when the eccentricity
is small enough. This expectation is again confirmed by our numerical
time-evolutions.

\begin{figure}[ht]
\centering
\includegraphics[height=7.2cm,angle=-90]{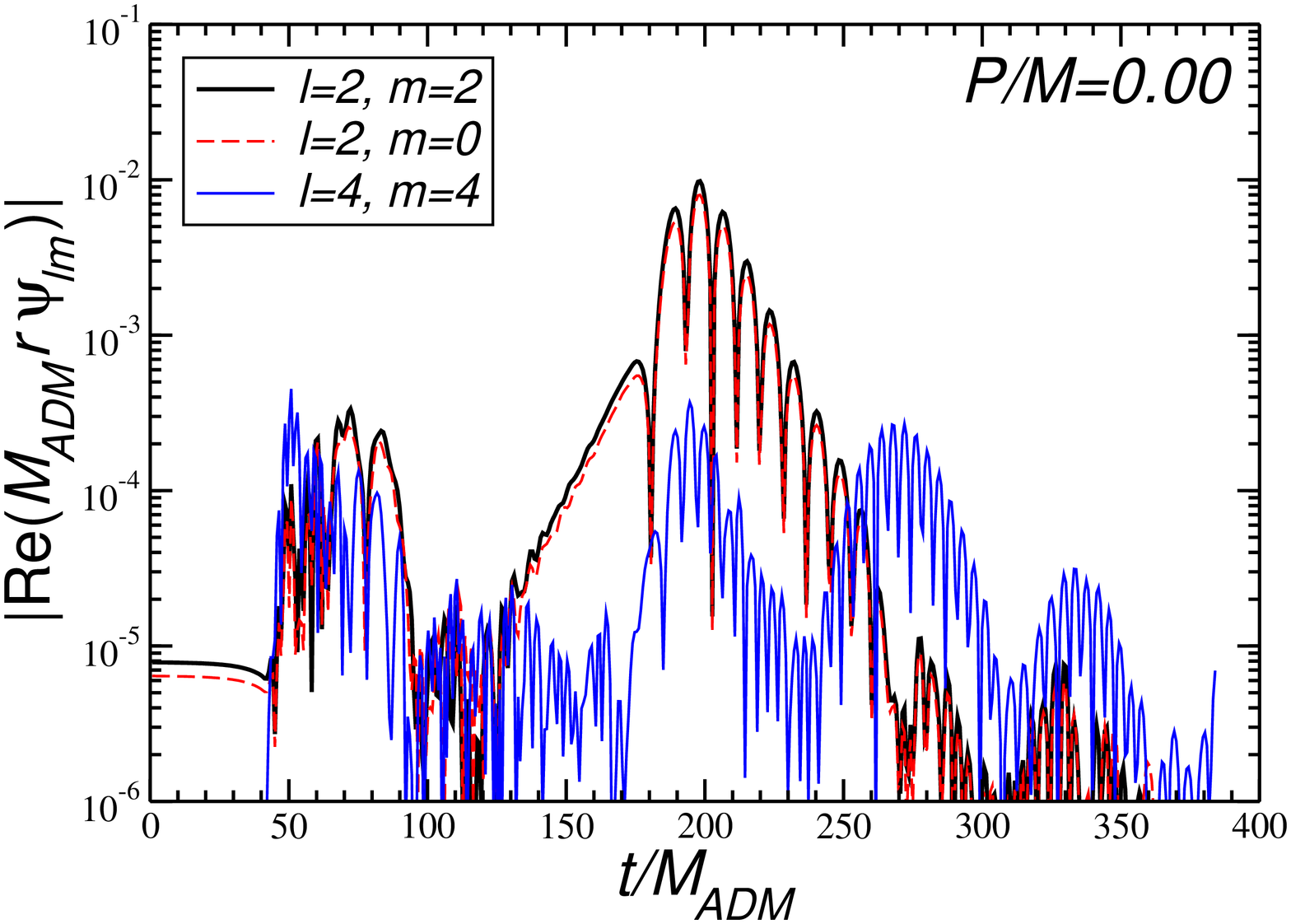}
\includegraphics[height=7.2cm,angle=-90]{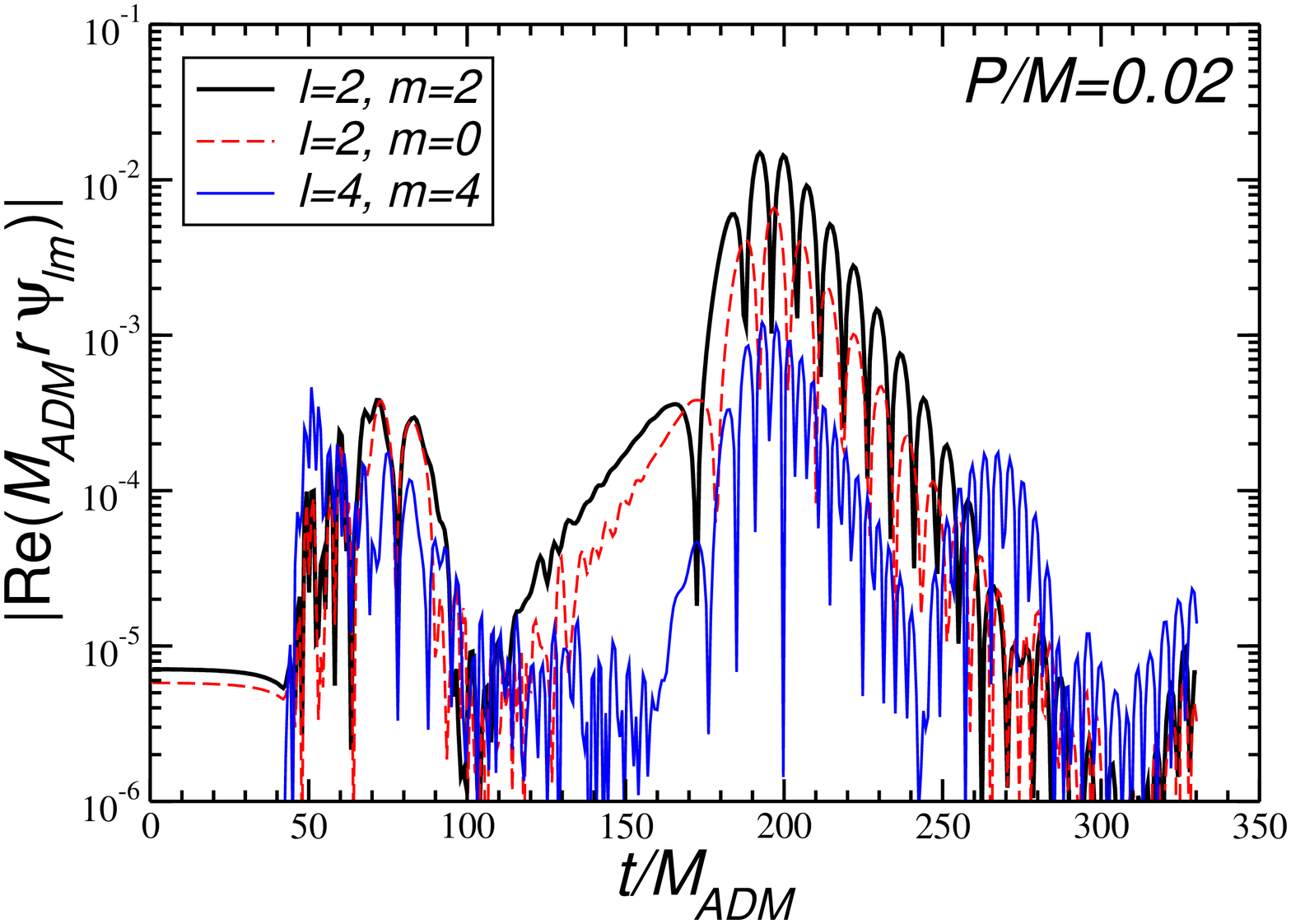}
\includegraphics[height=7.2cm,angle=-90]{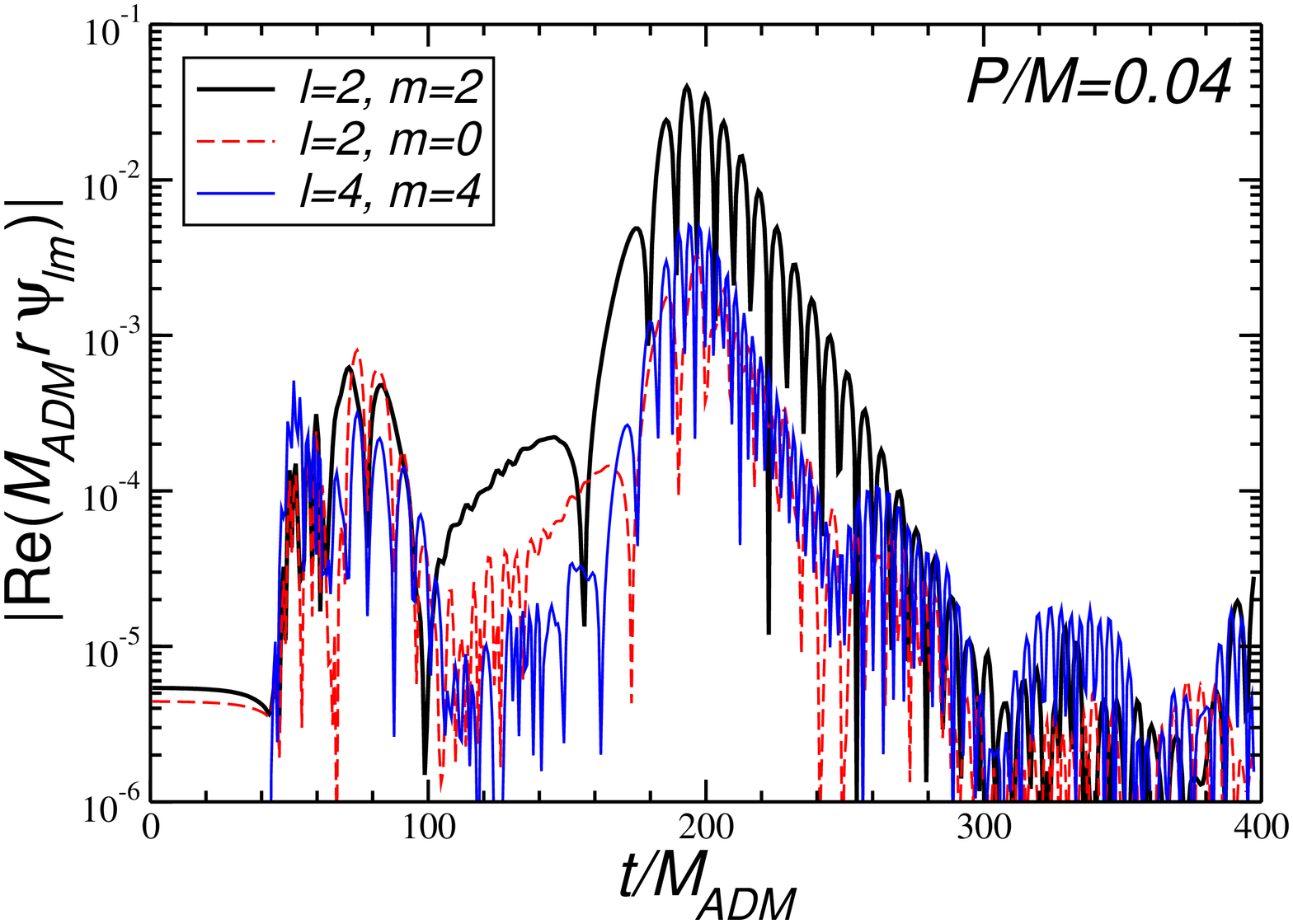}
\includegraphics[height=7.2cm,angle=-90]{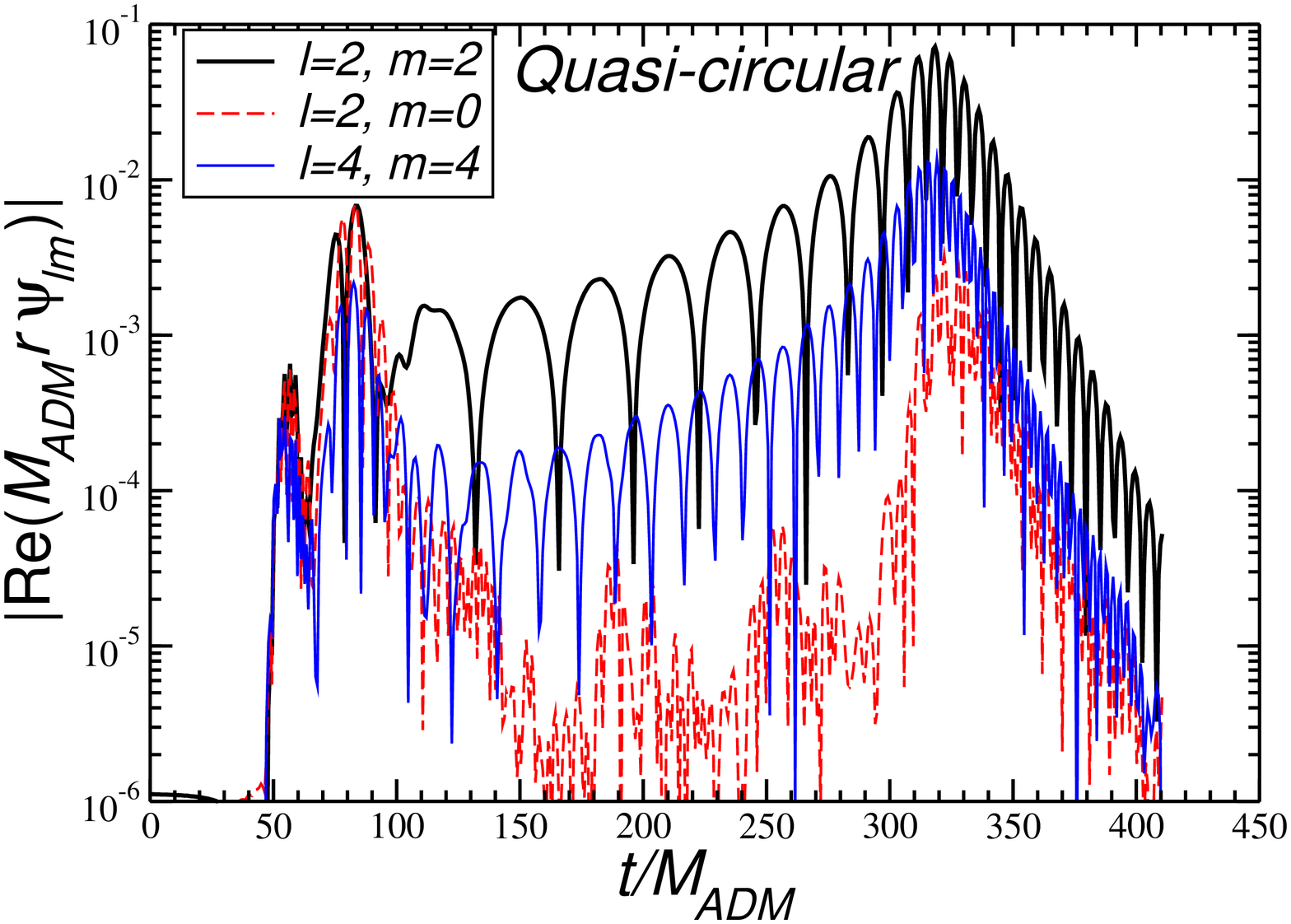}
\caption{Modulus of the real part of the $(\ell=2,~m=2)$, $(\ell=2,~m=0)$ and
  $(\ell=4,~m=4)$ components of the waveforms. We show waveforms from four
  representative simulations: sequence 2 runs with $P/M=0$ (head-on limit),
  $P/M=0.02$, $P/M=0.04$ and $P/M=0.1247$ (quasi-circular case).}
\label{fig: ampl}
\end{figure}

To be more quantitative, in Fig.~\ref{fig: ampl} we show the modulus of the
real part of the $(\ell=2,~m=2)$, $(\ell=2,~m=0)$ and $(\ell=4,~m=4)$
components of the Weyl scalar for some representative runs. From these plots
it is clear that the $(\ell=2,~m=0)$ component contributes significantly for
$P/M\lesssim 0.05$.  That the $(\ell=2,~m=0)$ component should be significantly
excited in the head-on limit $P/M \to 0$ is known from previous numerical
simulations: in fact, the $m=0$ mode would by far be dominant if we had chosen
the collision to happen along the $z$-axis (see
eg.~\cite{Sperhake:2005uf,Sperhake:2007ks}).  Notice that in this paper GWs
have always been extracted assuming that the $z$-axis is perpendicular to the
orbital plane. In the case of head-on collisions this is contrary to most
previous studies, where to take full advantage of the symmetry of the problem
the axis of reference for the angular coordinates is identified with the axis
of collision. In consequence, we find the radiated energy of head-on
collisions to be quadrupole dominated, but to contain $m=\pm2$ and $m=0$
contributions of comparable magnitude rather than almost exclusively an
$\ell=2$, $m=0$ contribution as is the case for alignment of the two axes. Our
choice is entirely motivated by using identical angular coordinates throughout
the sequence of models. A detailed analysis of the transformation
properties of multipolar components of the radiation under rotations,
translations and boosts can be found in Ref.~\cite{Gualtieri:2008ux}.

As $P/M$ increases and we approach quasi-circular motion, the low-amplitude
portion of the $(\ell=2,~m=0)$ mode decreases in amplitude, and it is
significantly contaminated by noise. As expected from our previous study
\cite{Berti:2007fi}, in the same limit the amplitude of the $(\ell=4,~m=4)$ mode
grows.  Unfortunately, Fig.~\ref{fig: ampl} illustrates that even for
$(\ell=4,~m=4)$, which is the largest of the subdominant radiation modes, the
ringdown signal is strongly distorted by either non-linear effects or boundary
reflection noise. For this reason it is difficult to use higher multipoles to
improve spin estimates from quasinormal mode (QNM) fittings, as proposed in
\cite{Berti:2007fi}. This problem will be discussed in more detail in
Sec.~\ref{QNMs} below.

\begin{figure}[ht]
\centering
\includegraphics[height=8cm,angle=-90]{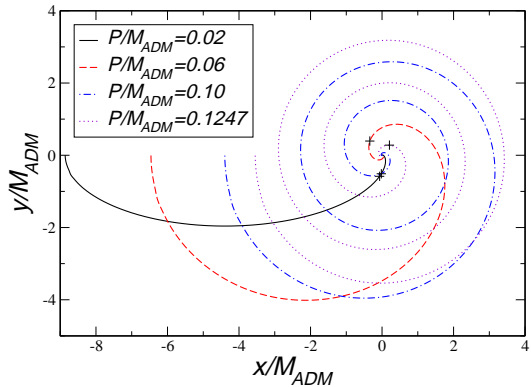}
\caption{Trajectories of the models with $P/M=0.02$, $0.06$, $0.10$ and
  $0.1247$ of sequence 2. The trajectory of one hole only is shown in each
  case. The positions of the respective second holes follow from symmetry
  across the origin. The $+$ denote the locations of the individual holes at
  the time of common apparent horizon formation.  }
\label{fig: traj}
\end{figure}

We can obtain an estimate of the number of orbits in our simulations from the
puncture trajectories, calculated according to $dx^i/dt=-\beta^i$.
For illustration, in Fig.~\ref{fig: traj} we plot
the trajectories of four models of sequence 2 with $P/M=0.02$, $0.06$, $0.10$
and $0.1247$. The figure demonstrates the inspiralling nature of the
simulations with large initial momentum, whereas those with small momentum
rather represent plunging configurations.

We define a phase $\phi_{\rm punc}$ of these trajectories
by expressing the puncture's
position in spherical polar coordinates
\begin{equation}
  (x,y) = (r_{\rm punc}\cos \phi_{\rm punc},~r_{\rm punc}\sin \phi_{\rm punc}).
\end{equation}
Then we consider the phase difference $\Delta \phi_{\rm punc} = \phi_{\rm
  punc}(t_{\rm cah}) - \phi_{\rm punc}(t_0)$, where $t_{\rm cah}$ is the time
of formation of a common apparent horizon and we choose $t_0=50M_{\rm ADM}$ to
cut off the initial data burst (this is consistent with the derivation of the
frequency from the GW signal, discussed below). The number of orbits then
follows from $N_{\rm punc}=\Delta \phi_{\rm punc}/(2\pi)$.

The gravitational wave signal $\Psi_4$ also serves to estimate the number of
orbits completed by the binary prior to merger. For this purpose we focus on
the $(\ell=2,~m=2)$ multipole and decompose the mode coefficient into a
time-dependent amplitude $A(t)$ and phase $\phi(t)$ according to
\begin{equation}
  \psi_{22}(t) = A(t)e^{i\phi(t)}.
\end{equation}
We next calculate the phase difference
\begin{equation}
  \Delta \phi = \phi(t_{\rm cah}+r_{\rm ex}) - \phi(t_0+r_{\rm ex})\,,
\end{equation}
where $r_{\rm ex}$ takes (approximately) into account the time it takes for
the waves to propagate to the extraction radius. The number of orbits
completed by the binary is then estimated as $N_{\rm waves}=\Delta
\phi/(4\pi)$, where the additional factor of $2$ compensates for the
difference between the orbital frequency and that of a multipole with $m=2$.
Both estimates of the number of orbits, together with the time of formation of
the common apparent horizon, are given in Table \ref{tab: models}.  Small
differences in these numbers are due to the fact that the approximate relation
$\omega_{\rm waves}=m\omega_{\rm punc}$ breaks down near the merger of the holes.

\begin{figure}[ht]
\centering
\includegraphics[height=7.2cm,angle=-90]{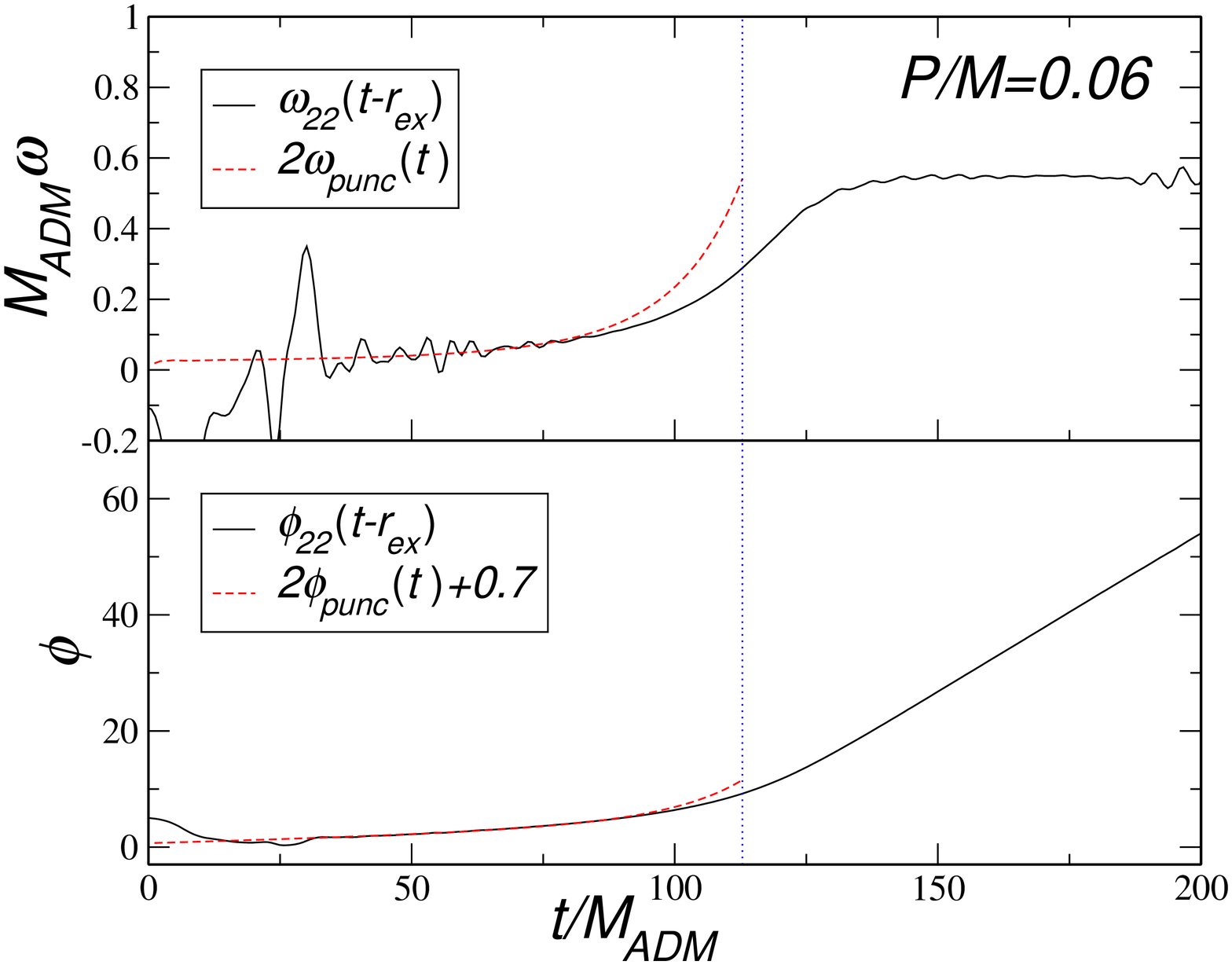}
\includegraphics[height=7.2cm,angle=-90]{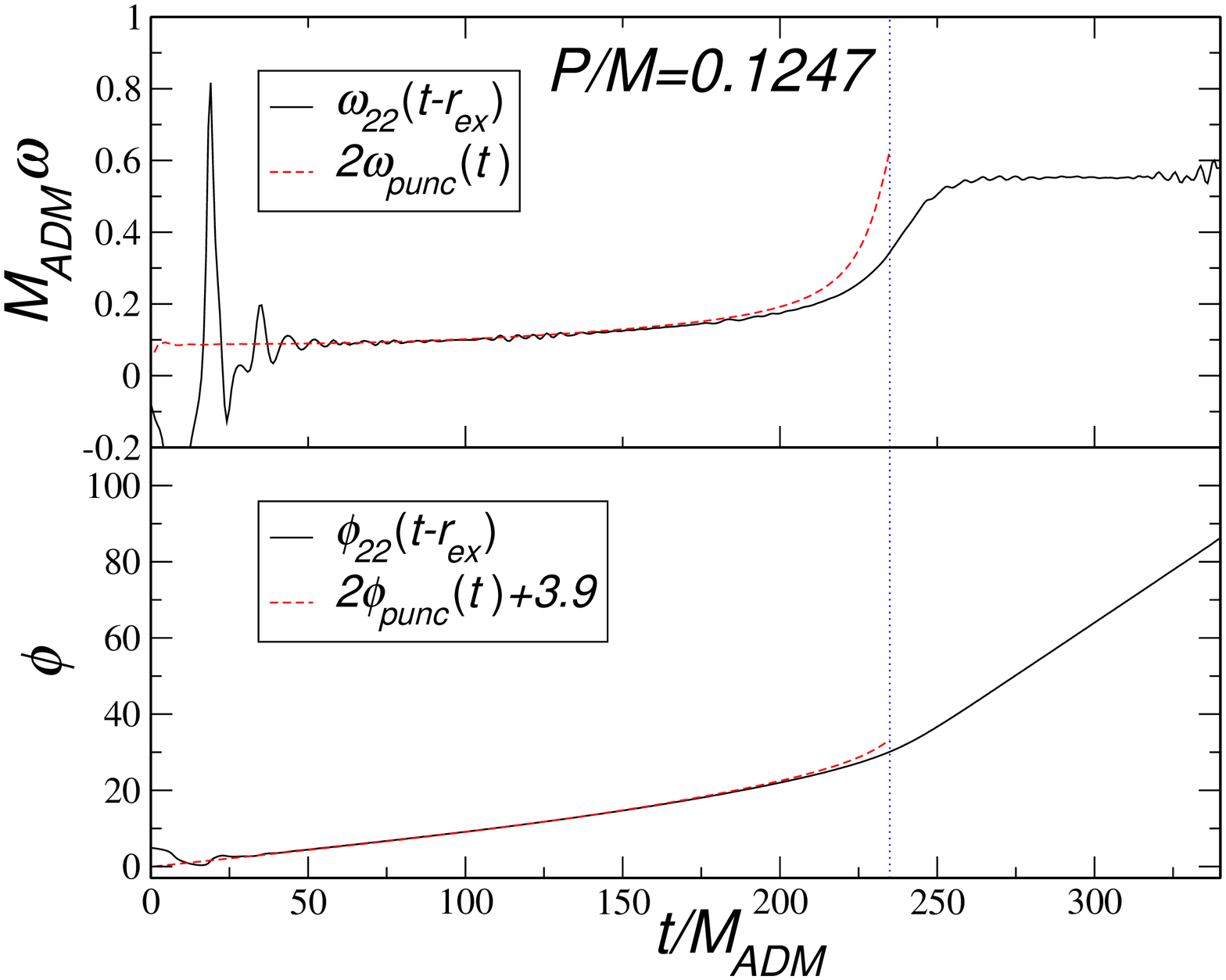}
\includegraphics[height=7.2cm,angle=-90]{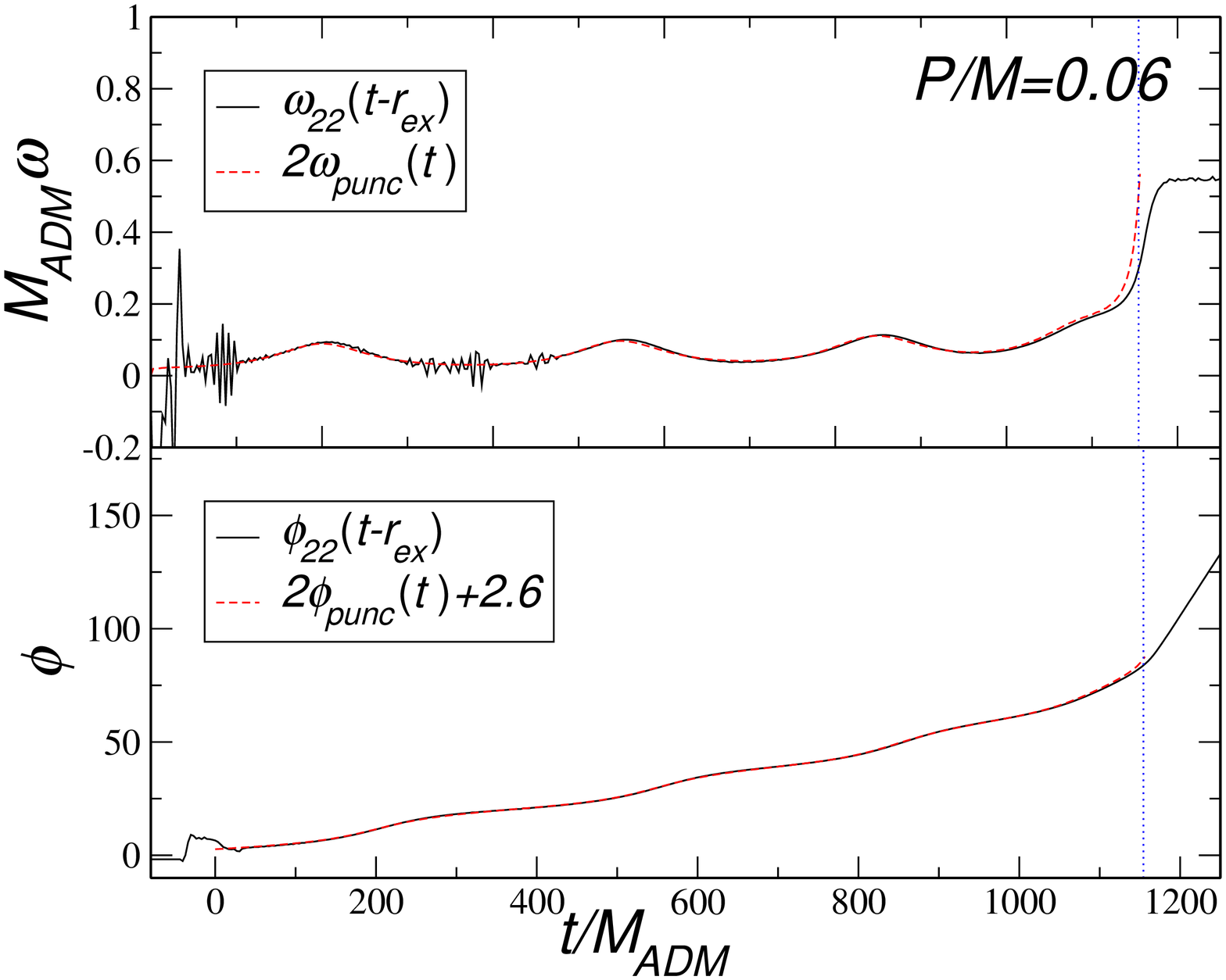}
\includegraphics[height=7.2cm,angle=-90]{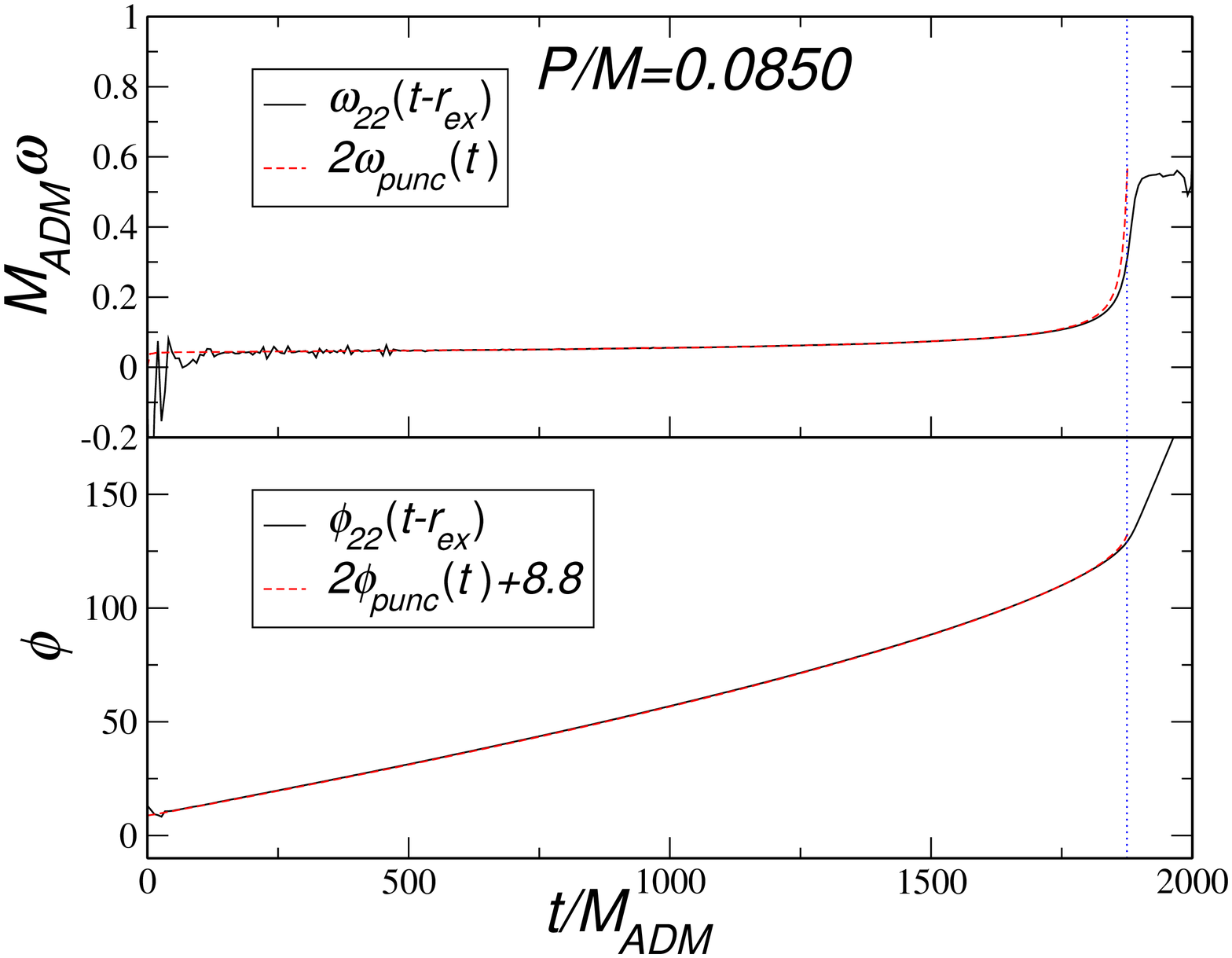}
\caption{Frequency and phase obtained from the
  $\ell=2$, $m=2$ multipole of the gravitational radiation as well as the
  puncture trajectory for models $P/M=0.06$ and $0.1247$ of sequence 2
  (upper panels) and $P/M=0.06$ and $0.0850$ of sequence 3 (lower panels).
  The dotted vertical lines mark the formation of the apparent horizon.  }
\label{fig: om_phi}
\end{figure}

This is illustrated in Fig.~\ref{fig: om_phi}. There we plot the frequency
$\omega_{22}$ and the phase $\phi_{22}$ obtained from $\psi_{22}$, comparing
with the frequency $m\omega_{\rm punc}$ and phase $\phi_{\rm punc}$ obtained
from the puncture trajectory (see \cite{Berti:2007fi} for details). Estimates
from the punctures' motion are physically irrelevant after formation of the
apparent horizon. The upper panels refer to different models from sequence 2: a
nearly plunging motion with $P/M=0.06$ (left) and a quasi-circular orbit with
$P/M=0.1247$ (right). It is clear from the figure that deviations between
$\omega_{\rm waves}$ and $m\omega_{\rm punc}$ grow significantly near merger.
% for near-plunging orbits.
The same holds for the two models of sequence 3 shown in the bottom panels.

From Fig.~\ref{fig: traj}, and from the data for $N_{\rm punc}$ and $N_{\rm
  waves}$ listed in Table \ref{tab: models}, we deduce that the simulations
with $L\lesssim L_{\rm crit}\simeq 0.08M^2$ complete significantly
less than one cycle, so
they are effectively plunging trajectories.  Consistently with this
interpretation, we have see in Sec.~\ref{eccentricity} below that PN estimates
of the eccentricity of the orbit fail for these plunging configurations.

\subsection{Polarization}

In Appendix D of \cite{Berti:2007fi} we proposed to measure the polarization
of a waveform using an ``elliptical component of polarization'' $P_E$. This
quantity has the property that $P_E=1$ for circular polarization, and $P_E=0$
for linear polarization. By looking at the dominant $(\ell=2,~m=2)$ component
of the radiation we also showed that, with the exception of the (unphysical)
initial data burst and of the final part of the signal, which is dominated by
noise, the polarization of a binary moving in a quasi-circular orbit is
circular to a very good approximation.

\begin{figure}[hbt]
\centering
\includegraphics[height=7.2cm,angle=-90]{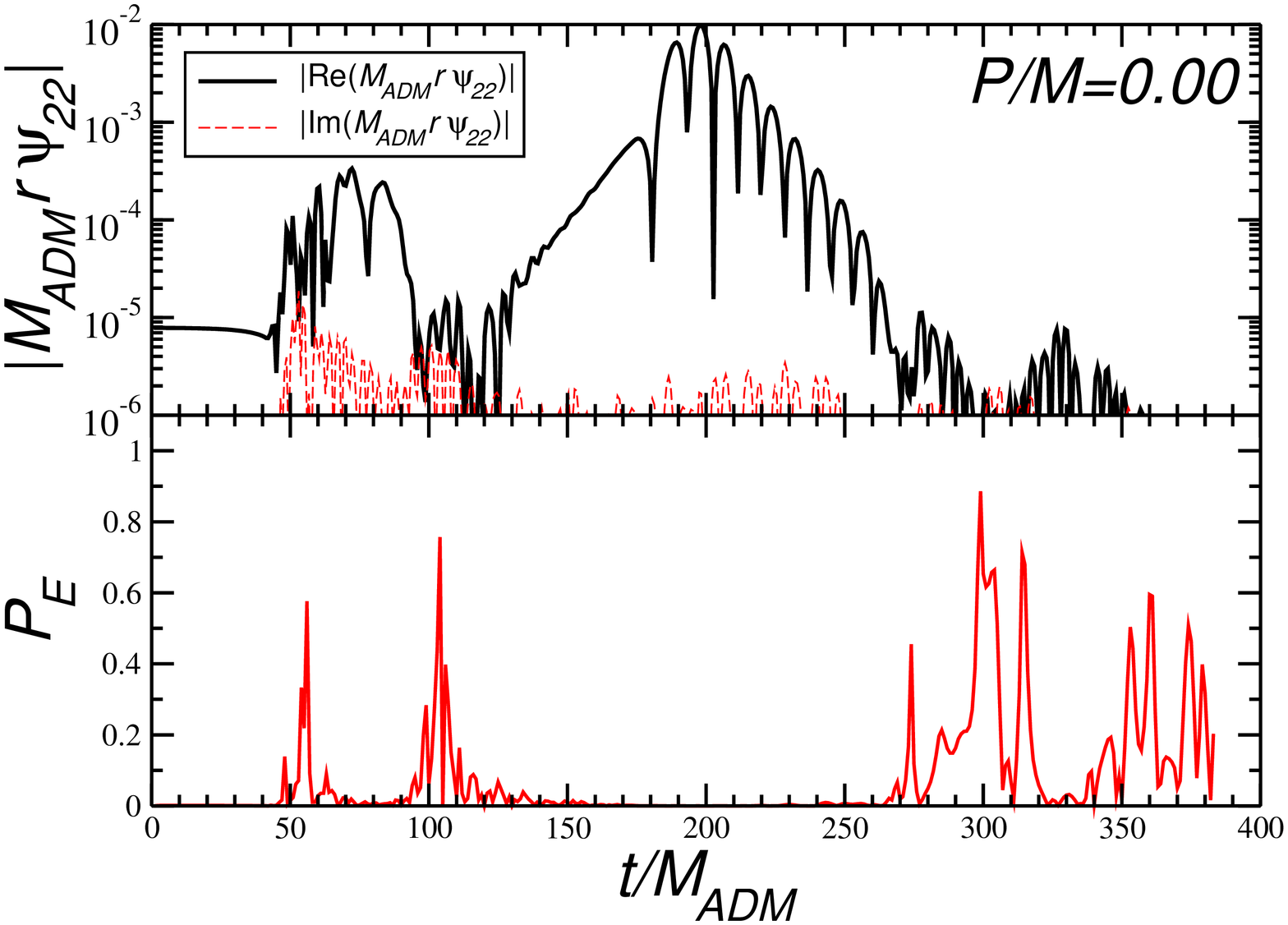}
\includegraphics[height=7.2cm,angle=-90]{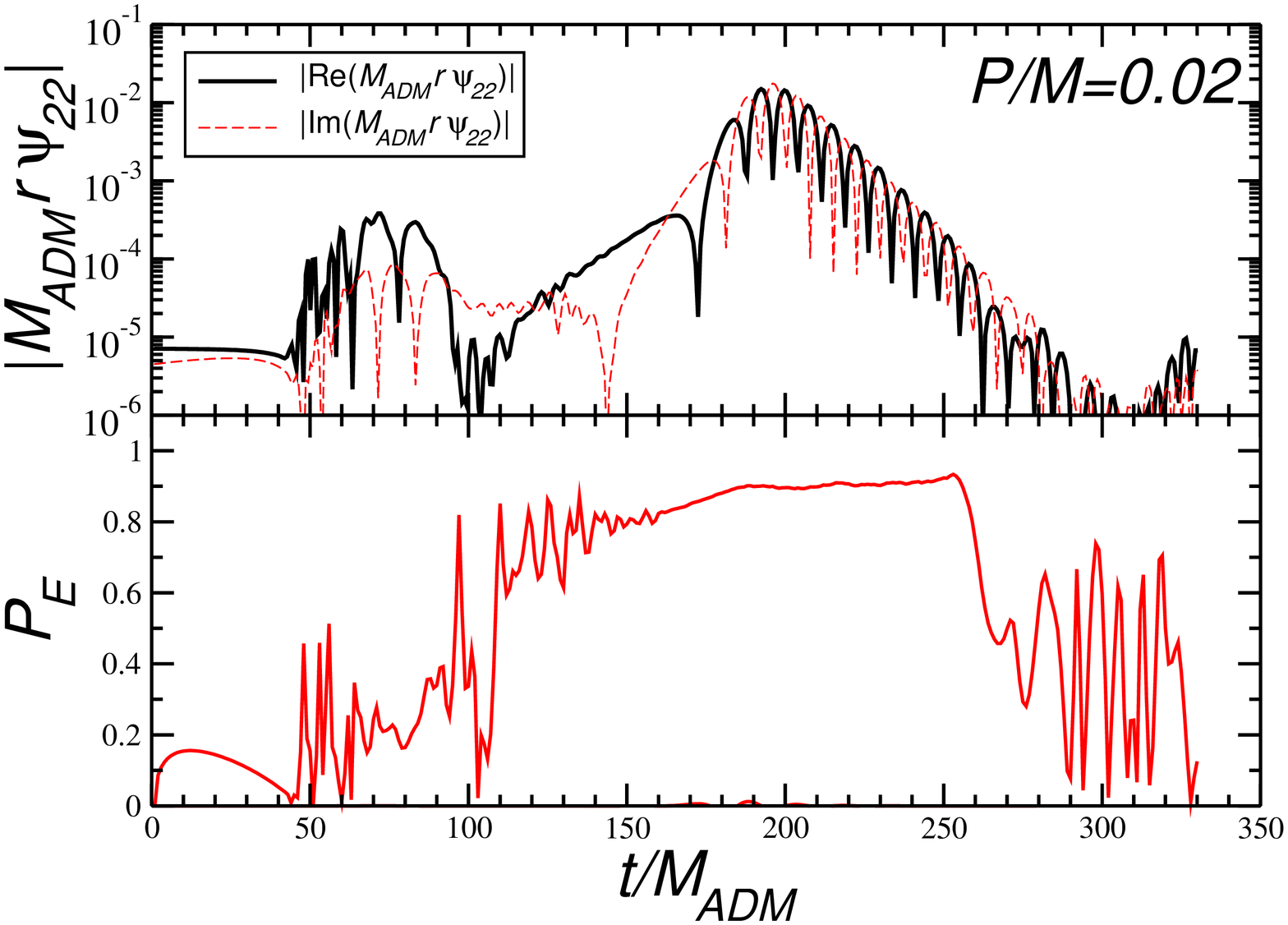}
\includegraphics[height=7.2cm,angle=-90]{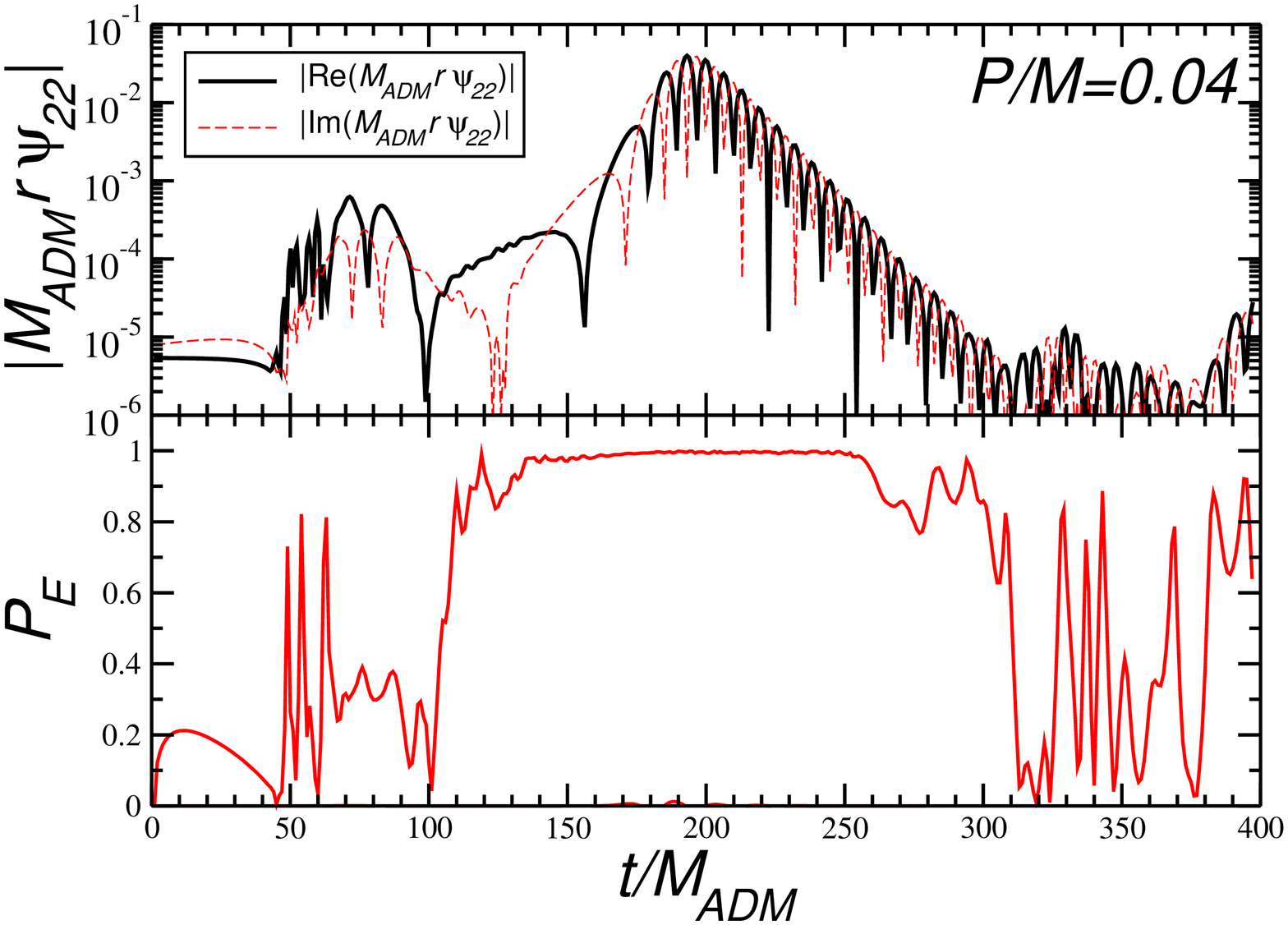}
\includegraphics[height=7.2cm,angle=-90]{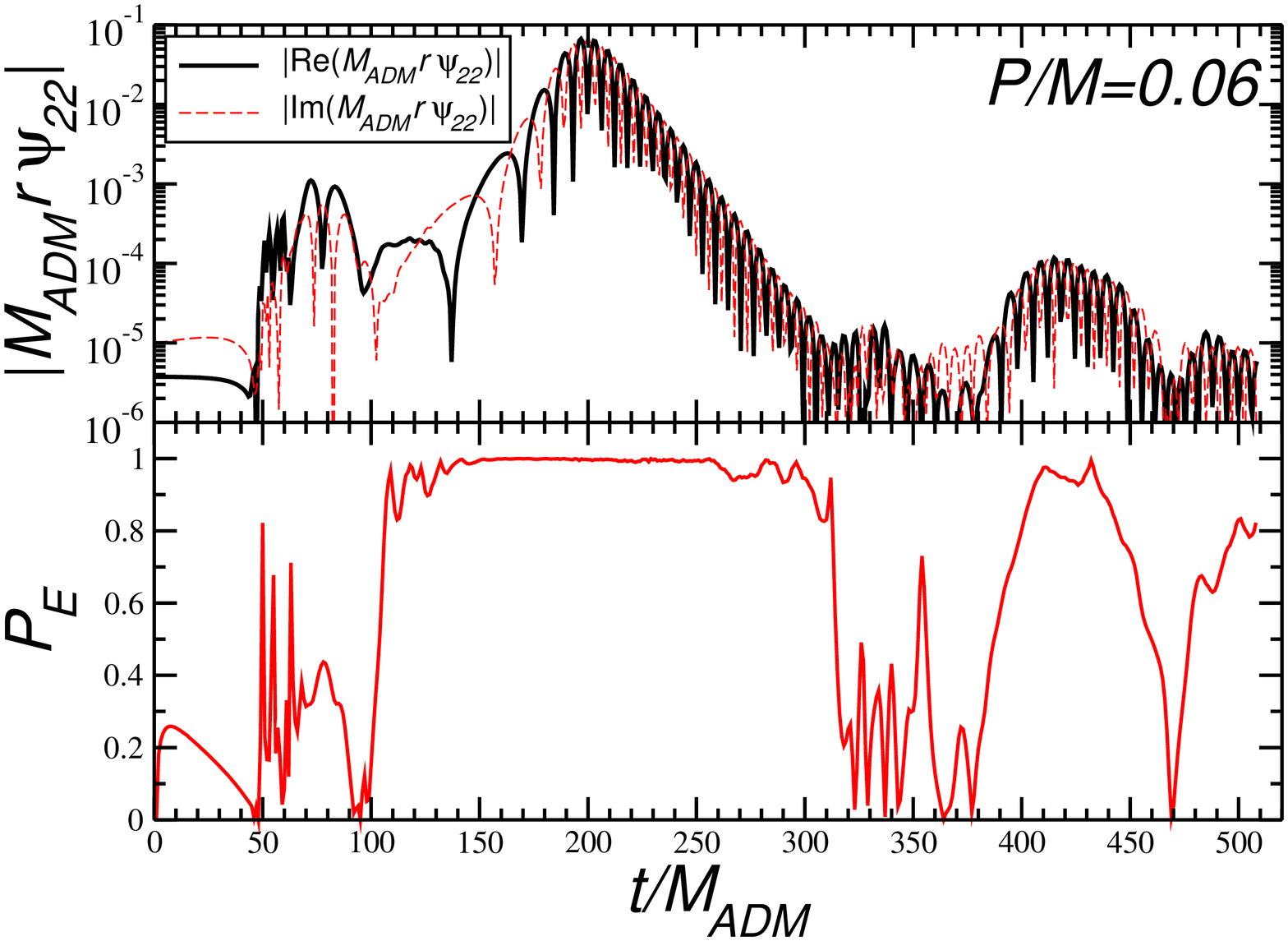}
\caption{$\ell=m=2$ components of the low-$L$ (or low-$P$) waveforms and their
  polarization for various models of sequence 2.
  For $L=0$ the imaginary part of the waveform is zero within
  the noise level (i.e., the cross component is zero for symmetry reasons). As
  $L$ increases, the polarization becomes circular. Spikes in $P_E$ at early
  times are due to the inital data burst, and spikes at late times are due to
  boundary reflection noise and low strength of the signal.}
\label{fig: pol}
\end{figure}

In Fig.~\ref{fig: pol}, to be compared with Fig.~28 of
Ref.~\cite{Berti:2007fi}, we show the real and imaginary parts of the dominant
$(\ell=2,~m=2)$ multipolar component of the radiation emitted by equal-mass
binaries with different values of $P/M$ as obtained from our sequence 2.
In the bottom panel of each plot we
compute the elliptical component of polarization. The results clearly
illustrate that the polarization is linear in the head-on limit, where the
imaginary component of the radiation vanishes, and that $P_E\to 1$ as the
orbit becomes circular. It is also worth noticing that the ringdown part of
the signal is circularly polarized even when $P_E$ is slightly less than one
in the inspiral part (see eg. the plot for $P/M=0.02$).

\section{\label{energyj}Radiated energy and final angular momentum}

In this section we study the multipolar energy distribution and the final
angular momentum for the three sequences of simulations listed in Table
\ref{tab: models}.

\begin{figure}[ht]
\centering
\includegraphics[height=7.2cm,angle=-90]{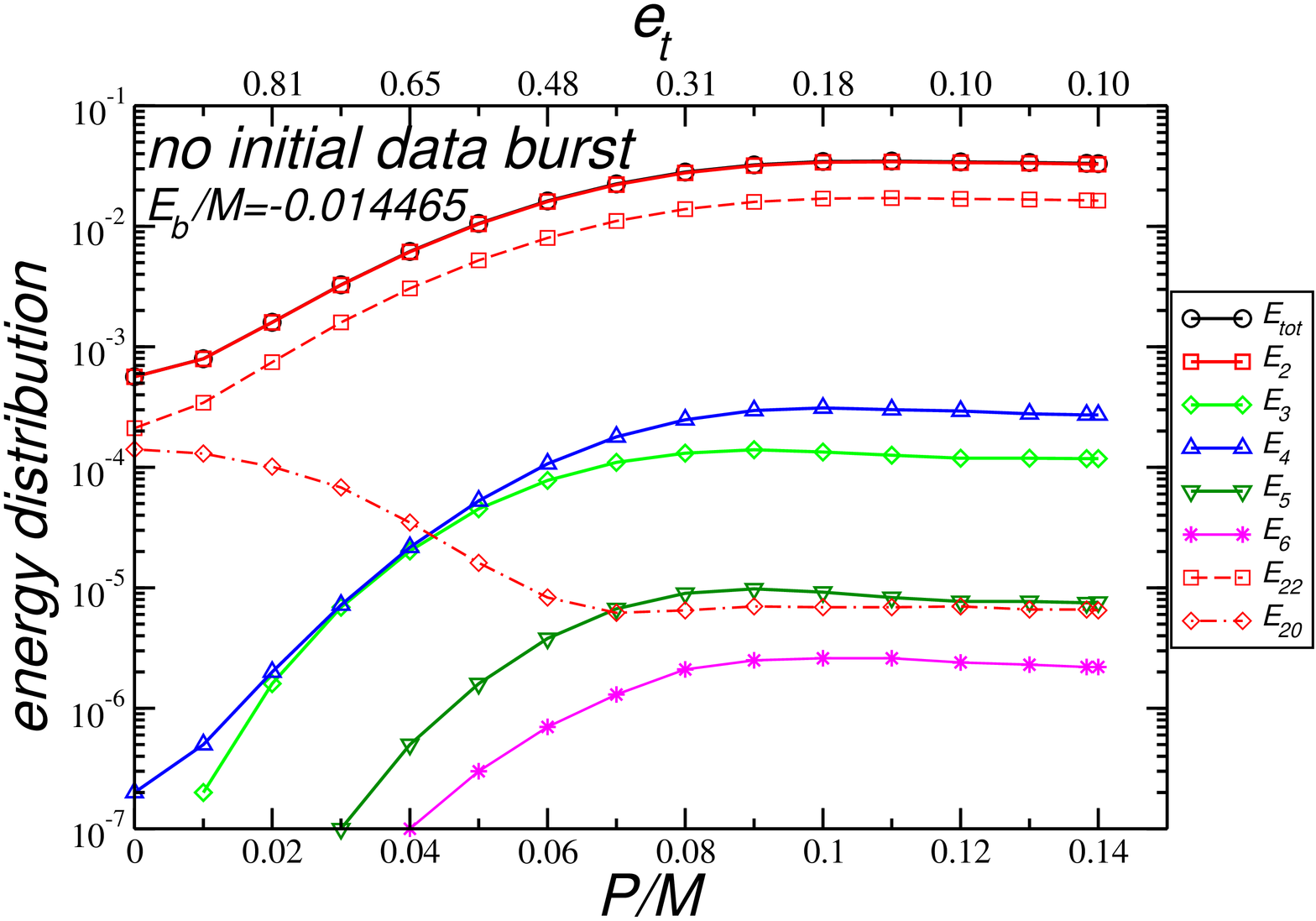}
\includegraphics[height=7.2cm,angle=-90]{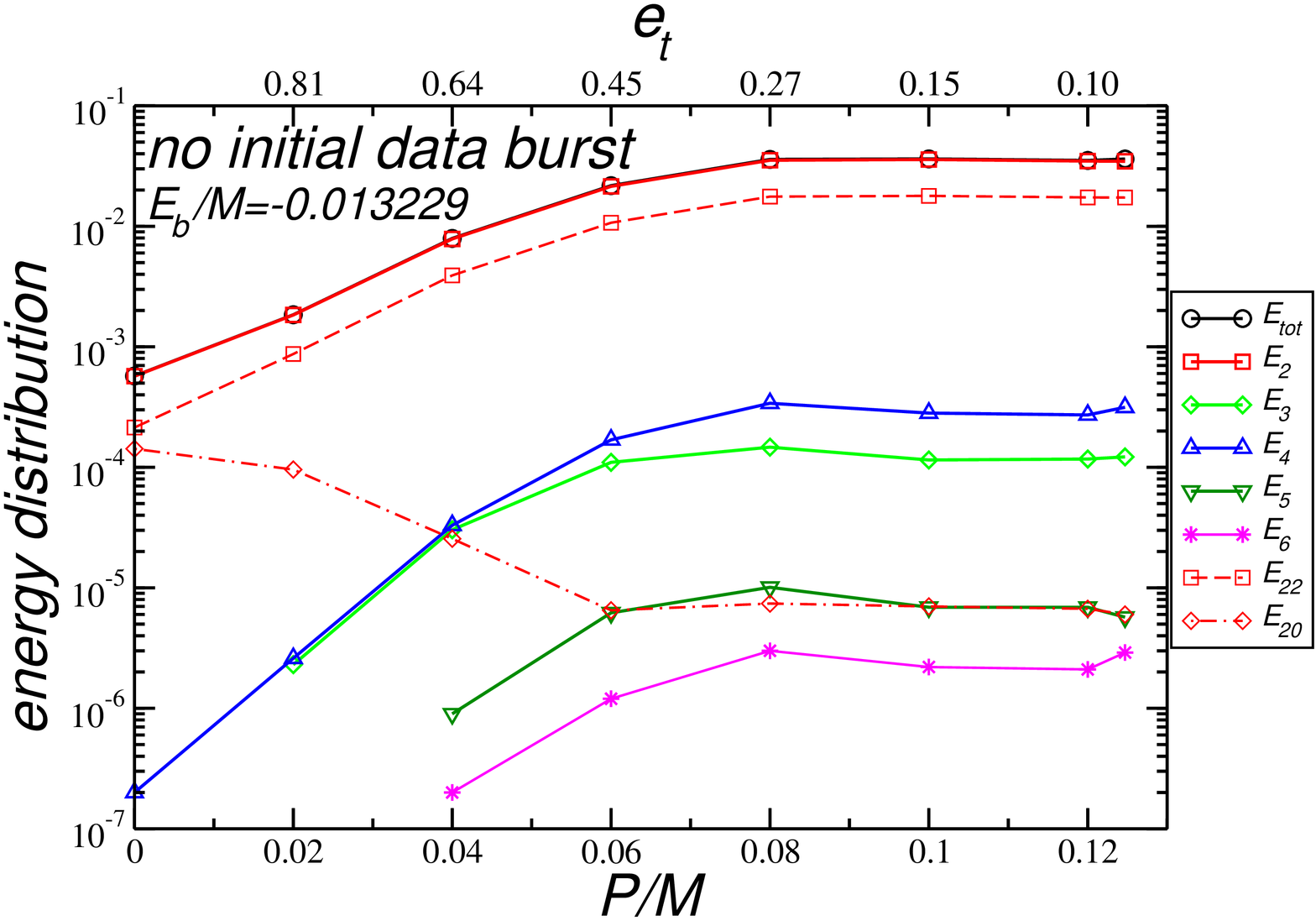}
\includegraphics[height=7.2cm,angle=-90]{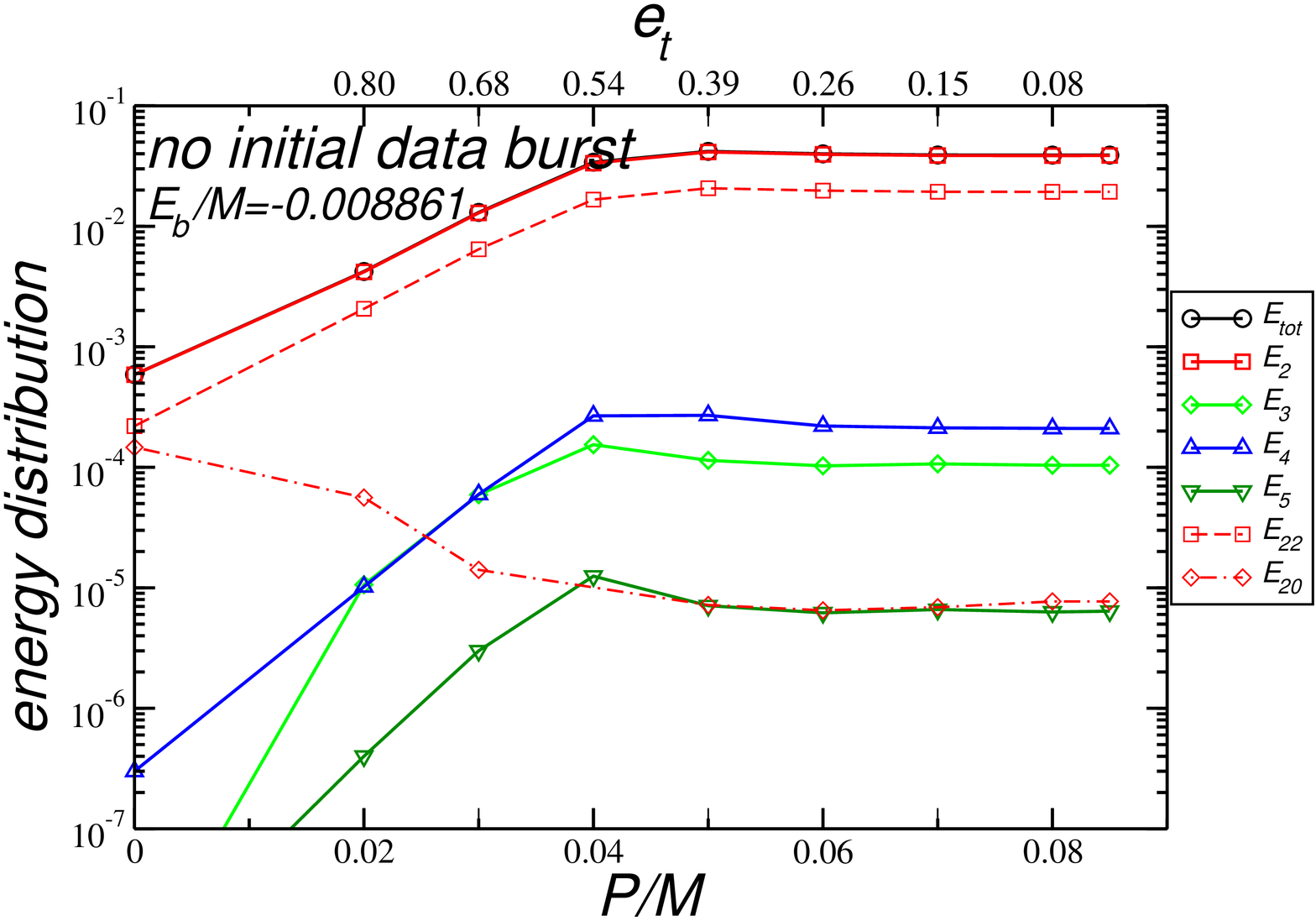}
\caption{Multipolar energy distribution as a function of the initial momentum
  $P$ for sequence 1 (upper left), sequence 2 (upper right) and sequence 3
  (lower). $E_{\ell}$ denotes the energy radiated in all multipoles with
  indices $\ell$ and $m=-\ell,...,\ell$.
  We remove the initial
  data burst by ignoring all data with $t<r_{\rm ext}+50M$.
%  {\bf TODO: Uli will add one more point for a seq.3 head-on collision}
}
\label{fig: El}
\end{figure}

The multipolar distribution of radiated energy for sequence 1 is shown in the
upper left panel of Fig.~\ref{fig: El}. We have excluded radiation due to the
initial burst by ignoring contributions at simulation times $t<50~M+r_{\rm
  ex}$.  The results demonstrate a relatively weak dependence of the radiated
energy in each multipole on initial linear momenta $P/M\gtrsim 0.08$,
corresponding to an angular momentum of $L/M^2\sim 0.8$.

For smaller linear (or angular) momenta, we observe that: (1) the total
radiated energy decreases almost exponentially, (2) the relative contribution
of multipoles with $\ell>2$ becomes weaker, and (3) the contribution of the
$(\ell=2,~m=0)$ mode increases to approximately the same level as the
$(\ell=2,~m=\pm2)$ modes. All of these features are quite insensitive to the
inclusion of the spurious initial wave burst: only the $(\ell=2,~m=0)$ mode is
significantly contaminated by the initial radiation when $P/M\gtrsim 0.08$.
The same observations also apply to the models of sequence 2, shown in the
right panel of Fig.~\ref{fig: El}.  Here the transition seems to occur at a
linear momentum slightly below $P/M=0.08$.  From Table \ref{tab: models} we
see that this again corresponds to an angular momentum of $L/M^2\approx 0.8$.
Similarly, the transition occurs just below $P=0.04$ in the case of sequence 3,
which again corresponds to an orbital angular momentum $L/M^2\approx 0.8$.
A fit of sequence 1 runs by a polynomial in $L/M^2$ yields
\begin{equation}
\frac{E_{\rm rad}}{M_{\rm ADM}}=
0.0212\left(\frac{L}{M^2}\right) -
0.1020 \left(\frac{L}{M^2}\right)^2 +
0.1478 \left(\frac{L}{M^2}\right)^3.
\end{equation}

\begin{figure}[b]
\centering
\includegraphics[height=7.2cm,angle=-90]{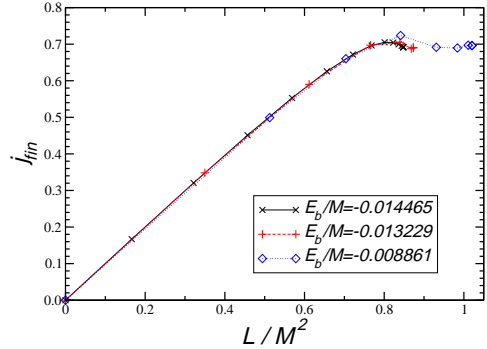}
\caption{Final angular momentum as a function of the initial momentum $P$ for
  sequence 1 (black solid, cross), 2 (red dashed, plus)
  and 3 (blue dotted curve, diamond).
  %{\bf [Todo: Uli, caption and $y$-scale here are wrong.]}
}
\label{fig: jfin}
\end{figure}

A similar behavior is found for the angular momentum of the final black hole,
shown in Fig.~\ref{fig: jfin}. We have calculated the final spin from balance
arguments: the final black hole mass $M_{\rm fin}$ is obtained by subtracting
the total radiated energy from the initial ADM mass, and the final black hole
angular momentum $J_{\rm fin}$ is similarly given by the initial orbital
angular momentum minus the momentum radiated in GWs. The figure shows the
dimensionless Kerr parameter $j_{\rm fin}=J_{\rm fin}/M_{\rm fin}^2$.  Again,
small increases in eccentricity only lead to a mild increase in the final
spin. The Kerr parameter has a local maximum, and then it decreases rapidly
for $L/M^2\lesssim 0.8$. The results for the three sequences show remarkably
good agreement below $L/M^2\lesssim 0.8$
and merely differ at large angular momenta, as initial
configurations with such large $L$ only exist when choosing sufficiently
large separations.
%The results in Fig.~\ref{fig: jfin} are not easy to fit by a polynomial.
By experimenting with sequence 1 runs, we found the following, reasonably
accurate three-parameter polynomial fit:
\begin{equation}
\frac{J_{\rm rad}}{M^2_{\rm ADM}}=
0.0225\left(\frac{L}{M^2}\right) +
0.0381\left(\frac{L}{M^2}\right)^2 +
0.5589\left(\frac{L}{M^2}\right)^7.
\end{equation}
The large exponent in the final term is an artifact of the
phenomenological nature of the fit and yields optimal agreement
with the data. We found that the following alternative fitting
functions, with two unknowns, also perform very well, yielding
$E_{\rm rad}/M_{\rm ADM}=3.18\times 10^{-4}\times e^{5.54 \, L/M}$
and $J_{\rm rad}/M^2_{\rm ADM}=7.34 \times 10^{-4}e^{6.74\, L/M}$.

An understanding of the apparent threshold in the orbital angular momentum
$L/M^2$ separating inspiraling and plunging orbits can be obtained by
considering geodesic orbits in a Schwarzschild background. The key observation
here is that particles of mass $m_p$ in closed orbits must satisfy the
condition $L/m_p>2\sqrt{3}M$.  A rough extrapolation of this result to
comparable-mass binaries would yield $L/(\eta M)>2\sqrt{3}M$, or
$L/M^2>0.866$. There is no firm theoretical justification for this
extrapolation, and yet the point-particle threshold $L/M^2=0.866$ is
remarkably close to the observed transition range of $L/M^2\approx 0.8$.

The agreement gets even better if, in the spirit of
Ref.~\cite{Buonanno:2007sv}, we use a slightly improved perturbative model,
considering particle orbits around a Kerr black hole with spin given by the
final Kerr parameter of our simulations (see Sec.~\ref{BKL} and Appendix
\ref{sec:appendixsep} for details).  For eccentric inspirals, the minimum
allowed angular momentum should be attained at the so-called {\em separatrix}
corresponding to the maximal allowed eccentricity, $e_p=1$, where $e_p$ is the
eccentricity defined in Eq.~(\ref{tpoints}) for point particle geodesics. The
orbital angular momentum at the separatrix is also a function of the final
black hole's spin, $L_{\rm sep}(j_{\rm fin}, e_p)$, and it is defined below in
Eq.~(\ref{finalae}). An explicit calculation yields striking agreement with
numerical simulations: $L_{\rm sep}(j_{\rm fin}=0.69, e_p=1)=0.778 M^2$ when
we consider the final spin $j_{\rm fin}=0.69$ produced by a quasi-circular
inspiral, and $L_{\rm sep}(j_{\rm fin}=0.724, e_p=1)= 0.763 M^2$ when the
final hole has the largest spin observed in our simulations.

Our interpretation of this similarity is that simulations with $L/M^2\lesssim
0.8$ can no longer be viewed as eccentric orbiting black-hole binaries, but
rather represent plunging configurations or grazing collisions. The change in
character of the energy distribution visible in Fig.~\ref{fig: El} when
$L/M^2\approx 0.8$ thus demonstrates the transition from orbiting to plunging
binary systems.

\begin{table}[b]
  \centering \caption{\label{tab: t21} The transition times $t_{32}$
    ($t_{21}$) it takes a
    sequence 3 (2) model to radiate an amount of gravitational wave energy
    corresponding to the difference in binding energy from sequence 2 (1). The
    amount of angular momentum $\Delta J_{\rm rad}/M_{\rm ADM}^2$
    for models near the critical orbital angular momentum $L_{\rm crit}$
    radiated up to this time is small compared with the initial
    orbital angular momentum $L/M^2$ listed in Table \ref{tab: models}.}
\begin{tabular}{ccc|ccc}
\hline  \hline
\multicolumn{3}{c}{seq3 $\rightarrow$ seq2} &
\multicolumn{3}{c}{seq2 $\rightarrow$ seq1} \\
\hline
$L/M^2$ & $T_{32}/M_{\rm ADM}$ & $\Delta J_{\rm rad}/M_{\rm ADM}^2(t_{32})$ &
$L/M^2$ & $t_{21}/M_{\rm ADM}$ & $\Delta J_{\rm rad}/M_{\rm ADM}^2(t_{21})$ \\
\hline
1.020  & 1646 & 0.145 & 0.873        &  91 & 0.020   \\
1.021  & 1653 & 0.145 & 0.873        &  98 & 0.022   \\
1.011  & 1450 & 0.130 & 0.868        & 106 & 0.023   \\
0.984  &  926 & 0.112 & 0.839        & 106 & 0.021   \\
0.931  &  444 & 0.082 & 0.765        & 108 & 0.017   \\
0.841  &  203 & 0.055 & 0.612        & 114 & 0.014   \\
0.704  &  201 & 0.042 & 0.350        & 127 & 0.010   \\
\hline \hline
\end{tabular}
\end{table}

Our simulations starting at different initial separation show some degree of
universality in this transition. This can be understood in terms of the
relatively low amount of angular momentum the binaries emit in the earlier
stages of the inspiral (or plunge). In fact, consider some arbitrary
configuration of sequence 3 with binding energy $E_{\rm b}^{(3)}/M=-0.008861$.
As the binary evolves, it emits gravitational radiation and the binding energy
decreases. Of particular interest for our argument is the moment in time when
the binding energy decreases to the value $E_{\rm b}^{(2)}/M=-0.013229$
corresponding to sequence 2.  In order to estimate this time, we consider the
radiated energy $\Delta E_{\rm rad}(t)$ measured at $r_{\rm ex}$, integrated
up to time $t$.  The {\em transition time} $t_{32}$ from sequence 3 to
sequence 2 is then approximated by the relation
\begin{equation}
  \Delta E_{\rm rad}(t_{32}+r_{\rm ex}) = E_{\rm b}^{(3)} - E_{\rm b}^{(2)}.
\end{equation}
We next calculate the total amount of angular momentum $\Delta J_{\rm
  rad}/M_{\rm ADM}^2(t_{32}+r_{\rm ex})$ radiated away by the system during
its transition from sequence 3 to sequence 2. The same procedure is applied
for the transition from sequence 2 to 1. Results are shown in Table
\ref{tab: t21}.

By comparison with the initial orbital angular momentum $L/M^2$ listed in the
third column of Table \ref{tab: models}
%
%{\bf [Todo: Uli, I would replace $P/M$ by $L/M^2$ in the Table, or maybe add
%  two more columns with $L/M^2$ - it's quite annoying to flip back the pages
%  to look at the other Table. Also, can we show $\Delta J_{\rm rad}/M^2$
%  instead of normalizing to $M_{\rm ADM}$? Seems more reasonable to me.]}
%
we see that all binary models starting with orbital angular momentum near the
critical value $L/M^2\simeq 0.8$ radiate only a few per cent of their angular
momentum until they reach a more strongly bound state. This explains the
approximate universality of models belonging to different sequences, but
having the same initial orbital angular momentum.
%  It will be very interesting in future work
%to confirm this ``near-universal'' behavior by studying radiated energy and
%angular momentum for binaries starting from much larger initial separations.
%
%Alternative fits to the energy and angular momentum:
%
%\begin{eqnarray}
%J&=&7.336\times 10^{-4}\,\exp(6.7386L)\\
%E&=&3.177\times 10^{-4}\,\exp(5.5388L)\,.
%\end{eqnarray}

\section{\label{BKL} Perturbative estimates of the final angular momentum of eccentric binary black hole coalescence}

A simple, surprisingly successful model to compute the spin of the final black
hole for binaries in quasicircular orbits has recently been proposed by
Buonanno, Kidder and Lehner \cite{Buonanno:2007sv}. Here we discuss extensions
of that model to eccentric binaries. The model introduced in
Ref.~\cite{Buonanno:2007sv} is based on three main assumptions: (i) the
gravitational energy radiated after the formation of a common apparent horizon
is a second order quantity, and has a small effect on the spin of the final
hole; (ii) the magnitude of the individual spins remains constant during the
inspiral, and (iii) most of the angular momentum is radiated during the long
inspiral stage until the system reaches the innermost stable circular orbit
(ISCO).  A crucial final ingredient of the model is to estimate the
contribution of the orbital angular momentum to the final spin by associating
it with the orbital angular momentum of a point particle orbiting a Kerr black
hole with spin parameter $j_{\rm fin}$ corresponding to the {\it final} hole.

The generalization of this model to eccentric orbits is straightforward, if
one identifies the ISCO with the separatrix, or innermost stable bound orbit
\cite{Cutler:1994pb,Glampedakis:2002ya} (see Appendix \ref{sec:appendixsep}
for details). Following \cite{Buonanno:2007sv}, in the general case of
eccentric orbits we write the (approximate) conservation equation
\begin{equation}
j_{\rm fin}=\frac{L_{\rm sep}(j_{\rm fin}, e_p)}{M^2}
+\frac{M_1a_1}{M^2}+\frac{M_2a_2}{M^2}\,, \label{finalae}
\end{equation}
where $a_i$ is the Kerr parameter of each hole, the eccentricity $e_p$ for
point particle geodesics is defined in Eq.~(\ref{tpoints}) and $L_{\rm
  sep}(j_{\rm fin}, e_p)$ is the orbital angular momentum at the separatrix,
computed within a point-particle framework. A prescription for the computation
of this quantity is given in Appendix \ref{sec:appendixsep}. Note that $e_p$
is defined in terms of coordinate positions of the particle, and therefore
there is no direct physical relation between $e_p$ and the eccentricity
quantifiers discussed above. Assuming without loss of generality that $M_1\geq
M_2$, we can rewrite Eq. (\ref{finalae}) as
\begin{equation} j_{\rm fin}=\frac{L_{\rm sep}(j_{\rm fin},e_p)}{M^2}
+\frac{j_1}{4}(1+\sqrt{1-4\eta})^2+\frac{j_2}{4}(1-\sqrt{1-4\eta})^2\,,\label{finalae2}
\end{equation}
where $j_i=a_i/M_i$ and $\eta$, as usual, denotes the symmetric mass ratio.

\begin{figure}[b]
\centering
\includegraphics[height=7.2cm,angle=-90]{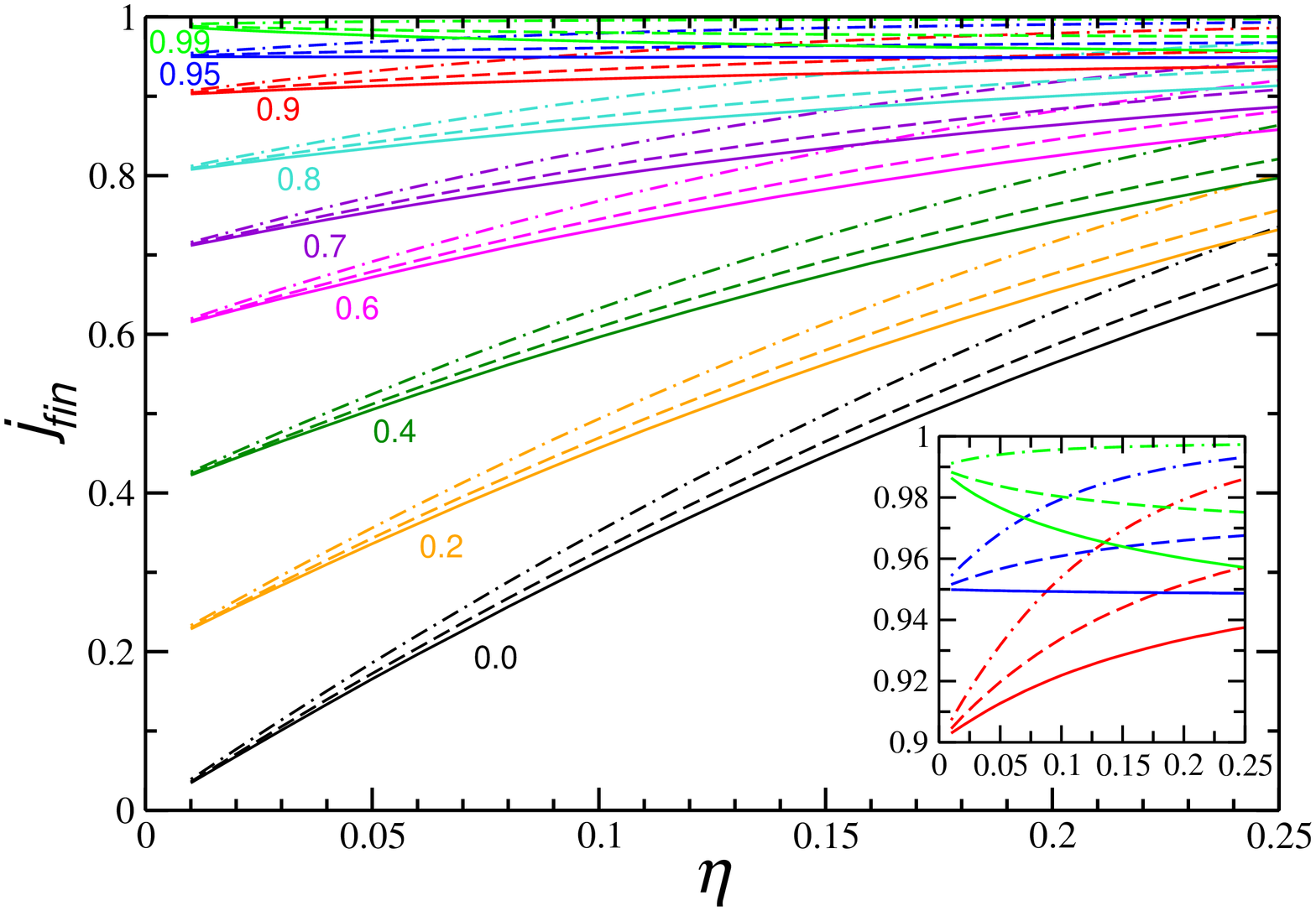}
\includegraphics[height=7.2cm,angle=-90]{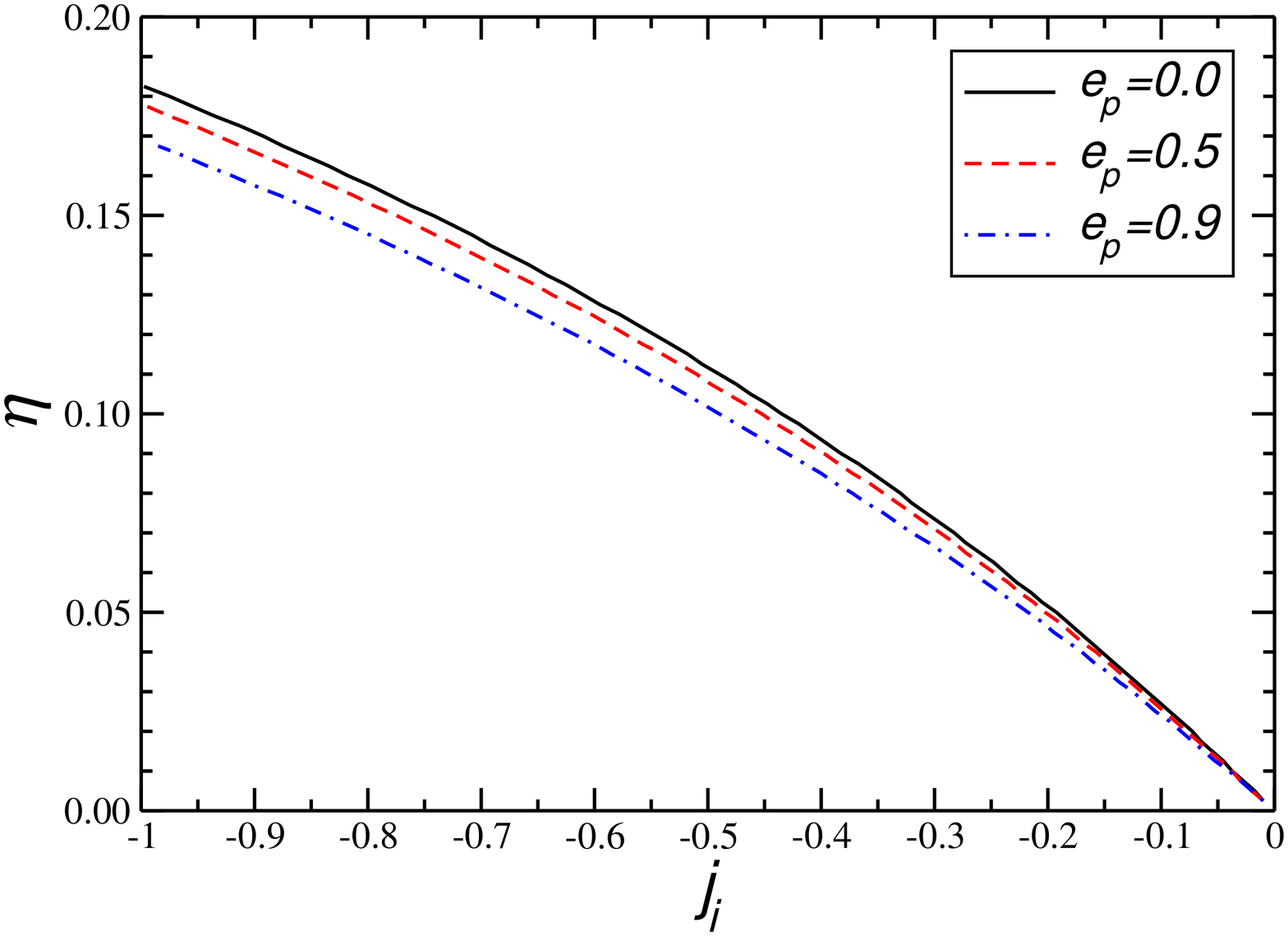}
\caption{Left panel: predicted angular momentum by point-particle
  extrapolation of results from perturbation theory. Solid, dashed and
  dash-dotted lines assume a point-particle eccentricity $e_p=0,~0.5$ and
  $0.9$, respectively (see Appendix \ref{sec:appendixsep} for a definition of
  this eccentricity parameter, not to be confused with the Newtonian
  eccentricity).  Numbers next to each set of lines denote different values of
  $j_i=j_1=j_2$, and the initial spin on each hole is assumed to be aligned
  with the orbital angular momentum. Right panel: we assume the initial spin
  on each hole to be antialigned with the orbital angular momentum
  ($j_i=j_1=j_2<0$). For various eccentricities (as indicated in the legend)
  we show the functional dependence between mass ratio and $j_i$ needed to
  produce a final non-spinning hole.}
\label{fig:PointParticle}
\end{figure}

We will focus on the special case where the initial spins have equal
magnitude, $j_1=j_2$. In Fig.~\ref{fig:PointParticle} (left panel) we show the
predicted spin of the final hole as a function of the symmetric mass ratio
$\eta$ when both initial spins are aligned with the orbital angular momentum.
The effect of eccentricity is always mild, and for all values of the initial
spin magnitude, eccentricity tends to {\it increase} the spin of the final
hole.

Ref.~\cite{Buonanno:2007sv} also predicted that, for circular orbits, the final
spin of the remnant should increase as the mass ratio $q\to 1$ (i.e., as
$\eta\to 1/4$) for $j_i\lesssim j_{\rm crit}$, while it should decrease as
$q\to 1$ if $j_i\gtrsim j_{\rm crit}$. They estimated $j_{\rm crit}\simeq
0.948$. At the critical value, any merger will leave the final spin
essentially unchanged, irrespective of the mass ratio $q$.  This expectation
is confirmed by our calculations (see the inset of the left panel of
Fig.~\ref{fig:PointParticle}). However, since a non-zero eccentricity always
tends to increase the final spin, $j_{\rm crit}$ grows with eccentricity.

\begin{table}[htb]
  \centering \caption{\label{tab:criticspin} Estimated critical value of the
    Kerr parameter $j_{\rm crit}$ (see text) for selected values of the
    eccentricity $e_p$, as defined perturbatively (see Appendix
    \ref{sec:appendixsep}).}
\begin{tabular}{cccccc}
\hline  \hline
                         &$e_p=0.0$&$e_p=0.1$&$e_p=0.2$&$e_p=0.5$&$e_p=0.9$  \\
\hline $j_i^{\rm crit}$  &$0.948$&$0.950$&$0.953$&$0.972$&$0.998$\\
\hline \hline
\end{tabular}
\end{table}

Table \ref{tab:criticspin} shows the critical value of the initial
Kerr parameter $j_i^{\rm crit}=j_1=j_2$ as a function of the
eccentricity (as defined perturbatively). For large eccentricities
$e_p\sim 1$ the critical Kerr parameter is very close to the maximum
possible value, $j_i^{\rm crit}\sim 1$: in other words, for large
eccentricities the final spin should always {\em
  increase} as we approach the equal-mass limit. Therefore we conjecture that
the maximum spin $j_{\rm fin}\simeq 0.724$ that we found in our sequence of
equal-mass merger simulations should be an {\em absolute upper limit} on the
spin that can be achieved as the end-product of non-spinning black hole
mergers: unequal-mass mergers should always produce smaller final spins.

An interesting question explored in \cite{Buonanno:2007sv} concerns spin-flip
configurations.  Suppose that initially both black holes have equal Kerr
parameters $j_i=j_1=j_2$, and spins {\em antialigned} with respect to the
orbital angular momentum. What is the value of the symmetric mass ratio $\eta$
for which the inspiral produces a Schwarzschild black hole, as a function of
$j_i$?  As argued in \cite{Buonanno:2007sv} these ``critical'' configurations
could be particularly interesting, since mild variations of the parameters
around the critical values may produce interesting orbital dynamics (eg.  spin
flips) and complex gravitational waveforms.

The critical curve predicted by the model has been shown to yield good
agreement with numerical simulations, especially for spins aligned with the
orbital angular momentum \cite{Rezzolla:2007rd, Berti:2007nw}.  In particular,
our study \cite{Berti:2007nw} obtains for a mass ratio $q=4$ and initial
dimensionless Kerr parameters of the individual holes $j_i=-0.75$, $-0.8$ and
$-0.85$ a final Kerr parameter $j_{\rm fin}=0.0533$, $0.0237$ and $-0.0038$,
respectively, thus bracketing the formation of a Schwarzschild hole.  For
comparison, for zero eccentricity Eq.~(\ref{finalae2}) yields $j_i=-0.815083$,
while the fitting formula in \cite{Rezzolla:2007rd} yields
$j_i=-0.823462$. Details of this study, together with a multipolar analysis of
several spinning configurations, are given in Ref.~\cite{Berti:2007nw}. As
shown in the right panel of Fig.~\ref{fig:PointParticle}, the effect of
eccentricity on the location of these spin-flip configurations is very mild.
For a fixed mass-ratio, we predict that the magnitude of the (antialigned)
spins required to produce a Schwarzschild black hole as the result of the
merger should increase with eccentricity.

Considering the remarkable success of point-particle extrapolations, some
words of caution are required. First of all, when we extrapolate results to
comparable masses we should not attach any special meaning to the actual
values of the point-particle eccentricity, since we do not know how that notion
translates into the different PN definitions discussed in Section
\ref{eccentricity}. At best, we can only expect that the PN eccentricities are
related to the perturbative eccentricity, as implicitly defined in
Eq.~(\ref{tpoints}), by some monotonically increasing functional dependence.

Furthermore, we should not forget that the model proposed in
\cite{Buonanno:2007sv} is only an approximation: for example, for a merger of
equal-mass, non-spinning holes the model predicts a Kerr parameter $j_{\rm
  fin}=0.663$ for the final hole, to be compared with $j_{\rm fin}\simeq
0.691$ from our numerical simulations (see also \cite{Berti:2007fi}).

Despite these ambiguities and limitations, the point-particle model can still
make remarkable predictions. For example, we can ask the following question:
if we consider a non-spinning black hole binary, how much can we increase the
Kerr parameter of the final hole by setting the binary in an eccentric orbit?
We showed in Fig.~\ref{fig:PointParticle} that, according to point-particle
extrapolations, the final Kerr parameter should increase monotonically with
eccentricity.  Despite the unclear meaning of the eccentricity parameter, it
is reasonable to assume that the maximum increase in the final Kerr parameter
for non-spinning binaries should be $\Delta j_{\rm fin}^{\rm max}=j_{\rm
  fin}(e_p=1)-j_{\rm fin}(e_p=0)\simeq 0.750-0.663\simeq 0.087$. As the motion
turns into a plunge, $j_{\rm fin}$ should again decrease. From the results
listed in Table \ref{tab: models} we can read out the ``true'', general
relativistic prediction for the maximal spin increase induced by orbital
eccentricity: $\Delta j_{\rm fin}^{\rm max}=0.724-0.691\simeq 0.033$. The
level of agreement with simple extrapolations from perturbation theory is,
once again, really surprising.

Finally, it is interesting to ask if we can use the
eccentricity-induced increase of the final Kerr parameter to violate
the cosmic censorship conjecture, by producing a final object with
$j_{\rm fin}>1$. In principle this could be possible: even for zero
eccentricity, if we consider maximally spinning black holes
($j_1=j_2=1$) with spins aligned to the orbital angular momentum,
according to the point-particle extrapolation the final spin
resulting from an equal-mass merger is $j_{\rm fin}=0.959$, which is
very close to the Kerr bound \cite{Buonanno:2007sv}.  In fact,
extrapolation of point-particle results suggests that cosmic
censorship will {\it not} be violated. From the previous discussion
it should be clear that the most ``dangerous'' situation corresponds
to maximal eccentricities ($e_p=1$). In this case, the extrapolation
of point-particle results can be performed analytically. The
expression for general initial spins and general mass ratio is too
cumbersome to display here. If we consider equal-mass binaries, we
get
\be
j_{\rm fin}=
\frac{j_1+j_2}{4}+\frac{3+\sqrt{9-4(j_1+j_2)}}{8}\,.
\ee
From this equation we are led to conjecture that {\em Kerr black holes
  produced as the result of a merger always satisfy the Kerr bound}. The bound
is only saturated when $e_p=1$ and $j_1=j_2=1$.

If instead we drop the equal-mass assumption, but we consider
$j_i=j_1=j_2$ (and again $e_p=1$), we get the following expression
for the final spin:
\be
j_{\rm fin}=
j_i+2\eta\left[(1-j_i-\eta)+\sqrt{(1-2\eta)(1-j_i)+\eta^2}\right]\,.
\ee
This suggests that in the $j_i\rightarrow 1$ limit, $j_{\rm fin}\rightarrow 1$
for {\it any} mass ratio.

Early numerical simulations of spinning black hole binary coalescence support
our conjecture: quasi-circular black hole binaries with spins aligned to the
orbital angular momentum tend to radiate excess angular momentum by undergoing
a longer inspiral, or ``orbital hang-up,'' and they never seem to produce
naked singularities
\cite{Campanelli:2006uy,Rezzolla:2007xa,Rezzolla:2007rd,Marronetti:2007wz}.
It will be interesting to see if this conclusion holds true, as we predict,
when the black hole binary is both spinning {\it and} highly eccentric.

%\clearpage

\section{\label{QNMs} The ringdown waveform}

It is well known that frequencies and quality factors of ringdown waveforms
encode information about the properties of the Kerr black hole produced as a
result of the merger (see \cite{Berti:2005ys} and references therein). Here we
use the matrix pencil method to estimate time variations in the frequencies
and quality factors of the dominant multipole $(\ell=2,~m=2)$ as we reduce the
orbital angular momentum of the orbit. This and other methods have been
discussed at length in Refs.~\cite{Berti:2007dg,Berti:2007fi}, and we will not
repeat that discussion here.  An introduction to Prony methods to estimate
parameters of complex exponentials in noise, in the context of black hole
perturbation theory, can be found in Ref.~\cite{Berti:2007dg}. To remove
boundary reflection noise from our simulations, we use the procedure described
in \cite{Berti:2007fi} for quasi-circular inspiral simulations.

\begin{figure}[ht]
\centering
\includegraphics[height=7.2cm,angle=-90]{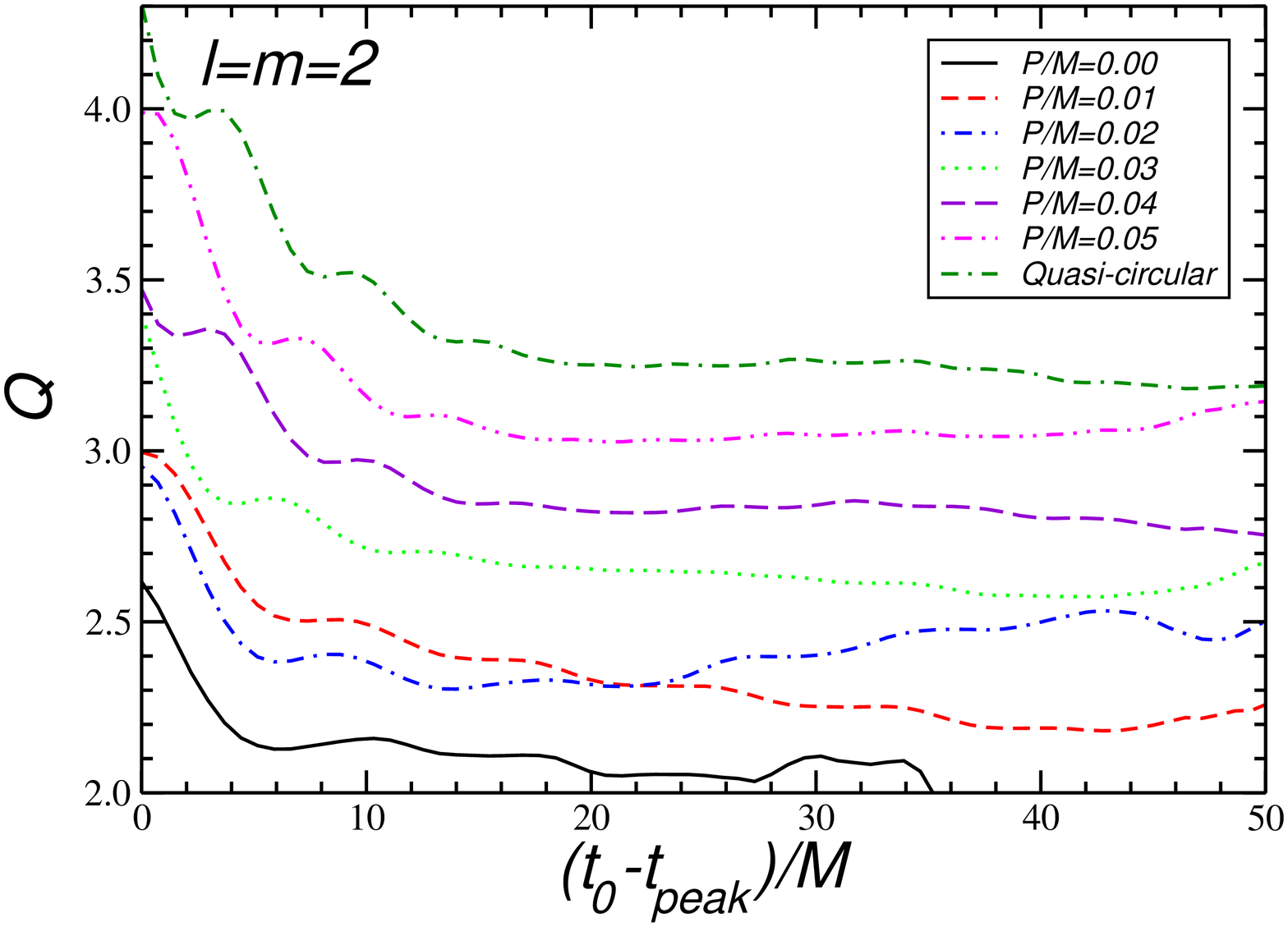}
\includegraphics[height=7.2cm,angle=-90]{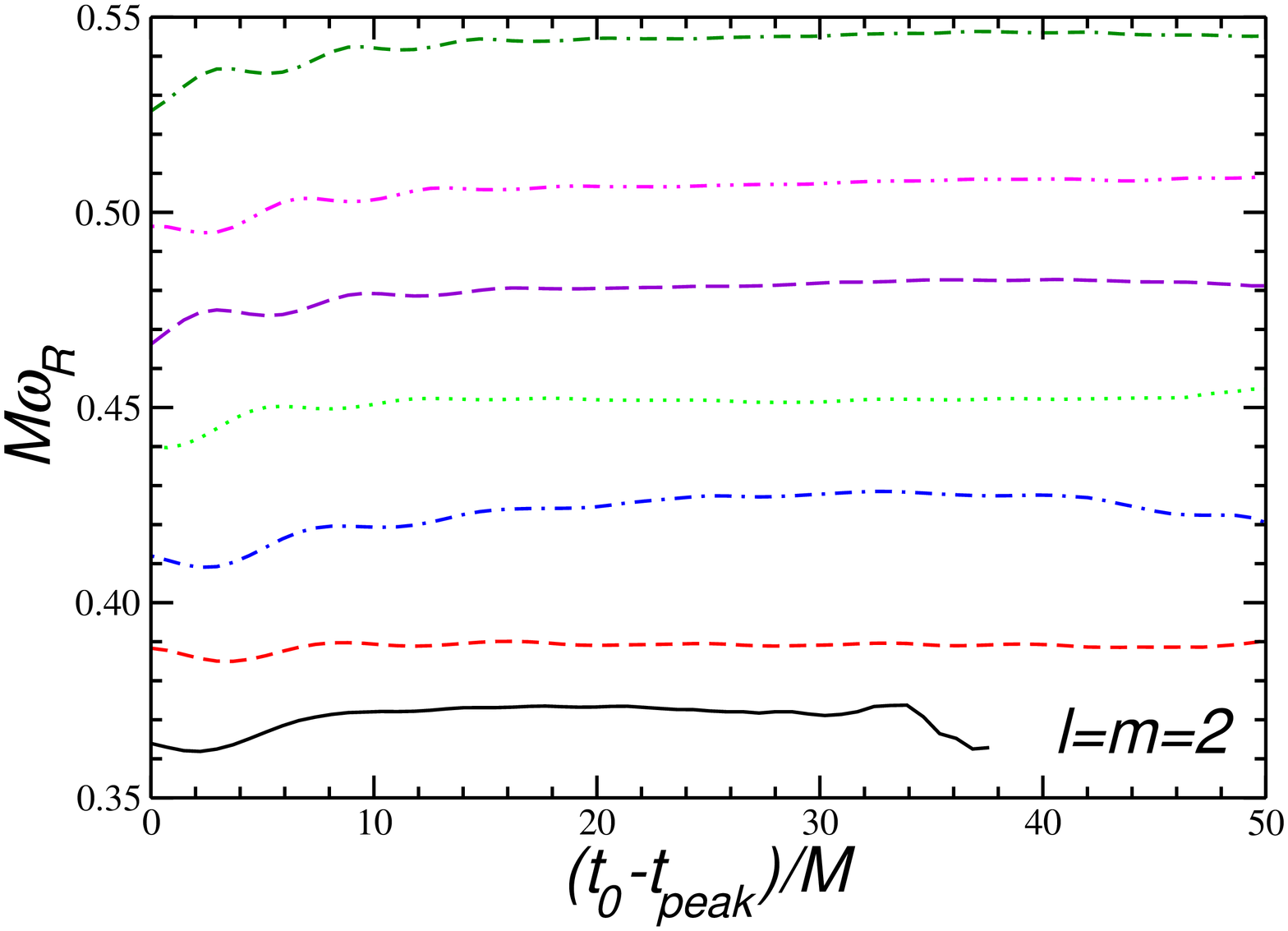}
\caption{Prony estimates for the quality factor $Q$ (left) and dimensionless
  oscillation frequency $M\omega_R$ (right) of the fundamental mode with
  $\ell=m=2$. Different linestyles (see the legend in the left panel) refer to
  different models in sequence 1.}
\label{fig:PronyQw}
\end{figure}

In Fig.~\ref{fig:PronyQw} we show the time dependence of the estimated
(dimensionless) QNM frequency $M\omega_R$ and of the quality factor $Q$ for
the dominant multipolar component ($\ell=m=2$) of the radiation.  All
quantities are plotted as functions of the starting time for the fit $t_0$, as
measured from the time $t_{\rm peak}$ corresponding to the maximum in the
modulus of the waveform's amplitude $|\psi_{22}(t_{\rm peak})|$. We see the
same features we observed in \cite{Berti:2007dg,Berti:2007fi}. After a short
transient for $t_0-t_{\rm peak}\lesssim 10M$, frequencies and quality factors
become roughly constant, except at very late times, where the ringdown signal
is very small and noise contaminates the estimates.  Frequency estimates are
usually better than quality factor estimates, and temporal variations in both
quantities (which are probably induced by noise in the simulations) become
more significant for the shorter, plunging configurations with low values of
$P/M$.

\begin{figure}[ht]
\centering
\includegraphics[height=7.2cm,angle=-90]{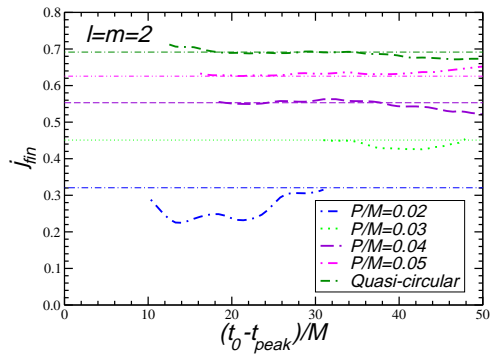}
\caption{Angular momentum estimate $j_{\rm fin}$ from Prony fits. Horizontal
  lines are estimates of the final Kerr parameter from energy balance
  arguments (see Table \ref{tab: models}).}
\label{fig:PronyJ}
\end{figure}

The quality factor is a dimensionless quantity, and as such it can only depend
on the dimensionless Kerr parameter of the final hole: $Q=Q(j_{\rm fin})$. The
Kerr parameter can then be obtained by inverting this relation. We perform the
inversion by interpolating QNM tables \cite{Berti:2005ys}. Results are shown
in Fig.~\ref{fig:PronyJ}. There we discard all points for which we get
unphysical black hole parameters (eg. points for which the final estimated
black hole mass $M_{\rm fin}>M_{\rm ADM}$).

Unfortunately, estimates of the final angular momentum become less reliable as
$L/M^2\to 0$.  The reason is apparent by looking, for example, at the right
panel of Fig.~5 in Ref.~\cite{Berti:2005ys}. As the spin of the final hole
decreases, and in particular for $j_{\rm fin}\lesssim 0.5$, relatively large
variations in the Kerr parameter produce mild variations in the $\ell=m=2$
quality factor.  Conversely, small oscillations in $Q$ produce large
variations in $j_{\rm fin}$.  From Fig.~\ref{fig:PronyQw} we see that
noise-induced variations in $Q$ actually {\em increase} for near head-on
collisions. Therefore, QNM based estimates of the Kerr parameter are sensibly
affected by noise when $j_{\rm fin}\lesssim 0.5$. The situation becomes even
worse when we consider higher multipoles. However, it is clear from
Fig.~\ref{fig:PronyJ} that QNM estimates are consistent (within the errors,
which increase as $P/M\to 0$) with estimates obtained from angular momentum
balance arguments.

%\clearpage

\subsection{Energy-Maximized Orthogonal Projection}

As discussed thoroughly in \cite{Berti:2007fi}, there is no unique definition
of the ringdown starting time, and correspondingly there is no unique way to
define the fraction of energy radiated into ringdown for a given waveform. A
useful measure of the ringdown energy is provided by the ``Energy Maximized
Orthogonal Projection'' (EMOP). In \cite{Berti:2007fi} the EMOP was shown to
yield reasonable estimates of the ringdown radiation from non-spinning,
unequal-mass black hole binaries, and therefore we shall perform the same
analysis on the present waveforms.

\begin{table}[ht]
  \centering \caption{\label{tab:emop} EMOP data for $l=2$. Numbers
    separated by a comma correspond to the $+$ and $\times$ polarizations,
    respectively. The fraction of the total energy in ringdown is strongly
    dependent on the eccentricity. We find that, independently of the run, the
    value of $t_{\rm EMOP}$ for a given polarization is generally at a fixed
    position relative to the maximum of the waveform's amplitude $t_{\rm
      peak}$. We measure this relative difference by $\Delta t_{\rm
      EMOP}\equiv t_{\rm peak}-t_{\rm EMOP}$. Angular brackets denote an
    average over the two polarizations.}
\begin{tabular}{ccccccc}
\hline  \hline
$P/M$  &$\frac{E_{\rm EMOP}}{E_{\rm rad}}$ &$\frac{\langle{t}_{\rm
EMOP}\rangle}{M_{\rm ADM}}$ & $\frac{\Delta t_{\rm EMOP}}{M_{\rm
ADM}}$ &$\frac{\langle{\Delta t_{\rm EMOP}}\rangle}{M_{\rm ADM}}$
&$\frac{\langle{10^2 E}_{\rm EMOP}\rangle}{M_{\rm ADM}}$  \\
\hline
0.14  & $0.42,0.41$    &  $230.5$     & $9.1,6.1$     &7.6    &0.0164\\
0.1383  & $0.40,0.41$    &  $312.2$     & $4.9,8.4$     &6.6     &0.0149\\
0.13  & $0.43,0.40$    &  $230.2$     & $8.1,4.6$     &6.3     &0.0158\\
0.12  & $0.42,0.43$    &  $222.2$     & $5.2,8.7$     &6.9     &0.0181\\
0.11  & $0.44,0.42$    &  $214.2$     & $9.3,5.8$     &7.6     &0.0171 \\
0.10  & $0.46,0.45$    &  $198.2$     & $10.0,13.6$   &11.8    &0.0222 \\
0.09  & $0.48,0.52$    &  $190.0$     & $10.5,7.5$    &9.0     &0.0207 \\
0.08  & $0.56,0.58$    &  $183.8$     & $6.1,9.6$     &7.9     &0.0183  \\
0.07  & $0.61,0.58$    &  $179.8$     & $9.2,5.7$     &7.4     &0.0147 \\
0.06  & $0.61,0.65$    &  $177.0$     & $5.0,8.0$     &6.5     &0.0111 \\
0.05  & $0.68,0.65$    &  $175.0$     & $9.0,5.0$     &7.0     &0.00775 \\
0.04  & $0.69,0.66$    &  $174.0$     & $6.0,10.0$    &8.0     &0.00483 \\
0.03  & $0.66,0.75$    &  $176.5$     & $4.3,8.3$     &6.3     &0.00234 \\
0.02  & $0.72,0.75$    &  $175.2$     & $11.2,6.8$    &9.0     &0.00131 \\
0.01  & $0.73,0.75$    &  $178.0$     & $9.2,6.2$     &7.7     &0.000617 \\
0  & $0.77,-$       &  $177.5$     & $8.2,-$       &8.2     &0.000475 \\
\hline \hline
\end{tabular}
\end{table}

Our results are summarized in Table \ref{tab:emop} for sequence 1. Results for
sequence 2 and 3 are very similar, and we do not show them here.
Ref.~\cite{Berti:2007fi} found that ringdown accounts for about $42\%$ of the
total energy emitted in the merger of non-spinning black hole binaries in
quasi-circular orbits, i.e., $E_{\rm EMOP}/E_{\rm rad}\simeq 42\%$.  In the
second column of Table \ref{tab:emop} we show the corresponding results for
the present simulations. As a general trend, higher eccentricity means that
the black holes spend less time orbiting each other before merger.  Therefore
one expects that the relative ringdown content, as measured by $E_{\rm
  EMOP}/E_{\rm rad}$, should increase with eccentricity, since less energy is
radiated during the pre-merger phase. This trend is clearly visible in Table
\ref{tab:emop}. On the other hand, the ringdown starting time $\langle{t}_{\rm
  EMOP}\rangle$, as measured from $t_{\rm peak}$, is insensitive to the
eccentricity of the run. This is consistent with the idea that ringdown is
associated with the formation of a deformed common apparent horizon, and it
should not depend on the details of the process leading to the formation of
the horizon.  Finally, in the last entry of Table \ref{tab:emop} we show the
fraction of the total ADM mass radiated in ringdown waves. This number is
important for estimates of ringdown detectability with both Earth-based GW
detectors and LISA \cite{Berti:2005ys,Berti:2007zu}. A polynomial fit as a
function of the dimensionless orbital angular momentum yields
\begin{equation}
\frac{\langle{E}_{\rm EMOP}\rangle}{M_{\rm ADM}}
=
10^{-2}\left[
0.11-
1.55\left(\frac{L}{M^2}\right)+
4.22\left(\frac{L}{M^2}\right)^2
\right]\,,
\end{equation}
where (as in Table \ref{tab:emop}) angular brackets denote an average over the
two polarizations.

%%%%%%%%%%%%%%%%%%%%%%%%%%%%%%%%%%%%%%%%%%%%%%%%%%%%%%%%%%%%%%%%%%%%%%%%%%%%%%%
\section{Conclusions and outlook}

We have presented a study of the gravitational waveforms produced by sequences
of equal-mass, non-spinning black hole binaries. For each sequence, the
binding energy of the system is kept constant and the orbital angular momentum
is progressively reduced to zero, producing orbits of increasing eccentricity
and eventually a head-on collision. We find that the motion transitions from
inspiral to plunge when the orbital angular momentum $L=L_{\rm crit}\simeq
0.8M$. For $L<L_{\rm crit}$ the binary always completes less than $\sim 1$
orbit, and PN estimates of the orbital eccentricity are no longer meaningful.
As the initial momentum of the holes $P/M\to 0$ the polarization quickly
becomes linear, rather than circular, and the radiated energy drops (roughly)
exponentially. For equal-mass, non-spinning binaries, orbits with $L\simeq
L_{\rm crit}$ produce the largest dimensionless Kerr parameter for the
remnant, $j_{\rm fin}=J/M^2\simeq 0.724\pm0.13$ (to be compared with the Kerr
parameter $j_{\rm fin}\simeq 0.69$ resulting from quasi-circular inspirals).
These results are quite insensitive to the initial separation of the holes,
and they can be understood using extrapolations from black hole perturbation
theory.
% It will be interesting to improve the accuracy of our predictions by
%using higher resolution, larger extraction radii and especially larger initial
%orbital separations.
Larger separations, as used in sequence 3, will be necessary to perform
accurate comparisons with PN predictions for the evolution of eccentric
binaries (see \cite{Hinder2008}). Such an analysis is beyond the scope of
this work, however, and a corresponding analysis of the sequence 3 data will be
presented elsewhere.

For equal masses, we found that eccentric binary mergers with $L\simeq L_{\rm
  crit}$ maximize the Kerr parameter of the final black hole. Using arguments
based on point-particle extrapolations, we proposed two conjectures: (1)
$j_{\rm fin}\simeq 0.724\pm 0.13$
should be close to the largest Kerr parameter that can be
produced by {\it any} non-spinning black hole binary merger, independently of
the binary's mass ratio; and (2) even if we consider maximally spinning holes
with spins aligned with the orbital angular momentum, orbital eccentricity
should {\it not} lead to violations of the cosmic censorship conjecture. It
will be interesting to check these conjectures using numerical simulations of
eccentric binaries with unequal masses and non-zero spins.

%%%%%%%%%%%%%%%%%%%%%%%%%%%%%%%%%%%%%%%%%%%%%%%%%%%%%%%%%%%%%%%%%%%%%%%%%%%%%%%
\section*{Acknowledgments}
%%%%%%%%%%%%%%%%%%%%%%%%%%%%%%%%%%%%%%%%%%%%%%%%%%%%%%%%%%%%%%%%%%%%%%%%%%%%%%%
We thank Clifford Will, Achamveedu Gopakumar, Christian
K\"onigsd\"orffer and Gerhard Sch\"afer for discussions about
eccentricity in the PN formalism, and Luciano Rezzolla and Michele
Vallisneri for comments on the manuscript.  This work was supported
in part by DFG grant SFB/Transregio~7 ``Gravitational Wave
Astronomy''.  E.B.'s research was supported by an appointment to the
NASA Postdoctoral Program at the Jet Propulsion Laboratory,
California Institute of Technology, administered by Oak Ridge
Associated Universities through a contract with NASA; by the
National Science Foundation, under grant number PHY 03-53180; and by
NASA, under grant number NNG06GI60 to Washington University.  V.C.'s
work was partially funded by Funda\c c\~ao para a Ci\^encia e
Tecnologia (FCT) –- Portugal through projects PTDC/FIS/64175/2006
and POCI/FP/81915/2007. We thank the DEISA Consortium (co-funded by
the EU, FP6 project 508830), for support within the DEISA Extreme
Computing Initiative (\url{www.deisa.org}). Computations were
performed at LRZ Munich and the Doppler and Kepler clusters at the
Institute of Theoretical Physics of the University of Jena. We are
grateful to the Center for Computational Physics (CFC) in Coimbra
for granting us access to the Milipeia cluster.

\appendix

\section{\label{sec:appendixsep} Orbits of point particles in the Kerr geometry and the angular momentum at the separatrix}

Here we consider a test body of mass $m_p$ moving in a Kerr spacetime, and we
neglect radiation reaction (therefore we focus on purely geodesic motion). For
simplicity, we only consider equatorial orbits.  The equations of motion are
given in Boyer-Lindquist coordinates by
\begin{eqnarray}
r^2 \frac{dr}{d\tau}&=& \pm (V_r)^{1/2} \;, \label{eom1}
\\
\nonumber \\
r^2 \frac{d\phi}{d\tau}&=& V_{\rm \phi} \equiv -(Mj_{\rm
fin}\tilde{E} -\tilde{L}) + \frac{Mj_{\rm fin} T}{\Delta} \;,
\label{eom2}
\\
\nonumber \\
r^2 \frac{dt}{d\tau}&=& V_{\rm t} \equiv -Mj_{\rm fin}(Mj_{\rm
fin}\tilde{E} -\tilde{L}) + \frac{(r^2 +M^2j_{\rm fin}^2)T}{\Delta}
\;, \label{eom3}
\\
\nonumber \\
\theta(\tau)&=& \pi/2 \;,
\end{eqnarray}
where
\begin{math}
T=E(r^2+M^2j_{\rm fin}^2) -\tilde{L}Mj_{\rm fin} , \ \ V_r= T^2
-\Delta(r^2 + (\tilde{L} -Mj_{\rm fin}\tilde{E})^2 ), \ \ \Delta=
r^2 -2Mr + M^2j_{\rm fin}^2 .
\end{math}
The two constants of motion
\begin{equation}
\tilde{E}\equiv E/{m_p}\,,\quad \tilde{L}\equiv L/m_p\,,
\end{equation}
denote the orbit's specific energy and $z$-component of angular
momentum respectively. We have prograde (retrograde) orbits
according to whether $\tilde{L} > 0 ~(< 0)~$. Bound orbits require $
0<\tilde{E}<1 $. A general bound equatorial orbit can equivalently
be described either by the constants $\tilde{E}$ and $\tilde{L}$, or
by a semi-latus rectum $p$ and an eccentricity $e_p$ (with $ 0 \leq
e_p < 1$). The semi-latus rectum measures the size of the orbit, and
the eccentricity measures the degree of non-circularity. We define
these parameters in terms of the two turning points of the orbit
($r_p$ is the periastron and $r_{a}$ the apastron):
\begin{equation}
r_p = \frac{p}{1+e_p}, \ \ r_a = \frac{p}{1-e_p} \;. \label{tpoints}
\end{equation}
The specific angular momentum and energy can be computed by solving
the simultaneous equations for $\tilde{E}$ and $\tilde{L}$
\begin{eqnarray}
\tilde{E}&=& \left[ 1 -\left (\frac{M}{p}\right ) (1-e_p^2) \left \{
1 -\frac{x^2}{p^2}(1-e_p^2) \right  \} \right ]^{1/2} \,,
\label{energy}\\
x^2&=& \frac{ -N(p,e_p) \mp \Delta^{1/2}_x(p,e_p)
}{2F(p,e_p)}\,,\label{x2}
\end{eqnarray}
where the upper (lower) sign corresponds to prograde (retrograde)
motion, and we have defined
\begin{eqnarray} x&=& \tilde{L} -Mj_{\rm fin}
\tilde{E}\,.
\end{eqnarray}
The explicit form of the functions $F$, $N$ and $\Delta_x$ is
\begin{eqnarray}
F(p,e_p)&=& \frac{1}{p^3}[p^3 -2M(3 +e_p^2)p^2 + M^2(3 +e_p^2)^2p
-4M^3j_{\rm fin}^2(1-e_p^2)^2 ],
\\
\nonumber \\
N(p,e_p)&=& \frac{2}{p}[-Mp^2 + M^2((3+e_p^2) -j_{\rm fin}^2)p
-M^3j_{\rm fin}^2(1 +3e_p^2) ],
\\
\nonumber \\
\Delta_x(p,e_p)&=& N^2 -4FM^2(j_{\rm fin}^2 -p/M)^2.  \label{xcoeff}
\end{eqnarray}

In general there are three different radial turning points (i.e., roots of
$V_r=0$): the periastron at $r=r_p$, the apastron at $r=r_{a}$ and a third
root $r=r_3$, defining a forbidden region. The case with $r_p=r_3$ corresponds
to a marginally stable orbit: once at the periastron, the particle will enter
into a circular orbit of radius $r_p$. This location has been called the
innermost stable bound orbit \cite{Glampedakis:2002ya}, and it is the
generalization of the ISCO for general eccentric orbits.  At this stage the
orbit has become unstable, so that a slight inwards ``push'' will drive the
particle to plunge into the black hole. Therefore, stable bound orbits should
satisfy $r_3 < r_p$.  This translates to the inequality
\begin{equation}
x^2(1+e_p)(3-e_p) < p^2 \;. \label{sepax}
\end{equation}
The boundary curve defined by
\begin{equation}
p_s^2=x^2(1+e_p)(3-e_p) \label{sepa1}
\end{equation}
defines the separatrix of bound orbits.

Equations (\ref{energy})-(\ref{sepa1}) allow one to compute the energy and
orbital angular momentum at the separatrix. Strictly speaking, these equations
are only valid for point particles. The substitution
\begin{equation}
\tilde{L}\rightarrow \frac{L}{\eta M}
\end{equation}
should provide a reasonable extrapolation to the general finite mass ratio
case.

%\section*{References}
%\bibliographystyle{unsrt}
%\bibliography{paper-fix}

\end{document}